\documentclass[preprint]{aastex631}

\usepackage{ulem}

\usepackage{amsmath}

\submitjournal{AJ}

\shorttitle{DSA-110 FRB polarization sample}
\shortauthors{Sherman et al.}

\begin{document}

\title{Deep Synoptic Array Science: Polarimetry of 25 New Fast Radio Bursts Provides Insights into their Origins}

\correspondingauthor{Myles B. Sherman}
\email{msherman@caltech.edu}

\author{Myles B. Sherman}
\affiliation{Cahill Center for Astronomy and Astrophysics, MC 249-17 California Institute of Technology, Pasadena CA 91125, USA.}

\author{Liam Connor}
\affiliation{Cahill Center for Astronomy and Astrophysics, MC 249-17 California Institute of Technology, Pasadena CA 91125, USA.}

\author{Vikram Ravi}
\affiliation{Cahill Center for Astronomy and Astrophysics, MC 249-17 California Institute of Technology, Pasadena CA 91125, USA.}
\affiliation{Owens Valley Radio Observatory, California Institute of Technology, Big Pine CA 93513, USA.}

\author{Casey Law}
\affiliation{Cahill Center for Astronomy and Astrophysics, MC 249-17 California Institute of Technology, Pasadena CA 91125, USA.}
\affiliation{Owens Valley Radio Observatory, California Institute of Technology, Big Pine CA 93513, USA.}

\author{Ge Chen}
\affiliation{Cahill Center for Astronomy and Astrophysics, MC 249-17 California Institute of Technology, Pasadena CA 91125, USA.}

\author{Morgan Catha}
\affiliation{Owens Valley Radio Observatory, California Institute of Technology, Big Pine CA 93513, USA.}

\author{Jakob T. Faber}
\affiliation{Cahill Center for Astronomy and Astrophysics, MC 249-17 California Institute of Technology, Pasadena CA 91125, USA.}

\author{Gregg Hallinan}
\affiliation{Cahill Center for Astronomy and Astrophysics, MC 249-17 California Institute of Technology, Pasadena CA 91125, USA.}
\affiliation{Owens Valley Radio Observatory, California Institute of Technology, Big Pine CA 93513, USA.}

\author{Charlie Harnach}
\affiliation{Owens Valley Radio Observatory, California Institute of Technology, Big Pine CA 93513, USA.}

\author{Greg Hellbourg}
\affiliation{Cahill Center for Astronomy and Astrophysics, MC 249-17 California Institute of Technology, Pasadena CA 91125, USA.}
\affiliation{Owens Valley Radio Observatory, California Institute of Technology, Big Pine CA 93513, USA.}

\author{Rick Hobbs}
\affiliation{Owens Valley Radio Observatory, California Institute of Technology, Big Pine CA 93513, USA.}

\author{David Hodge}
\affiliation{Cahill Center for Astronomy and Astrophysics, MC 249-17 California Institute of Technology, Pasadena CA 91125, USA.}

\author{Mark Hodges}
\affiliation{Owens Valley Radio Observatory, California Institute of Technology, Big Pine CA 93513, USA.}

\author{James W. Lamb}
\affiliation{Owens Valley Radio Observatory, California Institute of Technology, Big Pine CA 93513, USA.}

\author{Paul Rasmussen}
\affiliation{Owens Valley Radio Observatory, California Institute of Technology, Big Pine CA 93513, USA.}

\author{Kritti Sharma}
\affiliation{Cahill Center for Astronomy and Astrophysics, MC 249-17 California Institute of Technology, Pasadena CA 91125, USA.}

\author{Jun Shi}
\affiliation{Cahill Center for Astronomy and Astrophysics, MC 249-17 California Institute of Technology, Pasadena CA 91125, USA.}

\author{Dana Simard}
\affiliation{Cahill Center for Astronomy and Astrophysics, MC 249-17 California Institute of Technology, Pasadena CA 91125, USA.}

\author{Jean Somalwar}
\affiliation{Cahill Center for Astronomy and Astrophysics, MC 249-17 California Institute of Technology, Pasadena CA 91125, USA.}

\author{Reynier Squillace}
\affiliation{Department of Astronomy, University of Virginia, 530 McCormick Rd, Charlottesville, VA 22904, USA.}

\author{Sander Weinreb}
\affiliation{Cahill Center for Astronomy and Astrophysics, MC 249-17 California Institute of Technology, Pasadena CA 91125, USA.}

\author{David P. Woody}
\affiliation{Owens Valley Radio Observatory, California Institute of Technology, Big Pine CA 93513, USA.}

\author{Nitika Yadlapalli}
\affiliation{Cahill Center for Astronomy and Astrophysics, MC 249-17 California Institute of Technology, Pasadena CA 91125, USA.}

\collaboration{200}{(The Deep Synoptic Array team)}


\begin{abstract}

We report on a full-polarization analysis of the first 25 as yet non-repeating FRBs detected at 1.4 GHz by the 110-antenna Deep Synoptic Array (DSA-110) during commissioning observations. We present details of the data-reduction, calibration, and analysis procedures developed for this novel instrument. Faraday rotation measures (RMs) are searched between $\pm10^6$\,rad\,m$^{-2}$ and detected for 20 FRBs with magnitudes ranging from $4-4670$\,rad\,m$^{-2}$. $15/25$ FRBs are consistent with 100\% polarization, 10 of which have high ($\ge70\%$) linear-polarization fractions and 2 of which have high ($\ge30\%$) circular-polarization fractions. Our results disfavor multipath RM scattering as a dominant depolarization mechanism. Polarization-state and possible RM variations are observed in the four FRBs with multiple sub-components. We combine the DSA-110 sample with polarimetry of previously published FRBs, and compare the polarization properties of FRB sub-populations and FRBs with Galactic pulsars. Although FRB polarization fractions are typically higher than those of Galactic pulsars, and cover a wider range than those of pulsar single pulses, they resemble those of the youngest (characteristic ages $<10^{5}$\,yr) pulsars. Our results support a scenario wherein FRB emission is intrinsically highly linearly polarized, and propagation effects can result in conversion to circular polarization and depolarization. Young pulsar emission and magnetospheric-propagation geometries may form a useful analogy for the origin of FRB polarization.
 
\end{abstract}


\section{Introduction}\label{introduction}

Fast Radio Bursts (FRBs) are energetic, millisecond-duration radio transients of extragalactic origin. There are now over 500 confirmed FRBs recorded in the Transient Name Server\footnote{\url{https://www.wis-tns.org/}} (TNS) catalog \citep{2020TNSAN.160....1P}. Several lines of evidence point to neutron-star progenitors for FRBs, although models with, e.g., black-hole engines remain viable \citep[e.g.,][]{zhang2022physics}. Their exact emission mechanism is unknown, though multiple theories have been proposed. Some models predict emission from maser processes within the immediate plasma environments of neutron stars \citep{lyubarsky2014model,2019MNRAS.485.4091M}. Others associate them with magnetospheric processes nearer to neutron-star surfaces, for example driven by crustal oscillations or magnetic reconnection \citep{beloborodov2021can,beniamini2020periodicity,2016ApJ...823L..28G}. Although both repeating and as yet one-off bursts can be described by these, other cataclysmic theories imply triggering in compact-object mergers and/or collapse that only produce one-off events \citep{2014A&A...562A.137F,2013PASJ...65L..12T}. It is possible that multiple mechanisms are responsible for the FRB phenomenon, possibly even from the same source \citep[e.g.][]{petroff2017fast,mckinven2022spectropolarimetry,2023arXiv230615505K,2023MNRAS.526.2039H}. Perhaps relatedly, a remarkable diversity of progenitor host-galaxy systems is observed, including globular clusters \citep{kirsten2022glob}, dwarf galaxies \citep{chatterjee2017r1,bhandari2023dwarf}, a variety of locations within galaxies on the star forming main sequence of galaxies \citep{mannings2021high,gordon2023}, and massive quiescent galaxies \citep{sharma2023massive,2023arXiv230703344L}. 

The polarization of FRBs can be used to identify viable emission theories. Coupled with the observed high brightness temperatures $T_B \gg 10^{12}\,$K, high linear polarization implies a coherent emission process such as curvature radiation, inverse-Compton scattering, or synchrotron-maser emission \citep[e.g.,][]{li2021hxmt,wang2022polarization,qu2023polarization}. Smooth variations in the polarization position angle (PPA) over the burst durations can be associated with moving emission sites, in analogy with the rotating vector model applied to pulsar emission, or with propagation effects within or immediately outside the source magnetosphere \citep{1969ApL.....3..225R,ravi16pol,luo2020diverse,cho2020spectropolarimetric,2023MNRAS.tmp.2169O}. Rapid jumps in PPA, observed in several pulsars, can indicate emission in different modes or from different locations within magnetospheres \citep[e.g.,][]{mth75,scr+84}. Circular polarization and depolarization from 100\% linearly polarized bursts typically require mode conversion or absorption during propagation, multiple emission sites, or Faraday depolarization in a strongly magnetized, possibly inhomogeneous medium \citep{beniamini2022faraday,qu2023polarization}. Faraday rotation measure (RM) variations observed in some FRBs also hint toward dynamic local environments of FRB progenitors \citep[e.g.][]{anna2023magnetic,mckinven2023revealing,li2022highly}. Each mechanism places stringent physical constraints on the sources and near-source environments.  

38 FRB sources have published polarization and/or RM data (see Appendix~\ref{app_lit} for a detailed compilation).  This small sample has limited the characterization of standard properties, although a few traits appear commonplace. Specifically, most repeating and non-repeating FRBs possess high linear polarization fractions ($\ge 70\%$) \citep[e.g.,][]{mckinven2023revealing, hilmarsson2021rotation, nimmo2021highly}. Some FRBs have exhibited non-negligible circular polarization \citep[$\ge 30\%$, e.g., FRB\,20190611B][]{day2020}. Repeaters appear non-uniform in their properties: some have stable polarization states like FRB\,20190208A while others vary from burst to burst, like FRB\,20121102A or FRB\,20190520B \citep[][]{mckinven2023revealing, feng2022circular, anna2023magnetic}. One-off bursts, for example FRB\,20181112A, can have complex time-domain morphologies and polarization properties that evolve over the burst duration \citep{cho2020spectropolarimetric}. Polarization properties so far do not appear to be a viable means of distinguishing repeaters from non-repeaters \citep[e.g.,][]{zhang2022physics}. Some repeaters exhibit smoothly time-varying RMs on week to month timescales, including FRB\,20121102A, FRB\,20190520B, and FRB\,20180301 \citep[][]{hilmarsson2021polarization, kumar2023spectro, anna2023magnetic}. Strong burst to burst variability in linear- and circular-polarization fractions, as well as frequency-dependent oscillations in these quantities, were observed from the repeating source FRB\,20201124A \citep{xu2022complex}. These were interpreted as originating during propagation through a strongly magnetized, evolving plasma environment within an AU of the FRB source. Frequency-dependent depolarization has also been observed in multiple FRBs \citep{feng2022frequency,mckinven2023revealing}. This may be described by the stochastic RM model, in which spatial RM variations within a scattering screen depolarize FRB emission as it propagates along multiple pathways \citep[][]{melrose1998stochastic, beniamini2022faraday, 2022ApJ...928L..16Y}. No systematic studies of the polarization properties of as yet non-repeating FRBs have been published. 

During the commissioning phase of the 110-antenna Deep Synoptic Array (DSA-110), an FRB survey at declination $\delta = 71^\circ$ at 1.4 GHz observing frequency was conducted. 
This paper reports on polarimetry of the first 25 FRBs detected with the DSA-110. The remainder of the paper will proceed as follows. Section~\ref{observations} will  describe the DSA-110 observations, polarization pipeline and methods for RM synthesis. We present linear- and circular-polarization fractions for 25 FRBs and RMs for 20 FRBs. Section~\ref{discussion} presents and discusses the physical implications of polarization subgroups based on the DSA-110 sample. In Section~\ref{analysis} we combine the DSA-110 sample with previously published FRBs to compare the polarization properties of repeaters and non-repeaters, and of FRBs with Galactic pulsars. The latter two sections include significant interpretation and discussion. We conclude in Section~\ref{conclusions}. We encourage the reader to peruse the extensive appendices for a detailed description of methodology in data reduction, analysis and interpretation, as well as a compilation of our data and literature data on FRB and pulsar polarization. We use a selection of symbols and common abbreviations summarized in Table~\ref{table:symtable}. Calibrated full-polarization filterbank data and derived parameters in RMTable format\footnote{\url{https://github.com/CIRADA-Tools/RMTable}} are made available on the CaltechDATA public repository\footnote{\url{https://data.caltech.edu/}}. These can be accessed through the DSA-110 archive \citep{Morrell_DSA-110_Event_Archive_2022}.\footnote{\url{https://code.deepsynoptic.org/dsa110-archive/}} Additional data will be made available upon request to the corresponding author.

\begin{deluxetable*}{  c | c | c }[ht]
\tabletypesize{\scriptsize}
\caption{Table of Frequently Used Symbols and Abbreviations}
\label{table:symtable}
\tablehead{\begin{tabular}{c} \textbf{Symbol/} \\ \textbf{Abbreviation} \end{tabular} & \textbf{Units} & \textbf{Definition}}
\startdata 
PA, $\chi$  & $^\circ$ & Observed linear-polarization position angle without correcting Faraday rotation \\
\hline
PPA, $\chi_0$  & $^\circ$ & Intrinsic linear polarization position angle after correcting for all Faraday rotation \\
\hline
$L/I$ & $\%$ & Frequency- and time-averaged debiased fractional linear polarization \\
\hline
$|V|/I$ & $\%$ & Frequency- and time-averaged fractional absolute value circular polarization \\
\hline
$V/I$ & $\%$ & Frequency and time-averaged fractional signed circular polarization \\
\hline
RVM &  & Rotating vector model \\
\hline
AP &  & Average profiles of Galactic pulsars \\
\hline
SP & & Single pulses from Galactic pulsars \\
\hline
S/N & & Signal to noise ratio \\
\hline
\enddata
\end{deluxetable*}

\section{Observations and data analysis}\label{observations}

\begin{figure*}[t]
\begin{center}
\includegraphics[width=0.49\textwidth]{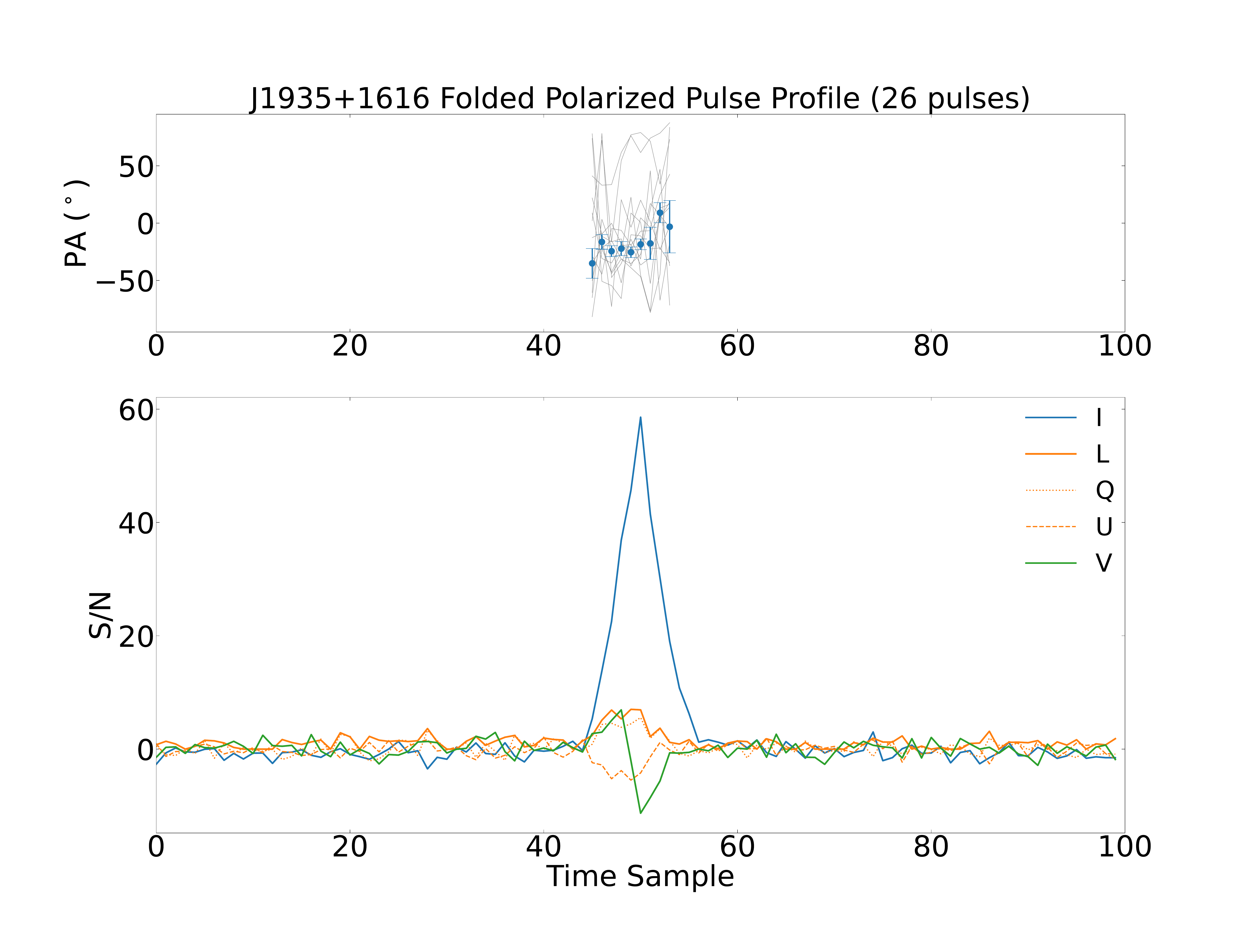} 
\includegraphics[width=0.49\textwidth]{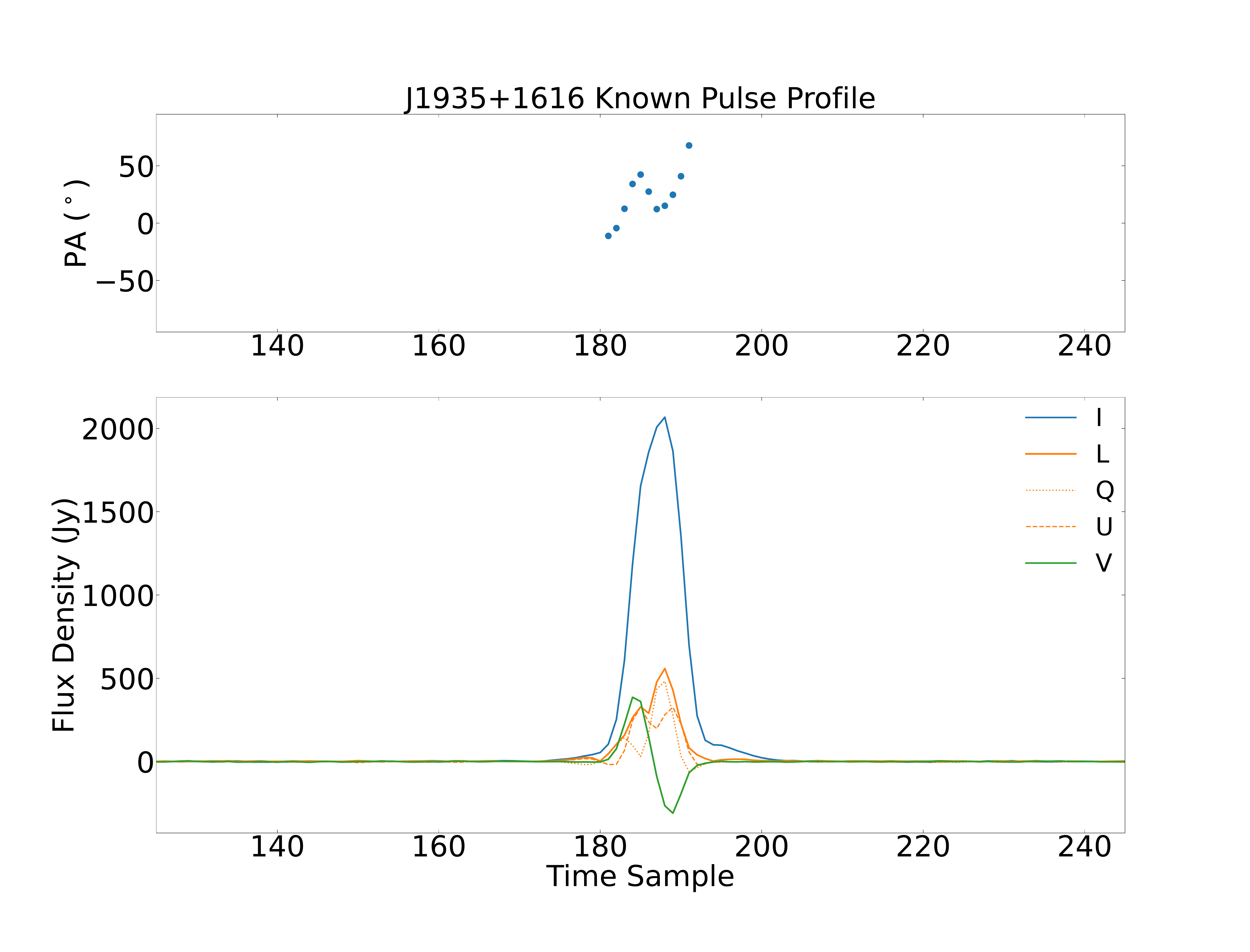} 
\caption{Comparison of the pulse profile of PSR\,J$1935+1616$ observed with the DSA-110 (left) to an archival profile from the European Pulsar Network (right). The observed profile was formed by centroiding and stacking 26 single-pulse detections automatically triggered and processed identically to DSA-110 FRBs. The top panels show the PA of the average profile (blue), and for the DSA-110 bursts, the PA of single pulses (grey). The bottom panels show the Stokes parameters and linear-polarization fraction ($L/I$) as labeled. Besides a PA offset of unknown origin, the profiles are otherwise similar.}
\label{fig:J1935}
\end{center}
\end{figure*}

The DSA-110 array is currently in its commissioning phase. During the observations reported here, spanning January--November 2022, the array consisted of 48 core antennas for FRB searching, and 15 outrigger antennas to assist in FRB localization. Each 4.65-m antenna is equipped with dual orthogonal linearly polarized receivers, operating between 1280--1530\,MHz. A linear basis using the IAU/IEEE handedness convention\footnote{The IAU/IEEE convention assumes that $V > 0$ is right-hand polarization, while $V < 0$ is left-hand polarization \citep{van2010psrchive}.} is  used for all polarimetry analysis discussed herein. The antennas are movable in elevation only, and are always pointed at the meridian. All FRBs we present here were detected during a survey at a declination of $71.6^{\circ}$. Ravi et al. in prep. will describe the instrument in detail, and some details have been included in previous publications on DSA-110 FRBs \citep[e.g.,][]{ravi2022deep,2023arXiv230703344L}.  None of the FRBs presented here have been observed to repeat by the DSA-110 in 150 hours of observing time during 2022 \citep{2023arXiv230703344L}, and we consider them as non-repeaters for the purposes of our analysis. Of the sample of 25 FRBs presented here, arcsecond-scale localizations, and secure host-galaxy association and characterization have been summarized for 11 events by \citet{2023arXiv230703344L}. As this paper is focused on the intrinsic polarization properties of the DSA-110 and other FRBs, we do not utilize information (e.g., redshifts) derived from host-galaxy data. A discussion of the RMs of the localized DSA-110 and other FRBs are presented in the companion paper, \citet{sherman2023deep}.

\begin{figure*}
    \centering
    \includegraphics[width=\textwidth]{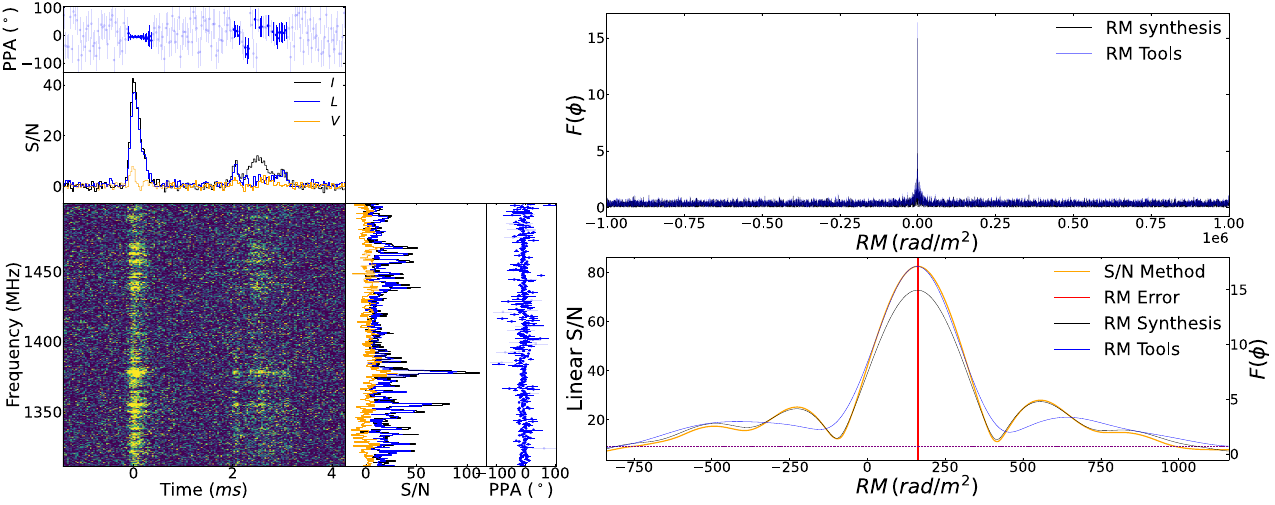}
    \caption{Summary of polarization and RM data on FRB\,20220207C (\textbf{\textit{Zach}}). Identical plots on all FRBs presented herein are shown in Appendix~\ref{app_data}. \textit{Left:} From top, we show the PPA (with $1\sigma$ errorbars) measured after correcting for RM, and frequency averaged time series of Stokes $I$, $L$, and $V$. The PPA is highlighted in bins with linear-polarization S/N $>3\sigma$. The dynamic spectrum is shown at center, to the right of which are shown optimally summed spectra (Appendix~\ref{app_weights}) and the spectrally resolved PPA with $>3\sigma$ bins highlighted. \textit{Right:} The top panel shows the RM spectra evaluated over the full range of sensitivity using two different methods (Appendix~\ref{app_RMderivation}), and the bottom panel shows a detailed analysis of the S/N of the peak.}
    \label{fig:zach_summ}
\end{figure*}

The calibration and analysis of the full-polarization properties of the 25 FRBs described herein were carried out using standard techniques. The chief novelty of the present analysis is that it is developed from the ground up using a custom implementation on a new telescope. Novel methods were developed for certain aspects of the analysis, such as the estimation of accurate RMs and the corresponding uncertainties. A summary of the calibration and analysis techniques can be found in Appendix~\ref{app_data_model}, and Appendices \ref{background} to \ref{app_RMderivation} justify them in detail. 

In this section we show a single demonstration of the functioning of the polarization calibration and analysis pipeline. Figure~\ref{fig:J1935} displays the calibrated, summed, polarized profile formed from 26 single-pulse detections of bright pulses from the PSR\,J1935$+$1616 (B1933). These pulses were detected and processed in exactly the same way as the DSA-110 FRBs. A comparison is made to the reported average profile from the European Pulsar Network (EPN) database of pulsar profiles \citep{gould1998multifrequency}.\footnote{\url{http://www.epta.eu.org/epndb/}} Good agreement between the DSA-110 polarization profile and the EPN profile is found: note particularly the swing in Stokes V. A time offset in the linear polarization peak is observed, which is attributed to low signal-to-noise of the DSA-110 profile, and pulse summation based on simple centroiding and stacking. An offset in PA is evident between the profiles: our method for deriving PAs is given in Appendix~\ref{app_derivation}, and it is unclear exactly what method was used in calibrating the EPN profile. Additional details on verification of the polarization pipeline with J1935+1616 are given in Appendix~\ref{app_J1935}.

We present polarization and RM results on all FRBs in Appendix~\ref{app_data}. Figures~\ref{fig:FRBStokes1} to \ref{fig:FRBStokes4} show calibrated and RM-corrected burst temporal profiles and dynamic spectra. Figures~\ref{fig:FRBRMs1} to \ref{fig:FRBRMs3} show RM spectra for all FRBs with detectable RMs. Table~\ref{table:PolTable} summarizes the various polarization fractions and RMs for each FRB. Throughout this paper we sometimes refer to DSA-110 FRBs by their internal natural names (e.g., FRB\,20220207C is referred to as \textbf{\textit{Zach}}). Correspondence between the natural and formal names is noted in Table~\ref{table:PolTable}. Example plots for FRB\,20220207C are shown in Figure~\ref{fig:zach_summ}.

\begin{rotatetable}
\movetableright=1mm
\begin{deluxetable*}{  ccccccccccc }
\tabletypesize{\scriptsize}
\tablewidth{0pt} 
\tablecaption{DSA-110 FRB Sample Polarization Properties\label{table:PolTable}}
\tablehead{\colhead{\textbf{FRB Names}} &         \colhead{\textbf{R.A. ($^\circ$)}} & \colhead{\textbf{Decl. ($^\circ$)}} &  \colhead{\textbf{$\mathbf{n_t}$}} &  \colhead{\textbf{$\mathbf{n_f}$}} & \colhead{\textbf{S/N}} &   \colhead{\textbf{$\mathbf{L/I}$}} &              \colhead{\textbf{$\mathbf{|V|/I}$}} &                \colhead{\textbf{ $\mathbf{V/I}$}} & \colhead{RM (rad\,m$^{-2}$)} & \colhead{DM (pc\,cm$^{-3}$)}}
\startdata 
FRB\,20220121B / ``Clare" &  $85.66^\circ$ &       $64.39^\circ$ &      8 &    128 & 13.36 &   $70.5 \pm 20.6 \%$ & $51.0 \pm 12.7 \%$ & $-51.0 \pm 12.7 \%$ & $-4.6 \pm 15.5$ & $313.5\pm0.2$\\
\hline
FRB\,20220204A / ``Fen" & $278.27^\circ$ &       $72.58^\circ$ &      6 &     32 & 17.68 &   $62.8 \pm 15.3 \%$ &    $7.5 \pm 7.7 \%$ &    $-6.7 \pm 7.7 \%$ & $-11.0 \pm 11.7$  & $612.2^{+0.005}_{-0.049}$\\
\hline
FRB\,20220207C / ``Zach" &  $310.2^\circ$ &       $72.88^\circ$ &      1 &     16 & 72.57 &     $74.9 \pm 4.5 \%$ &  $14.8 \pm 1.7 \%$ &   $11.1 \pm 1.7 \%$ & $162.48 \pm 0.04$  & $262.3\pm0.01$\\
\hline
FRB\,20220208A / ``Ishita" & $322.58^\circ$ &       $70.04^\circ$ &     32 &    256 & 10.70 & $114.3 \pm 29.1 \%$ & $11.4 \pm 15.5 \%$ & $-11.2 \pm 15.5 \%$ & $-23.3 \pm 9.7$  & $437.0^\pm0.6$\\
\hline
FRB\,20220307B / ``Alex" & $350.88^\circ$ &       $72.19^\circ$ &      1 &    128 & 15.72&  $69.8 \pm 16.0 \%$ &  $18.9 \pm 8.0 \%$ &  $-10.3 \pm 8.0 \%$ & $-947.2 \pm 12.3$  & $499.15\pm0.03$\\
\hline
FRB\,20220310F / ``Whitney" & $134.72^\circ$ &       $73.49^\circ$ &      1 &     32 & 98.32 &     $60.6 \pm 4.7 \%$ &   $16.6 \pm 1.6 \%$ &    $-7.4 \pm 1.6 \%$ & $11.4 \pm 0.2$ & $462.15\pm0.004$ \\
\hline
FRB\,20220319D / ``Mark" &  $32.18^\circ$ &       $71.04^\circ$ &      1 &     32 & 56.37 &     $16.1 \pm 4.0 \%$ &   $4.0 \pm 2.0 \%$ &   $-3.0 \pm 2.0 \%$ & $59.9 \pm 14.3$ & $110.95\pm0.01$ \\
\hline
FRB\,20220330D / ``Erdos" & $163.75^\circ$ &       $70.35^\circ$ &     40 &     64 & 17.98 &   $73.2 \pm 17.7 \%$ &  $15.8 \pm 9.0 \%$ &    $3.1 \pm 9.0 \%$ & $-122.2 \pm 4.5$  & $468.1^{+0.8}_{-0.9}$\\
\hline
FRB\,20220418A / ``Quincy" & $219.11^\circ$ &        $70.1^\circ$ &      2 &    128 & 16.84 &    $64.1 \pm 13.3 \%$ & $21.4 \pm 7.7 \%$ &   $2.7 \pm 7.7 \%$ & $6.1 \pm 7.5$  & $623.45\pm0.01$\\
\hline
FRB\,20220424E / ``Davina" &  $19.51^\circ$ &        $71.8^\circ$ &      2 &     64 & 14.60 & $59.9\pm16.2\%$ & $15.8\pm7.4 \%$   &  $15.8\pm7.4 \%$  &  $126.9\pm13.4$ & $863.48^{+0.022}_{-0.036}$\\
\hline
FRB\,20220506D / ``Oran" & $318.05^\circ$ &       $72.83^\circ$ &      1 &     32 & 24.17 &  $95.3 \pm 10.9 \%$ &  $31.4 \pm 4.6 \%$ &   $31.4 \pm 4.6 \%$ & $-32.4 \pm 3.6$  & $396.93\pm0.01$\\
\hline
FRB\,20220509G / ``Jackie" & $282.67^\circ$ &       $70.24^\circ$ &      1 &     32 & 40.32 &     $98.5 \pm 11.1 \%$ &   $6.0 \pm 2.8 \%$ &    $5.4 \pm 2.8 \%$ & $-109.0 \pm 1.2$  & $269.5\pm0.007$\\
\hline
FRB\,20220726A / ``Gertrude" &  $73.95^\circ$ &       $69.93^\circ$ &      2 &     64 & 18.25 &    $93.5 \pm 16.7 \%$ &  $13.8 \pm 6.2 \%$ &   $-3.7 \pm 6.2 \%$ & $499.8 \pm 7.2$  & $686.55\pm0.01$\\
\hline
FRB\,20220801A / ``Augustine" &  $54.96^\circ$ &       $70.18^\circ$ &      4 &     64 & 11.77 &    $30.1 \pm 18.0 \%$ & $35.8 \pm 13.7 \%$ &   $-1.3 \pm 13.7 \%$ & $-$ & $412.6\pm0.04$ \\
\hline
FRB\,20220825A / ``Ansel" & $311.98^\circ$ &       $72.59^\circ$ &      1 &     32 & 32.80 &   $53.7 \pm 11.4 \%$ &   $9.0 \pm 5.9 \%$ &   $-7.4 \pm 5.9 \%$ & $750.2 \pm 6.7$  & $651.2\pm0.02$\\
\hline
FRB\,20220831A / ``Ada" &  $338.7^\circ$ &       $70.54^\circ$ &      6 &     64 & 21.28 &   $84.6 \pm 12.8 \%$ &  $16.8 \pm 5.3 \%$ &    $1.7 \pm 5.3 \%$ & $772.1 \pm 7.2$  & $1146.25\pm0.2$\\
\hline
FRB\,20220914A / ``Elektra" & $282.06^\circ$ &       $73.34^\circ$ &      1 &     64 & 17.43 &   $16.3 \pm 14.0 \%$ &  $20.5 \pm 11.9 \%$ &   $11.9 \pm 11.9 \%$ & $-$  & $631.05\pm0.02$\\
\hline
FRB\,20220920A / ``Etienne" & $240.26^\circ$ &       $70.92^\circ$ &      2 &    128 & 18.88 &   $89.0 \pm 16.2 \%$ &  $19.7 \pm 8.1 \%$ &  $-14.9 \pm 8.1 \%$ & $-830.3 \pm 8.3$  & $315.0\pm0.06$\\
\hline
FRB\,20220926A / ``Celeste" & $328.73^\circ$ &       $72.86^\circ$ &      2 &    128 & 18.16 &      $17.5 \pm 21.0 \%$ &  $9.5 \pm 14.2 \%$ &   $2.0 \pm 14.2 \%$ & $-$ & $441.53\pm0.03$ \\
\hline
FRB\,20221002A / ``Arni" &  $281.0^\circ$ &        $71.6^\circ$ &     12 &    256 &  8.08 &    $26.8 \pm 22.7 \%$ & $19.7 \pm 18.1 \%$ & $-12.5 \pm 18.1 \%$ & $-$& $322.5\pm1.0$\\
\hline
FRB\,20221012A / ``Juan" $^\dagger$ &  $280.8^\circ$ &       $70.52^\circ$ &      8 &     32 &  9.33 &    $  63.0 \pm 21.4 \%$ & $25.7 \pm 11.0 \%$ &   $0.6 \pm 10.9 \%$ & $165.7 \pm 17.7$  & $441.2\pm0.26$\\
\hline
FRB\,20221027A / ``Koyaanisqatsi" &  $130.9^\circ$ &        $71.6^\circ$ &      1 &     32 & 21.82 &      $8.3 \pm 7.0 \%$ &  $18.4 \pm 5.2 \%$ &   $-2.2 \pm 5.2 \%$ & $-$  & $452.5^{\ddagger}$\\
\hline
FRB\,20221029A / ``Mifanshan" &  $143.8^\circ$ &        $71.6^\circ$ &      4 &    128 & 17.12 &   $85.7 \pm 16.1 \%$ &   $15.6 \pm 7.0 \%$ &     $6.8 \pm 7.0 \%$ & $-155.8 \pm 9.7$  & $1391.05^{\ddagger}$\\
\hline
FRB\,20221101B / ``Nina" &  $342.0^\circ$ &        $71.5^\circ$ &      4 &     32 & 14.21 &   $99.3 \pm 19.3 \%$ & $24.5 \pm 8.7 \%$ &  $-19.4 \pm 8.7 \%$ & $-32.2 \pm 9.1$ & $490.7^{\ddagger}$ \\
\hline
FRB\,20221101A / ``Ayo" &   $75.6^\circ$ &        $71.6^\circ$ &      2 &     64 & 19.27 &      $55.3 \pm 11.3 \%$ &  $11.6 \pm 6.2 \%$ &   $-5.5 \pm 6.2 \%$ & $4670.4 \pm 10.8$ & $1476.5^{\ddagger}$ \\
\hline
\enddata
\bigskip

\textbf{Notes:} $n_t$ and $n_f$ are the downsampling factors in time and frequency from the native DSA-110 resolution (see Appendix~\ref{app_data_model}) used to maximize S/N. The total, linear, absolute value of circular and signed circular polarization are given in columns 7, 8, 9, and 10. The right ascension and declination given are the ICRS frame coordinates of the beam in which the FRB was detected.

$^\dagger$ FRB\,20221012A's polarization is computed after derotating to the reported $\rm RM$. This detection was marginal, with low linear-polarization S/N below the $9\sigma$ threshold. However, this is taken as a reliable measurement given the clear peak the RM spectrum (Figure~\ref{fig:FRBRMs3}) and apparent coherence when compared to the RMSF.

$^{\ddagger}$Detailed burst structure analysis used to estimate DM errors has not yet been performed for these FRBs. See G. Chen et al., in preparation for details. 
\end{deluxetable*}
\end{rotatetable}

\section{Discussion of Individual Burst Polarization Properties}\label{discussion}

\subsection{Motivation for Polarization-based Classification}

The DSA-110 FRB sample spans a wide range of polarimetric and spectro-temporal properties. This is the largest uniform sample of FRB polarization, and motivates its use as a basis for a polarization-dependent classification scheme. The need for such classes stems from the ability of polarimetry to distinguish FRB emission and propagation models. The high brightness temperatures observed in FRBs require a coherent, highly polarized emission process \citep[e.g.][]{2014ApJ...785L..26L}. This could apply to pulsar-like `antenna' mechanisms driven by charged particle bunches within the magnetosphere or synchrotron `maser' emission via a shocked wind outside the magnetosphere \citep[e.g.][]{metzger2019fast,wang2022magnetospheric,wang2022polarization,zhang2022physics,Zhang_2022,qu2023polarization}. `Antenna' mechanisms appear more likely, since the observed circular polarization and de-polarization in FRBs require propagation through an extreme local plasma environment.
For example. Faraday Conversion (FC) and cyclotron absorption each result in circular polarization; their presence would represent evidence of a highly magnetized, dense plasma nearby the source \citep[e.g.][]{vedantham2019faraday,gruzinov2019conversion,suresh2019induced,kumar2022circularly,qu2023polarization}. Similarly, depolarization can indicate either multi-path scattering through a screen with non-uniform RM \citep[e.g.][]{melrose1998stochastic}, or the superposition of multiple emission modes \citep[e.g.][]{mmb23}; these can be investigated through multi-band, high time-resolution polarization studies of FRBs \citep[e.g.][]{hankins2003nanosecond,feng2022frequency,mckinven2023revealing}. A generalized classification scheme will help identify the characteristic FRB polarization properties to narrow the range of likely FRB theories. 

To date, the predominant FRB categorization method was offered by \citet{pleunis2021fast} based on observations of FRBs from the Canadian Hydrogen Intensity Mapping Experiment (CHIME). Their classes were based on the spectro-temporal morphology of FRBs detected in the 400-800 MHz band. However, we find these classes insufficient to describe the full DSA-110 sample's morphological properties. Furthermore, these classes do not include FRB polarization, for which we would like a straightforward and replicable classification framework. The DSA-110, whose polarization calibration scheme is verified in Appendices~\ref{app_stability}, \ref{app_leakage}, and \ref{app_J1935}, is well-equipped to define polarization sub-classes using the uniform sample of 25 FRBs presented here. However, we acknowledge that such a classification scheme would be limited by the DSA-110 sensitivity, and furthermore cannot be generalized beyond L-band detections. While we hope that these sub-groups will form the groundwork for a more general, polarization- and frequency-dependent classification scheme, we make no claims on its universality in the current iteration. Polarization properties are also difficult to distinguish from being intrinsic or the result of propagation effects. Therefore, while we identify classes based on total, linear, and circular polarization, we use them primarily to motivate a more detailed discussion of linear and circular polarization in the following section.

We define four unique polarization sub-groups which are applied in turn to the  total, linear, and circular polarization fractions of each DSA-110 FRB:

\begin{enumerate}
    \item \textbf{Consistent with 100\% Polarized}: The polarization fraction falls within $3\sigma$ of 100\% and beyond $3\sigma$ of 0\%.
    \item \textbf{Consistent with 0\% Polarized}: The polarization fraction falls within $3\sigma$ of 0\% and beyond $3\sigma$ of 100\%.
    \item \textbf{Intermediate}: The polarization fraction falls beyond $3\sigma$ of both 100\% and 0\%.
    \item \textbf{Unconstrained}: The polarization fraction falls within $3\sigma$ of both 100\% and 0\%. This is equivalent to stating the polarization is not well-defined.
\end{enumerate}

\noindent We then define two orthogonal sub-groups which generalize and place more quantitative limits on the morphological classes of \citet{pleunis2021fast}:

\begin{enumerate}
    \item \textbf{Single Component}: Bursts exhibit one component (on approx. millisecond timescales).
    \item \textbf{Multi-Component}: Bursts exhibit multiple components separated on millisecond timescales.
\end{enumerate}

\begin{figure*}
    \centering
    \includegraphics[width=\textwidth]{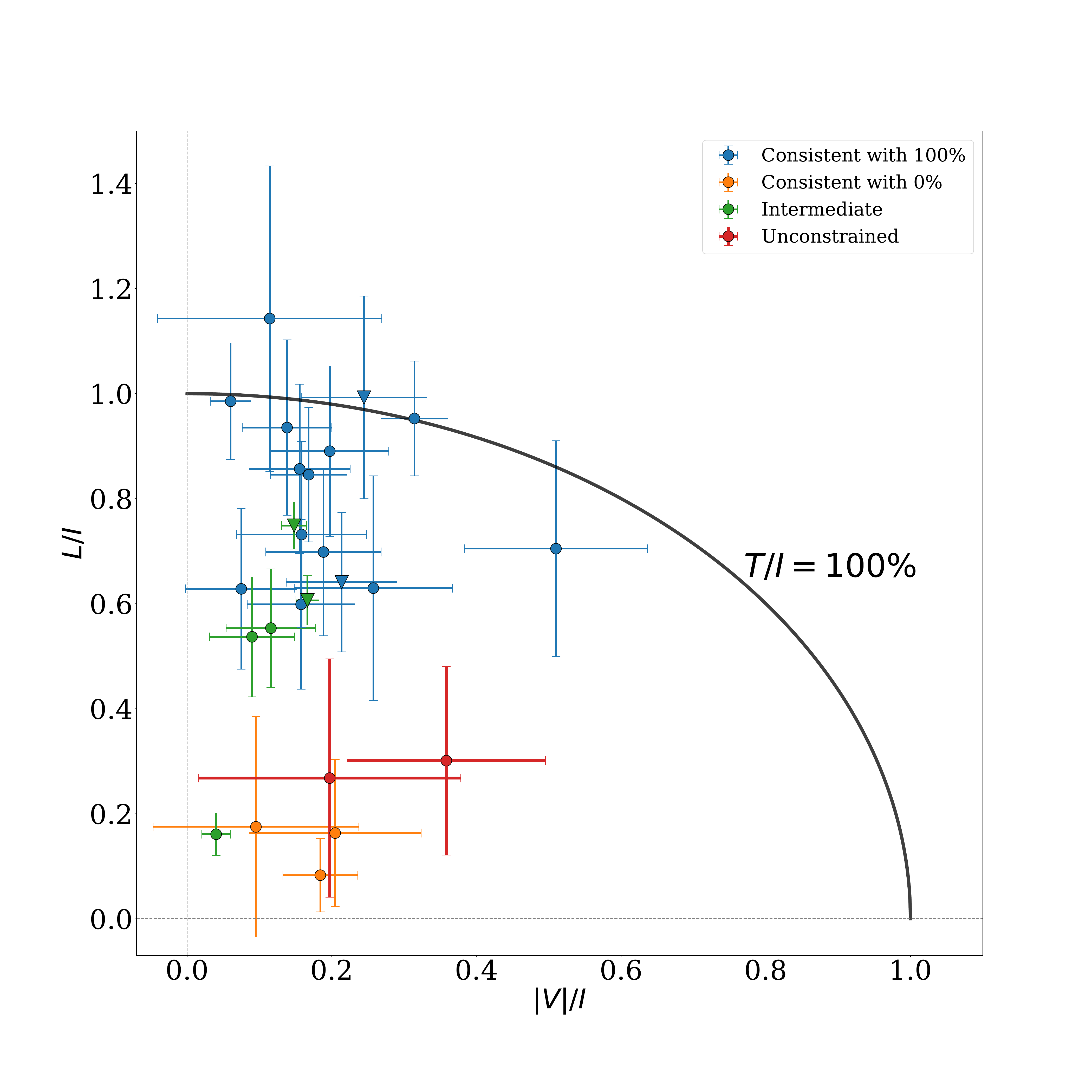}
    \caption{Linear and Absolute Value Circular Polarization for DSA-110 FRBs. Each FRB is color-coded by its polarization class as reported in Table~\ref{table:categories}. \textbf{Single Component} FRBs are shown as circles and \textbf{Multi-Component} FRBs are shown as triangles. The errorbars shown are $1\sigma$ uncertainties. The black line indicates where the total-polarization fraction $T/I = \sqrt{(|V|/I)^2 + (L/I)^2} = 100\%$. While all \textbf{Consistent with 100\%} (blue) and most\textbf{ Intermediate} (green) FRBs have $L/I > 50\%$, FRB\,20220121B and FRB\,20220506D also show significant circular polarization. FRB\,20220319D is the only FRB in either class with less than 50\% total polarization. All \textbf{Consistent with 0\%} FRBs (orange) have $L/I$ and $|V|/I$ below 25\%.} 
    \label{fig:newpolclasses}
\end{figure*}

\noindent In Table~\ref{table:categories}, we sort the DSA-110 sample among these sub-classes using total, linear, and circular polarization fractions. For the remainder of the paper, the sub-groups \textbf{Consistent with 100\% Polarized}, \textbf{Consistent with 0\% Polarized}, \textbf{Intermediate}, and \textbf{Unconstrained} will refer only to the total polarization-based sub-groups unless otherwise noted. In the next section, we explore the specific properties of FRBs in each class. Figure~\ref{fig:newpolclasses} shows $L/I$ and $|V|/I$ for the DSA-110 FRBs, grouped by their polarization class.


\begin{deluxetable*}{ c|ccccc }[ht]
\tabletypesize{\scriptsize}
\tablewidth{0pt} 
\caption{Polarization-based FRB subgroups from DSA-110 data.}
\tablehead{ \colhead{} & \colhead{Consistent with 100\%} & \colhead{Consistent with 0\%} & \colhead{Intermediate} & \colhead{ Unconstrained  }   }
\startdata
\begin{tabular}{@{}c@{}} Total Polarization\\ $(\sqrt{(L/I)^2 + (V/I)^2})$ \end{tabular} & \begin{tabular} {@{}c@{}} FRB\,20220121B (1) \\ FRB\,20220204A (1i) \\ FRB\,20220208A (1a)\\ FRB\,20220307B (1) \\ FRB\,20220330D (1i) \\  FRB\,20220424E (1i) \\ FRB\,20220506D (1ia) \\ FRB\,20220509G (1ia)\\ FRB\,20220726A (1a) \\ FRB\,20220831A  (1a) \\ FRB\,20220920A (1) \\ FRB\,20221012A  (1)$^\ddagger$\\ FRB\,20221029A  (1i) \\ \textbf{FRB\,20220418A (3a)} \\ \textbf{FRB\,20221101B (2)} \end{tabular}   
& \begin{tabular}{@{}c@{}} FRB\,20220914A (1i)\\ FRB\,20220926A (1)\\ FRB\,20221027A (1i)$^\dagger$ \end{tabular}    
& \begin{tabular}{@{}c@{}}  FRB\,20220319D (1i) \\ FRB\,20220825A (1ia) \\ FRB\,20221101A (1a) \\ \textbf{FRB\,20220207C  (2ia)} \\ \textbf{FRB\,20220310F (2ia)}\end{tabular} 
&  \begin{tabular}{@{}c@{}}  FRB\,20220801A (1ia)\\ FRB\,20221002A (1)\end{tabular} \\
\hline
\hline
\begin{tabular}{@{}c@{}} Linear Polarization\\ $(L/I)$ \end{tabular} & \begin{tabular} {@{}c@{}} FRB\,20220121B (1) \\ FRB\,20220204A (1i) \\ FRB\,20220208A (1a)\\ FRB\,20220307B (1) \\ FRB\,20220330D (1i) \\  FRB\,20220424E (1i) \\ FRB\,20220506D (1ia) \\ FRB\,20220509G (1ia)\\ FRB\,20220726A (1a) \\ FRB\,20220831A  (1a) \\ FRB\,20220920A (1) \\  FRB\,20221029A  (1i) \\ \textbf{FRB\,20220418A (3a)} \\ \textbf{FRB\,20221101B (2)} \end{tabular}   
& \begin{tabular}{@{}c@{}}  FRB\,20220914A (1i)\\ FRB\,20220926A (1)\\ FRB\,20221027A (1i)$^\dagger$ \\ FRB\,20220801A (1ia)\\ FRB\,20221002A (1) \end{tabular}    
& \begin{tabular}{@{}c@{}}   FRB\,20220319D (1i) \\ FRB\,20220825A (1ia) \\ FRB\,20221101A (1a) \\  \textbf{FRB\,20220207C  (2ia)} \\ \textbf{FRB\,20220310F (2ia)}\end{tabular} 
&  FRB\,20221012A  (1)$^\ddagger$ \\
\hline
\begin{tabular}{@{}c@{}} Circular Polarization\\ $(|V|/I)$ \end{tabular} & -- & \begin{tabular} {@{}c@{}} FRB\,20220121B (1) \\ FRB\,20220204A (1i) \\ FRB\,20220208A (1a)\\ FRB\,20220307B (1) \\ FRB\,20220330D (1i) \\  FRB\,20220424E (1i) \\  FRB\,20220509G (1ia)\\ FRB\,20220726A (1a) \\ FRB\,20220920A (1) \\ FRB\,20221012A  (1)$^\ddagger$\\ FRB\,20221029A  (1i) \\ FRB\,20220801A (1ia)\\ FRB\,20221002A (1) \\ FRB\,20220914A (1i)\\ FRB\,20220926A (1)\\  FRB\,20220319D (1i) \\ FRB\,20220825A (1ia) \\ FRB\,20221101A (1a) \\ \textbf{FRB\,20220418A (3a)} \\ \textbf{FRB\,20221101B (2)} \end{tabular}  & \begin{tabular} {@{}c@{}} FRB\,20220121B (1) \\ FRB\,20220506D (1ia) \\ FRB\,20221027A (1i)$^\dagger$ \\ \textbf{FRB\,20220207C  (2ia)} \\ \textbf{FRB\,20220310F (2ia)} \\ FRB\,20220831A  (1a) \\ \end{tabular} &  --  \\
\hline
\enddata
\bigskip
\textbf{Notes:} DSA-110 FRBs sorted into subgroups based on polarization significance as described in Section~\ref{discussion}. Each burst is labeled with the number of resolved components, an ``i" if the burst shows significant spectral striation, and an ``a" if the burst shows significant scattering (G. Chen et al., in prep.). \textbf{Multi-Component} FRBs are boldfaced, while \textbf{Single-Component} FRBs are in regular font.

$^\dagger$FRB\,20221027A is a unique FRB whose total polarization is \textbf{Consistent with 0\%}, but which has \textbf{Intermediate} circular polarization. This is attributed to its insignificant linear polarization which adds noise to the total polarization estimate.

$^\ddagger$FRB\,20221012A has a total polarization that is \textbf{Consistent with 100\%} and a circular polarization that is \textbf{Consistent with 0\%}, but its linear polarization is \textbf{Unconstrained}. From RM synthesis, we marginally detect an RM, although the peak in the RM spectrum has linear S/N below the $9\sigma$ threshold as shown in Figure~\ref{fig:FRBRMs3}. This suggests there is significant linear polarization which contributes to the total, though we cannot make more constraining statements within our current classification scheme.
\label{table:categories}
\end{deluxetable*}

\subsection{Single Component FRBs}

A significant majority $21/25$ of the DSA-110 sample are described by a single component on millisecond timescales, with most (13) of these being \textbf{Consistent with 100\%} polarization. 3 \textbf{Single Component} FRBs fall into each of the \textbf{Consistent with 0\%} and \textbf{Intermediate} classes, while two have \textbf{Unconstrained} polarization properties. In this section we explore the characteristic polarization properties of \textbf{Single Component} FRBs.

\subsubsection{Linear Polarization}


As shown in Figure~\ref{fig:newpolclasses}, all FRBs that are \textbf{Consistent with 100\%} have $L/I>50\%$, including 13 \textbf{Single Component} bursts; two-thirds of these (10) have $L/I > 70\%$. This is not uncommon among FRBs, as 36\% of previously published repeaters and non-repeaters with L-band (1.4\,GHz) polarization measurements also have $L/I > 50\%$ . This large fraction of highly linearly polarized FRBs supports coherent radiation models such as curvature radiation from coherent charge bunches \citep{wang2022polarization, qu2023polarization}. While \citet{feng2022frequency} demonstrate that repeaters become depolarized at lower frequencies, they associate this with stochastic RM scattering rather than intrinsic emission. Coherent radiation models are further supported by the relatively flat PPA's observed in this group. A $\chi^2$ goodness-of-fit test was conducted to compare the PPA in time bins with linear-polarization S/N $>3\sigma$ to the weighted average PPA value (see Appendix~\ref{app_chi}). None of the 13 \textbf{Single Component} FRBs consistent with 100\% polarization were found with significant variations ($p < 10\%$) in position angle. This may imply that there is a single dominant polarization mode in most FRBs, and resembles the behavior of young, high spin-down pulsars \citep{weltevrede2008profile,rmh10,mitra2016meterwavelength}. In Section~\ref{pol_section}, we explore possible connections between FRBs and young pulsars in more detail. Notably, of FRBs in the \textbf{Consistent with 100\%} and \textbf{Intermediate} sub-groups, only FRB\,20220319D has a linear polarization fraction below 20\%, a trait shared only by the 3 FRBs that are \textbf{Consistent with 0\%}.

\subsubsection{Circular Polarization}

Circular polarization appears to be less common than linear polarization among the \textbf{Single Component} sample. Only two FRBs that are \textbf{Consistent with 100\%} or \textbf{Intermediate} have $|V|/I>30\%$: FRB\,20220121B and FRB\,20220506D. Both are \textbf{Consistent with 100\%} and show no clear splitting of different signs of Stokes V. This is consistent with circular polarization originating through the propagation effects discussed above. A third FRB, FRB\,20220801A, also has a high $|V|/I$, but falls into the \textbf{Unconstrained} subgroup, making any evident polarization difficult to confirm. We do not observe any changes in the sign of Stokes V within FRB\,20220121B and FRB\,20220506D, as are often observed at the centers of the main components in radio-pulsar APs \citep{rr90}. Furthermore, while most \textbf{Single Component} FRBs appear to have negligible circular polarization (evident from both $|V|/I$ fractions and polarized profiles in Figure\textbf{s}~\ref{fig:FRBStokes1}-\ref{fig:FRBStokes4}), all four \textbf{Multi-Component} bursts appear to have significant circular polarization. We will explore this further in the next section.

The large fraction of FRBs that exhibit high linear polarization, as opposed to circular polarization, suggests that the intrinsic emission mechanism is often highly linearly polarized. Circular polarization and depolarization may then arise primarily from propagation effects. In this scenario, depolarization from the intrinsic mechanism due to unresolved micro or nano-second scale structure may be less significant than propagation effects, in contrast to pulsar radio emission \citep{mitra2015polarized,cordes1976pulsar,hankins2003nanosecond}. However, this claim cannot be confirmed without higher time resolution observations of FRBs.

The limited 187\,MHz bandwidth and 972\,kHz resolution\footnote{Here we refer to the limited resolution of the spectra in Figures~\ref{fig:FRBStokes1}-\ref{fig:FRBStokes4}, which were binned by at least $32\times30.4$\,kHz channels to maximize frequency structure.} of the DSA-110 spectra prevent detailed fits of frequency-dependent FC models. Instead, by assuming FC is significant in the region of interest, we can estimate the magnetic-field strength from frequency-averaged quantities as follows \citep{gruzinov2019conversion}. One can estimate the Faraday Conversion Measure (CM) from the root mean squared circular polarization of each FRB, assuming the circular emission arises from FC at a single magnetic field reversal:
\begin{equation}
    \sqrt{<V(\lambda^2)^2>} = \sqrt{2(e^{-({\rm CM}\lambda^2)^2/2} - e^{-({\rm CM}\lambda^2)^2})}
\end{equation}
\noindent The CM can be related to the RM and total magnetic field strength, $B$, by:
\begin{equation}
    {\rm CM} \approx B \sqrt{{\rm RM} \left ( \frac{8\pi^3 e^2}{m_e^3 c^4} \right )} \sim 10^{-2} \left ( \frac{B}{1\,{\rm G}} \right ) \left( \frac{\rm RM}{1\,{\rm rad\,m^{-2}}} \right )^{1/2},
\end{equation}
where $e$ is the electron charge, $m_e$ is the electron mass, and $c$ is the vacuum speed of light. For FRB\,20220121B and FRB\,20220506D, we measure root mean squared circular polarization $\sqrt{<V(\lambda^2)^2>} \approx 50\%$ and $40\%$, corresponding to conversion measures $\rm CM \approx 10\,$m$^{-2}$ in both cases. Given their RMs of $-4.60\pm15.48\,$rad\,m$^{-2}$ and $-32.38\pm3.60\,$rad\,m$^{-2}$, we estimate magnetic field strengths of $B \approx 510\,$G and $150$\,G, respectively. These high field strengths are not consistent with supernova-remnant and plerionic environments, including those of the FRBs associated with persistent radio sources \citep[e.g.,][]{vedantham2019faraday}. However, motivated by the likely FC observed during eclipses of the binary-system PSR\,B$1744-24$A \citep{2023Natur.618..484L}, the magnetized wind of a companion to the FRB progenitors is a reasonable possibility for the conversion site.

\subsubsection{Depolarization}


The DSA-110 sample reported here contains 3 \textbf{Single Component} FRBs that are \textbf{Consistent with 0\%} and 3 that have \textbf{Intermediate} polarization. $9/16$ FRBs in the published literature that have reported errorbars would also fall in these subgroups. Considering the circular polarization sub-groups, only FRB\,20221027A has an \textbf{Intermediate} $|V|/I$ ($>3\sigma$) among FRBs with \textbf{Intermediate} total polarization\footnote{With $|V|/I = 18.4\pm5.2\%$, FRB\,20221027A's linear polarization fraction is noise-dominated; therefore, its total polarization class (\textbf{Consistent with 0\%}) is likely erroneous due to small errors in noise statistics. We will consider it as an \textbf{Intermediate} FRB for this discussion given its significant circular polarization fraction.}. The lack of circular polarization in bursts with low total polarization implies that FC effects are less common, which one expects to depolarize the burst while producing residual circular polarization.

RM variations across a scattering screen are also unlikely to be the dominant depolarization mechanism for the DSA-110 FRBs. The variation in RM is captured by $\sigma_{\rm RM}$, the standard deviation of RM throughout the screen, and depolarization would be highly frequency dependent with the linear-polarization fraction of the form:

\begin{equation}\label{eq:fsigRM}
    f_{\sigma_{\rm RM}} = 1 - e^{-2\sigma_{\rm RM}^2\lambda^4}
\end{equation}

\noindent This model has been found to match observations of 8 repeating FRBs from 0.1--10\,GHz, with $\sigma_{\rm RM}$ values from $0.12$\,rad\,m$^{-2}$ for FRB\,20180916B to $218.9$\,rad\,m$^{-2}$ for FRB\,20190520B \citep{feng2022frequency,mckinven2023revealing}. However, for $RM < 1000$\,rad\,m$^{-2}$, averaging over the DSA-110 frequency band ($1.3-1.5$\,GHz) without RM correction is the dominant depolarization effect, which we quantify by:

\begin{equation}\label{eq:favg}
    f_{\rm avg} = 1 - \sqrt{<{\rm cos}({\rm RM}\lambda^2)>^2 + <{\rm sin}({\rm RM}\lambda^2)>^2}
\end{equation}

\noindent where $<>$ denotes the mean over frequency. Therefore, if we assume an FRB is intrinsically 100\% polarized, we can write the observed polarization fraction as $L/I \approx f_{\sigma_{\rm RM}}(\rm RM)f_{\rm avg}(\sigma_{\rm RM})$; note that channel bandwidth depolarization is negligible in this RM range (see Appendix~\ref{app_RMderivation}). \citet{feng2022frequency} (see Figure 4) derives an empirical, linear relation between $\sigma_{\rm RM}$ and RM, which we adopt to write Equation~\ref{eq:favg} in terms of $\sigma_{\rm RM}$. We apply this model to the average polarization of the three bursts with insufficient $L/I$ for RM measurements: FRB\,20220914A, FRB\,20220926A, and FRB\,20221027A. We find that $\sigma_{\rm RM} \approx 0.3-1.0$\,rad\,m$^{-2}$ could produce the observed polarization fractions for FRB\,20220914A and FRB\,20220926A, while $\sigma_{\rm RM} \approx 0.3-3.0$\,rad\,m$^{-2}$ could produce the polarization of FRB\,20221027A. From \citet{feng2022frequency}, these correspond to $RM \approx 200-900$\,rad\,m$^{-2}$, which should be detectable with the DSA-110 RM synthesis pipeline; however, no significant RM was found for these three FRBs. Low $S/N$ could prevent detection, but the low $\sigma_{\rm RM}$ suggests that RM synthesis would recover nearly 100\% linear polarization, making the maximized linear S/N in each burst $\sim20\sigma$ which exceeds the $9\sigma$ detection threshold (see Appendix~\ref{app_RMderivation}). 

We can perform a similar analysis for \textbf{Intermediate} FRBs, which do have RM measurements ranging from $10-4670$\,rad\,m$^{2}$. We find that $\sigma_{\rm RM}$ in the range of 10--20\,rad\,m$^{-2}$ would be required for these FRBs. Apart from FRB\,20221101A, this is around an order of magnitude higher than expected from \citet{feng2022frequency} for these RM magnitudes, and the $\ll1$\,ms scattering timescales of the DSA-110 FRBs (G. Chen et al., in prep.). Furthermore, assuming a common origin for the scattering and depolarization, the implied magnetic-field variations in the scattering screen are likely to be uncomfortably high, of order Gauss \citep[][Section S4]{2022ApJ...928L..16Y, feng2022frequency}. A potential exception to these arguments is the event FRB\,20220319D, as discussed in Appendix~\ref{app_mark}.

The significant fraction of FRBs in the DSA-110 sample that are \textbf{Consistent with 0\%} or \textbf{Intermediate} poses an interesting challenge to theory \citep{qu2023polarization}. The fact that depolarization has long been observed and puzzled over in pulsars, more significantly at high frequencies ($>1$\,GHz) than low, \citep[e.g.,][]{1973ApJ...179L...7M,manchester1971observations} should provide some hope that a common explanation may be found.

\subsection{Multi-Component FRBs}

$4/25$ of the DSA-110 sample exhibit multiple components on millisecond timescales\footnote{We note that sub-microsecond-scale structures have been previously observed in FRBs \citep[e.g.,][]{2021ApJ...919L...6M,nimmo2021highly,nimmo2022burst}, and indeed interesting structure within individual burst components is observed in some DSA-110 FRBs. We defer a high time resolution analysis of DSA-110 FRBs to future work.}. FRB\,20220418A and FRB\,20221101B are \textbf{Consistent with 100\% Polarization}, while FRB\,20220207C and FRB\,20220310F are \textbf{Intermediate}. In this section we explore the polarization properties of \textbf{Multi-Component} FRBs.

\subsubsection{Stable Polarization States}

Both FRB\,20220207C and FRB\,20220310F have two sub-components exhibiting significant linear polarization with $L/I > 50\%$. Significant circular polarization ($>3\sigma$) is also evident, with $|V|/I\approx15\%$. In FRB\,20220207C, the first component has a flat PPA while the second fluctuates, although a $\chi^2_\angle$ test finds this variation is not significant (see Appendix~\ref{app_chi} for details). In contrast, FRB\,20220310F displays a flat PPA in both components, separated by $\sim 10^{\circ}$. If both components originate from a fixed region in a rotating magnetosphere, this suggests negligible rotation over the approximately millisecond burst durations. A potentially better way to quantify a constraint on the rotation periods of progenitor objects is by assuming emission from coherent charge bunches, as motivated by this specific categorization \citep{wang2022polarization, qu2023polarization}. For this analysis we assume a `thin' bunch of charges, which aligns with the wide range of polarization properties observed in FRBs\footnote{For details, see Figures 2, 4, 5, and 6 of \citet{wang2022polarization}}. In this scenario, the burst duration must be much less than $(\gamma \Omega)^{-1}$, where $\gamma\sim100$ is the bunch Lorentz factor\footnote{We assume $\gamma \mathbf{\sim} 100$ based on the DSA-110 observing frequency 1.4 GHz and the expected curvature radius of magnetic field lines around a neutron star, $\rho \sim 10^{6-8}$\,cm. Using Equation 18 of \citet{qu2023polarization}, $58 \lesssim \gamma \lesssim 270$ are reasonable approximations for a coherent curvature radiation scenario.}, and $\Omega$ is the rotation angular frequency. Applying this condition for the $\sim 0.3\,$ms duration of the first FRB\,20220207C burst, we find required rotation periods $\gg30$\,ms. Similarly, the sub-bursts in FRB\,20220310F span $\sim0.2\,$ms each, for which we find required rotation periods $\gg20$\,ms under the same curvature radiation assumptions. These estimates are of particular interest since $30\,$ms roughly demarcates the boundary between millisecond pulsar (MSPs) and slow canonical pulsar (CPs) rotation periods \citep[e.g.][]{lorimer2008binary}. 

\subsubsection{Polarization Variation}

What is most striking about these bursts is that the different components appear to have different polarization properties, an effect that persists for non-repeating FRBs from at least the Australian Square Kilometre Array Pathfinder \citep[ASKAP;][]{cho2020spectropolarimetric,day2020}. In contrast, none of the 21 single-component FRBs had statistically significant PPA variability, suggesting negligible polarization evolution across the burst. The variations between the polarization properties of the sub-components of FRB\,20220207C, FRB\,20220310F, FRB\,20220418A, and FRB\,20221101B defy a unified description. FRB\,20220418A and FRB\,20220207C exhibit multiple components with different polarization properties. FRB\,20220207C's sub-components have very different profiles, although a $\chi^2_\angle$ test finds the PPA variation is not significant. Beyond this, the first sub-component is \textbf{Consistent with 100\%} linear polarization while the second has $\sim 50\%$ linear polarization and $\sim 20\%$ circular polarization. FRB\,20220418A is less constrained: while the first sub-component has high linear polarization, the second is \textbf{Unconstrained} and the third is \textbf{Consistent with 0\%} linear and circular polarization.

\begin{figure}
    \centering
    \includegraphics[width=\textwidth]{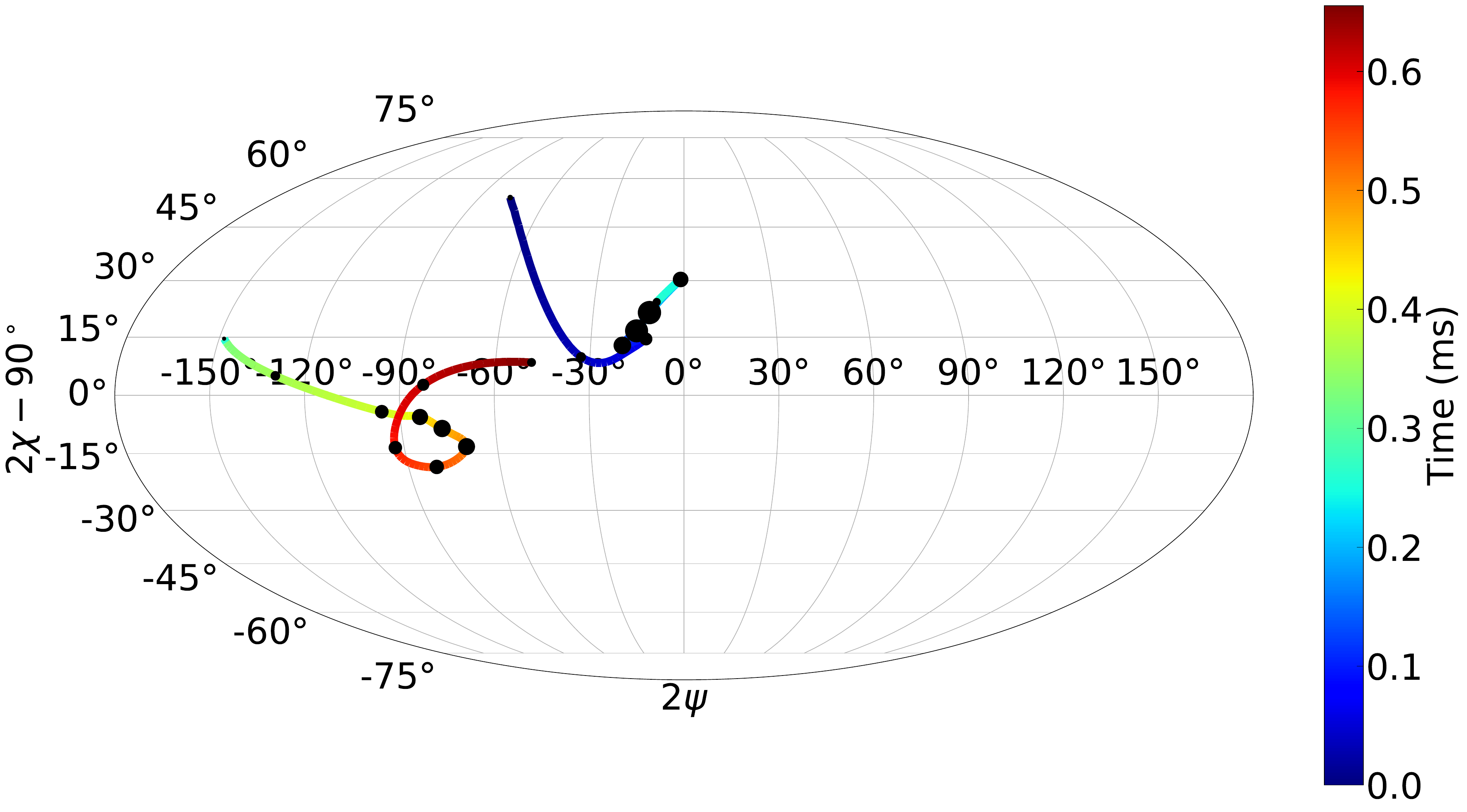}
    \caption{Projection of the polarization state of FRB\,20220310F onto the Poincar\'{e} sphere. The position angle ($\chi$) and longitude ($\psi$) are related to the Stokes parameters by $Q=\sin(2\psi)\sin(2\chi)$, $U=\sin(2\psi)\cos(2\chi)$, and $V=\cos(2\psi)$. Note the position angle axis has been shifted by $90^\circ$ for ease of viewing. Data are displayed in windows of eight and ten 32.768\,$\mu$s samples for the first and second components of the burst. }
    \label{fig:whitney}
\end{figure}

FRB\,20220310F's sub-components have more comparable properties, with $L/I$ of $54.8\pm6.0\%$ and $67.4\pm7.2\%$, but while the circular-polarization fractions have a similar magnitude, the signs are opposite: $V/I$ of $-16.7\pm1.3\%$ and $10.8\pm1.8\%$ respectively. Indeed, the polarization states of the two sub-components are individually well-defined, and the average locations of the two sub-components on the Poincar\'{e} sphere are almost orthogonal, with an angular separation of $71\pm2$\,degrees (Figure~\ref{fig:whitney}). These are reminiscent of emission in orthogonally polarized modes observed in several pulsars \citep{mth75,scr+84}. However, the change in sign of Stokes V between the two components is rarely if ever observed in multi-component pulsar APs \citep{han98}. Changes in the sign of Stokes V between components can potentially be interpreted as emission from different magnetic hemispheres, potentially implying rapid rotation of the progenitor \citep{jk19}. Alternatively, the opposite signs in circular polarization may indicate that birefringence could be the cause of the split sub-components in FRB\,20220310F \citep{suresh2019induced,qu2023polarization}. While birefringent dispersive delay would require $RM\approx10^7$\,rad\,m$^{-2}$ to produce the observed 0.3\,ms split in FRB\,20220310F, birefringent refraction may be more plausible, requiring only $RM\approx10^{1-2}$\,rad\,m$^{-2}$. The closed loop formed on the Poincar\'{e} sphere by the polarization vector of the second component of FRB\,20220310F is also reminiscent of the loop observed in the first component of FRB\,20181112A by \citet{cho2020spectropolarimetric}, which was potentially ascribed to FC in a relativistic plasma. From equation 2, we estimate a magnetic field strength $B\approx150$\,G would be required for FC, which is inconsistent with supernova-remnant and plerionic environments \citep[e.g.][]{vedantham2019faraday}.

\subsubsection{RM Variation}

We observe a significant $3.1\sigma$ variation in RM between the sub-components of FRB\,20221101B ($-13.1\pm10.5\,$rad\,m$^{-2}$ and $-75.9\pm17.5\,$rad\,m$^{-2}$), and marginally significant $1.7\sigma$ and $2.7\sigma$ variations in FRB\,20220207C ($161.5\pm 0.4\,$rad\,m$^{-2}$ and $173.5 \pm 7.2\,$rad\,m$^{-2}$)  and FRB\,20220310F ($10.1\pm0.6 \,$rad\,m$^{-2}$ and $13.7\pm1.2\,$rad\,m$^{-2}$), respectively. Figures~\ref{fig:rmtime1}-\ref{fig:rmtime4} in Appendix~\ref{app_rmtime} show the evolution of detected RM across across all four \textbf{Multi-Component} FRBs. No RM was detected for any but the strongest sub-component of  FRB\,20220418A, and intra-component RM variations were not observed for any other DSA-110 FRB. This effect is commonly observed in the APs of radio pulsars when sufficient S/N is available \citep{ilie19RM}, with magnitudes up to several tens of rad\,m$^{-2}$, and is attributed to propagation effects in pulsar magnetospheres \citep[e.g.,][Section 4.3]{2023MNRAS.520.2039Y}. Similar propagation effects, if responsible for the RM variations we observe, may also be responsible for the multiple emission components as we only observe these effects (and significant variations in polarization state) for the FRBs that exhibit sub-components. 

\section{Comparisons}\label{analysis}

The polarization properties of FRBs provide insight into the nature of FRB emission and FRB sources, as has been the case for pulsars. We therefore present comparisons between the polarization properties of FRB sub-samples, and between FRBs and Galactic pulsars, to search for similarities and differences. Although emission from FRBs and pulsars is vastly different in energy scale \citep[e.g.,][]{nimmo2021highly}, common propagation effects in magnetospheres and near-source environments may impact the polarization properties. We augment the DSA-110 FRB sample with a literature sample of 38 FRB sources with polarization data, described in detail in Appendix~\ref{app_lit}. The literature sample includes 19 repeating sources, for which we use the weighted mean of polarization measurements for the linear and circular fractions. When available, we use polarization measurements at L-band (1-2 GHz) for the literature sample in order to compare with the DSA-110 sample. A minimum $\rm{S/N} > 8\sigma$ is imposed on the literature sample of FRBs. For additional details, see Appendix~\ref{app_lit}. The DSA-110 sample thus more than doubles the total number of non-repeating FRBs with polarization measurements. In Section~\ref{selection_effects} we compare properties of different FRB sub-samples, and assess any frequency dependent or other selection effects. We then compare FRB polarization properties with those of various subsets of the Galactic pulsar population in Section~\ref{pol_section}. 

\subsection{Comparison between FRB sub-samples}\label{selection_effects}

\begin{figure}[t]
\centering
\includegraphics[width=0.64\textwidth]{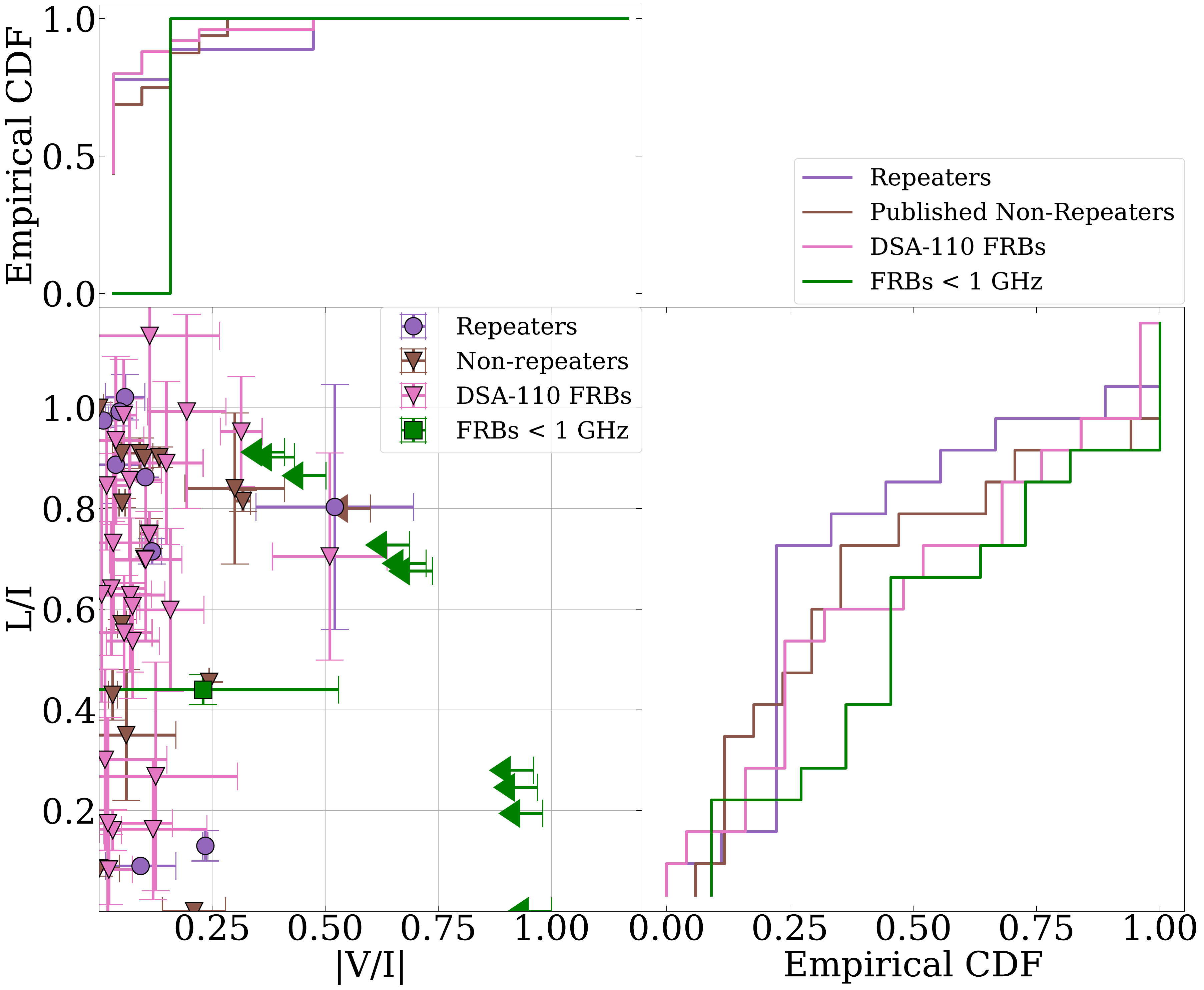} 
\caption{Comparison between the linear- and circular-polarization fractions of various FRB sub-samples. The top and right panels show empirical cumulative distribution functions (CDFs) for the DSA-110 sample (pink), published non-repeaters (brown), published repeaters (purple), and FRBs with measurements only at $<1$\,GHz (green). Arrows indicate upper limits in $|V/I|$ estimated from the measured values of $L/I$. See Appendix~\ref{app_lit} for a summary of the literature data. In general, no significant differences between sub-samples are found. 
}
\label{fig:pol_vother}
\end{figure}

In Figure~\ref{fig:pol_vother} we compare the distributions of various FRB sub-samples in $L/I$ and $|V|/I$. We consider the DSA-110 sample of 25 so far non-repeating events, along with published repeaters and non-repeaters, and measurements below 1\,GHz. The literature sample contains 17 non-repeaters and 9 repeaters with polarization measurements at L-band (1 - 2\,GHz), while there are 10 repeaters and 1 non-repeater with measurements only at lower frequencies ($< 1$\,GHz). 6 repeaters have measurements at multiple frequencies, for which we use the L-band observations where available. 16 non-repeaters (1 below 1 GHz) and 8 repeaters (1 below 1 GHz) in the literature sample have circular polarization measurements. Kolmogorov-Smirnov (K-S) and Anderson-Darling (A-D) tests at 90\% confidence (p-values below 10\% reject the null hypothesis that the samples are drawn from the same distribution) reveal no significant differences in the distributions of DSA-110 and other FRBs in $L/I$. This holds when comparing with both previously published repeaters and non-repeaters, which themselves have no evident differences. We acknowledge however that this test is incomplete, and biases due to the resolution, sky coverage, and bandwidth of the DSA-110 survey may still exist. Further exploration of this however is outside the scope of this work, and we proceed on the assumption that such biases are negligible given the consistency of the DSA-110 sample with the published sample of FRBs. 

A more detailed discussion is required to properly compare circular polarization. The circular polarization fraction can be defined either by averaging over the signed Stokes V fraction, $V/I$, or by taking the absolute value of Stokes V before averaging, $|V|/I$. The former preserves the sign (handedness) of circular polarization but neglects any significant sign changes across the burst. The latter accounts for sign changes, but introduces a bias which is not well-constrained. We provide both values for the DSA-110 bursts in Table~\ref{table:PolTable}, but most published FRBs use only $V/I$ \citep[e.g.,][]{hilmarsson2021polarization,cho2020spectropolarimetric}. Therefore, we proceed by using $V/I$ so that the sample is consistent, taking the absolute value ($|V/I|$) so that the already small sample is not further split by sign. With this definition, we find no significant difference when comparing the distribution of circular polarization of DSA-110 FRBs to previously detected non-repeaters. However, it is encouraged for future polarization surveys to report both $V/I$ and $|V|/I$, as the latter is a more accurate representation of the magnitude of circular polarization. Furthermore, $V/I$ and $|V|/I$ cannot be accurately compared due to the bias in $|V|/I$. For example, using $|V|/I$ for the DSA-110 sample and $|V/I|$ for the published sample, we find significant difference between the populations with p-value $1.2\%$. We conclude that the DSA-110 sample is not biased in its polarization properties with respect to published FRBs, and that polarization fractions do not appear to be a viable means of distinguishing repeaters and non-repeaters \citep[see also][]{qu2023polarization}. 

We compare the sample of FRBs with polarization data at L-band to those measured at lower frequencies. No significant difference is found. This may suggest that other effects besides frequency-dependent stochastic RM depolarization \citep{feng2022frequency,mckinven2023revealing} generally determine the polarization properties of the FRB population at different frequencies. Alternatively, if stochastic RM depolarization is dominant, this may imply that FRB progenitors can inhabit a wide variety of turbulent magneto-ionic environments. However, more robust conclusions would require a careful dissection of the population, for example into repeaters and non-repeaters; only two of the literature FRBs with polarization data reported below 1\,GHz are apparent non-repeaters. We further note this method is primarily effective in characterizing the frequency-dependence of non-repeaters and repeaters detected in only one frequency band by analyzing the population as a whole. \citet{feng2022frequency} and \citet{mckinven2023revealing} provide more detailed analyses of individual repeaters with measurements at multiple frequencies. For the remainder of the paper, we restrict the sample to the 18 non-repeaters and 9 repeaters with L-band polarization measurements.

\subsection{Comparison between FRBs and pulsars}\label{pol_section}

\begin{figure*}[t]
\centering
\includegraphics[width=0.64\textwidth]{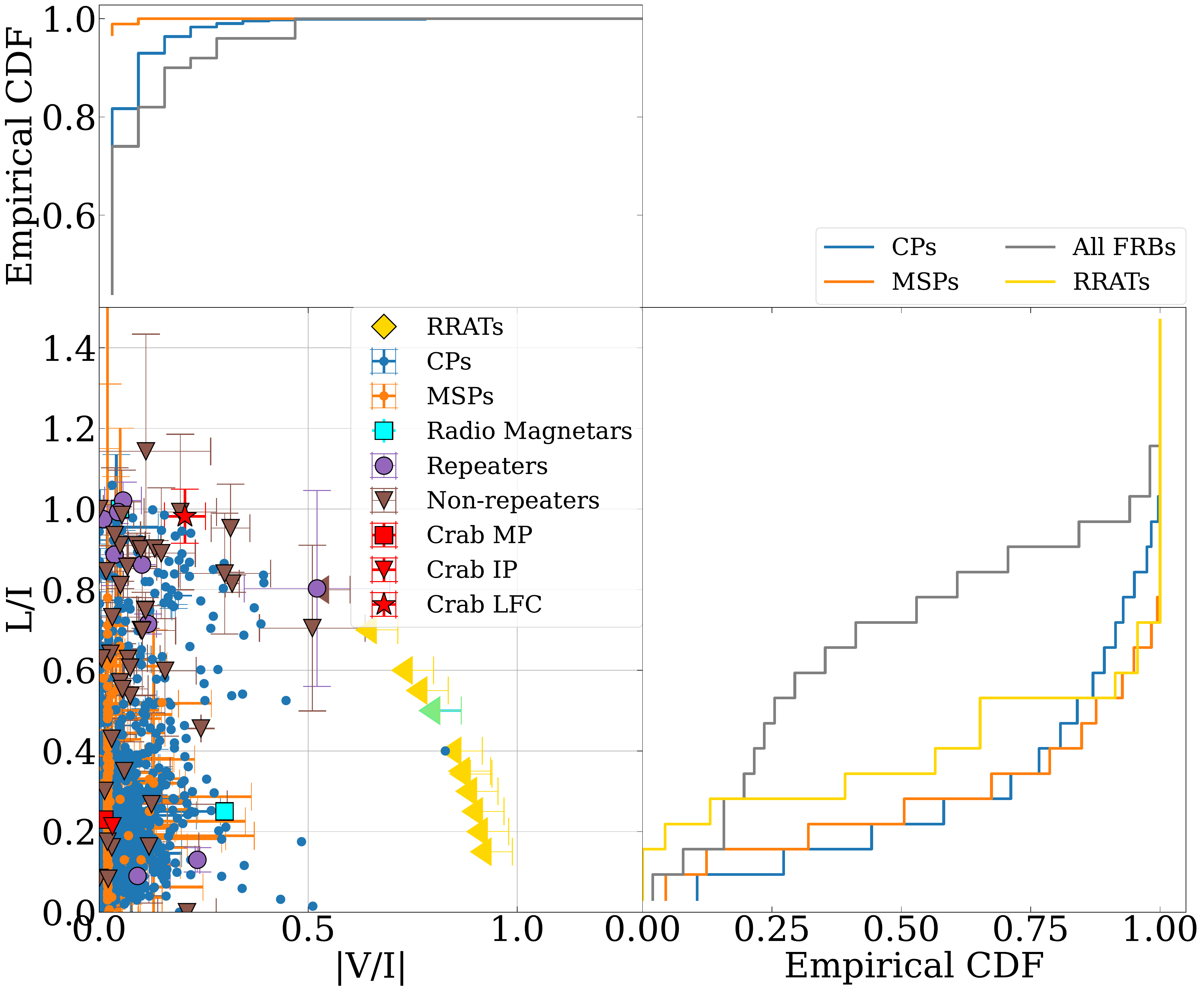} 
\includegraphics[width=0.64\textwidth]{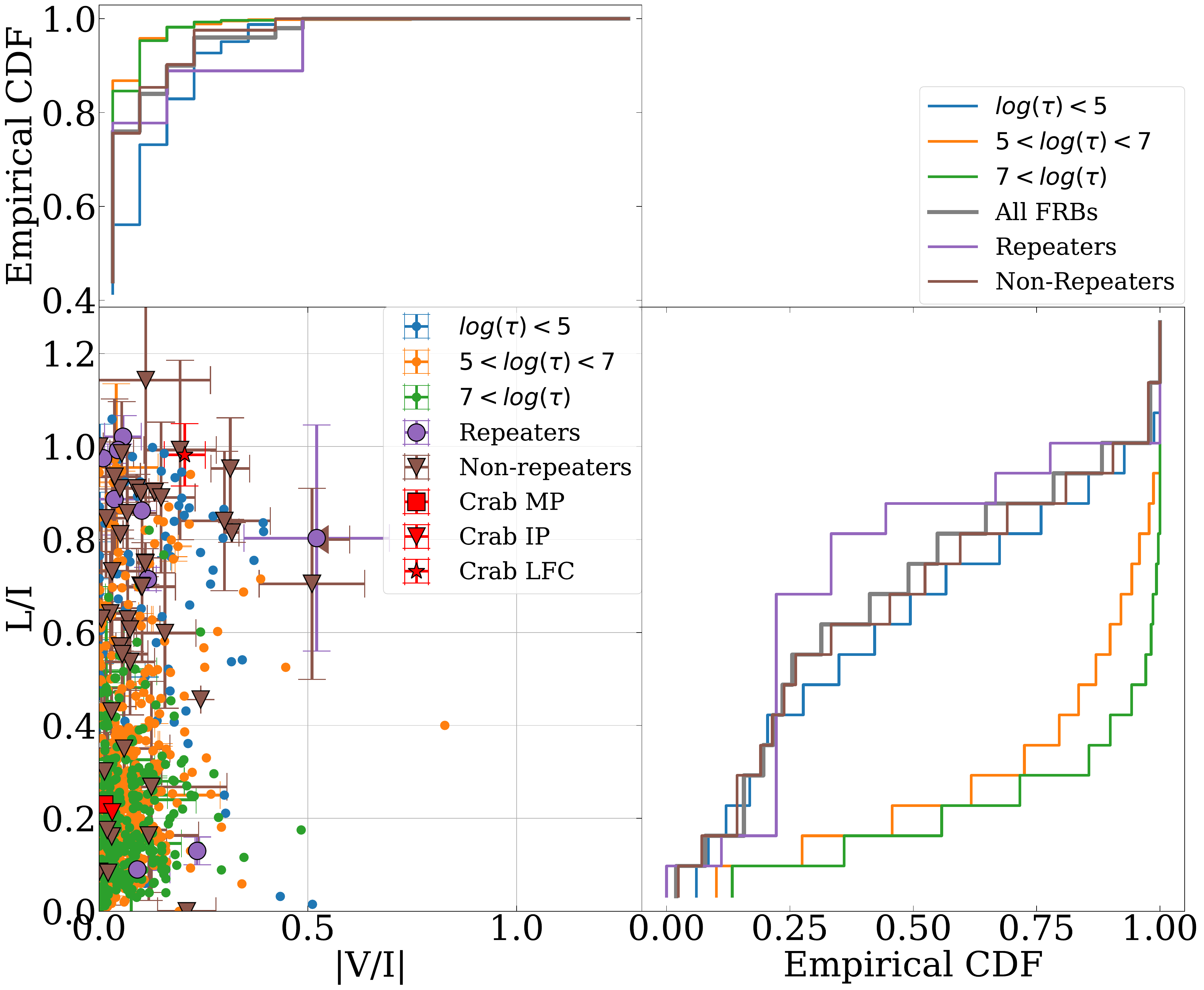}
\caption{Comparison between the linear- and circular- polarization fractions of FRBs, pulsar APs, and RRATs. \textit{Top:} FRBs (grey) are observed to have much higher linear-polarization fractions than MSPs (blue), CPs (orange) and RRATs (gold). Arrows indicate upper limits in $|V/I|$ estimated from the measured values of $L/I$. Somewhat higher circular-polarization fractions are also observed for FRBs. \textit{Bottom:} We compare FRBs with canonical pulsars ($P > 30$\,ms) with various characteristic ages. Young pulsars (blue) have remarkable similarity to FRBs (grey) in linear polarization, while older pulsars (orange, green) are more depolarized.}
\label{fig:pol_FRB_PSR}
\end{figure*}

We next compare the linear- and circular-polarization fractions of FRBs with the Galactic and Magellanic-Cloud pulsar populations. We include RRATs, noting that no circular polarization data are available for RRATs at the time of writing. Radio-loud magnetars are also included, and are grouped with Canonical Pulsars (CPs, $P > 30\,$ms) where necessary. We also consider both average-profile (AP) and single-pulse (SP) polarization fractions of pulsars. 

\begin{figure}[t]
\centering
\includegraphics[width=0.75\textwidth]{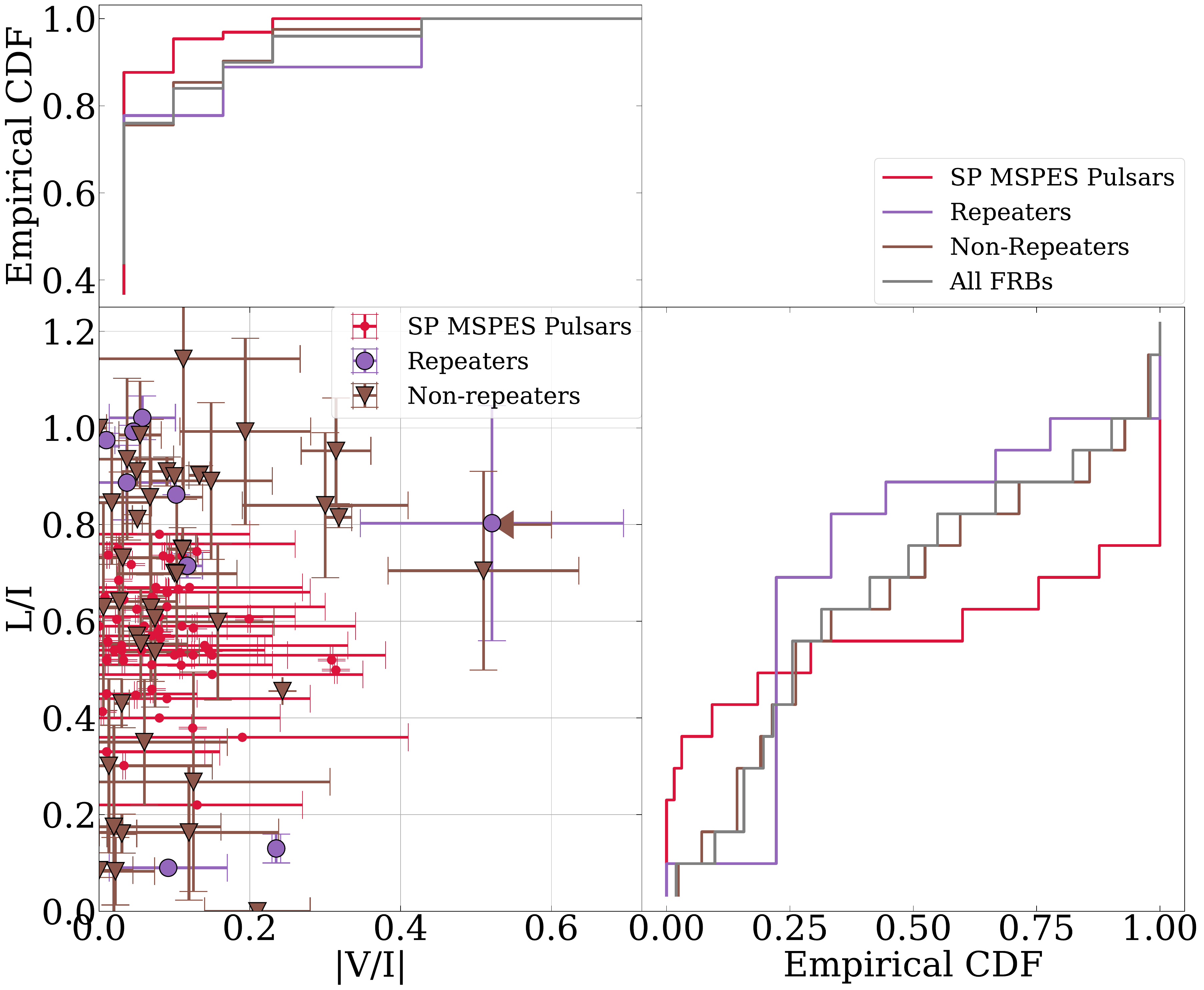}
\caption{Comparison between the linear- and circular- polarization fractions of FRBs with pulsar single pulses (SPs). Pulsar SPs (red) have similar average linear polarization fraction ($\sim60\%$), but cover a narrower range than FRBs (grey). Arrows indicate upper limits in $|V/I|$ estimated from the measured values of $L/I$.}
\label{fig:pol_FRB_SP}
\end{figure}

Although the polarization properties of APs across the $P-\dot{P}$ diagram are well characterized (see Appendix~\ref{app_lit} for a summary of existing observations), only a few SP polarization studies  of substantial pulsar samples exist \citep[e.g.,][]{mth75,scr+84,mitra2016meterwavelength,mmb23}. Among the non-recycled pulsar population, significant trends in SP and AP polarization are observed with the spin-down energy loss rate, $\dot{E}$. For high-$\dot{E}$ pulsars, both SPs and APs typically exhibit high polarization fractions, and valid rotating vector model \citep[RVM;][]{1969ApL.....3..225R} fits can sometimes even be made to SPs. For lower-$\dot{E}$ pulsars, while SP polarization fractions remain high, AP polarization fractions decrease. The depolarization can have multiple causes: for example, a wider variance of polarization states of SPs, and SPs clustering into orthogonally polarized modes. These trends are driven by $L/I$; $|V|/I$ remains consistent with $\dot{E}$ and between SPs and APs. Here we compare the FRB polarization fractions to a collation of AP polarization fractions obtained around 1.4\,GHz from the literature, and to the polarization fractions of SPs obtained with the Meterwavelength Single-pulse Polarimetric Emission Survey (MSPES) at 333 MHz and 618 MHz. The results are shown in Figures \ref{fig:pol_FRB_PSR} and \ref{fig:pol_FRB_SP}. 

The distributions of polarization fractions of FRBs are largely dissimilar to those of pulsars. Pulsar APs and RRATs show characteristically smaller linear and circular polarization fractions than the FRB sample, both repeaters and non-repeaters. Of the 9 L-band repeaters we consider, only two have small ($\lesssim30\%$) $L/I$, and the remainder have large ($\gtrsim70\%$) $L/I$. Non-repeating FRBs have similar $L/I$ and somewhat lower $|V/I|$ than repeaters, though small number statistics may limit concrete statements regarding repeaters. Although RRATs have slightly higher $L/I$ than pulsar APs, this difference is marginal in comparison to the FRB-AP difference. A KS test indicates that repeating FRBs, unlike the non-repeaters, are not significantly different from CPs in $|V/I|$. Although interpretation of this is limited by small-number statistics with only 7 repeaters with $|V/I|$ measurements.

Remarkably, FRBs (in particular non-repeating sources) have similar distributions of $L/I$ and $|V|/I$ to the youngest (characteristic ages $\tau_{c}<10^{5}$\,yr) pulsar APs. The polarization properties of young, high-$\dot{E}$ pulsars are known to be drastically different to the remainder of the pulsar population \citep{weltevrede2008profile}, with high $L/I$, a lack of jumps in polarization position-angle and orthogonally polarized modes, and a close similarity between typical-SP and AP polarization profiles. The lack of depolarization in young pulsars is attributed to either magnetospheric propagation effects that cause only one of the modes (likely the $O$-mode) to be observed, or to the dominance of curvature radiation at high $\dot{E}$ (resulting in $\gamma$-ray emission) that may change the radio emission mechanism. Young pulsars, with emission likely from high in the magnetosphere \citep{rmh10}, may prove to be useful analogs for the production of FRB radiation even if the specific radio emission mechanisms differ. As an example, we refer the reader to recent work by \citet{2023ApJ...945..115L} on the importance of both extreme emission heights and relativistic motion in giant pulses emitted by the Crab pulsar. 

Finally, we find that the MSPES SP distributions of $L/I$ and  $|V/I|$ are not consistent with the FRB distributions. The SP linear-polarization fractions are clustered around $\sim60\%$, whereas the FRBs span a wider range of $L/I$. The SPs typically exhibit slightly higher values of  $|V/I|$ than FRBs. The MSPES sample only contains seven SP measurements for $\dot{E}>10^{34}$\,erg\,s$^{-1}$, a threshold above which significant polarization changes are observed. 

\section{Conclusions}\label{conclusions}


We present a full-polarization analysis of the first 25 as yet non-repeating FRBs detected at 1.4 GHz with the DSA-110. This analysis includes a detailed description of data reduction and calibration for full-Stokes high time resolution data. We detect RMs for 20 FRBs, with magnitudes in the range $4-4670$\,rad\,m$^{-2}$. We propose a classification according to polarization, S/N, and temporal properties. $15/25$ DSA-110 FRBs are labeled as \textbf{Consistent with 100\% Polarization}, $3/25$ are labeled \textbf{Consistent with 0\% Polarization}, $5/25$ are \textbf{Intermediate}, and $2/25$ have \textbf{Unconstrained} polarization properties due to their low S/N. Four FRBs that show multiple components on millisecond timescales are also unique in showing time-variable polarization properties, including RM.

We combine the DSA-110 sample with polarization measurements of previously published FRBs, and compare FRB sub-samples and FRBs with Galactic pulsars. We find no significant differences in the distributions of $L/I$ and $|V/I|$ between the DSA-110 sample and the published samples; neither is there a significant difference between the repeater and non-repeater samples. We find that FRBs are typically far more polarized than the average profiles of pulsars and span a wider range in polarization fractions than pulsar single pulses, which makes them remarkably similar to young (characteristic ages $<10^{5}$\,yr) pulsars.

We conclude the following.
\begin{enumerate}

    \item The substantial fraction of highly linearly polarized FRBs ($15/25$ in the DSA-110 sample are \textbf{Consistent with 100\%} and have $L/I > 50\%$) suggests that the FRB emission mechanism is often intrinsically linearly polarized. In the full DSA-110 sample, statistically significant variations in PPA are observed in $3/25$ events ($2/13$ \textbf{Consistent with 100\%} events). This suggests modulation in some cases caused by a rapidly moving / rotating emission region. 

    \item Bursts with temporally separated components have variable polarization properties between the components. Of the $4/25$ such DSA-110 events, three have variable linear-polarization PPAs and fractions, two have variable signs and magnitudes of Stokes V, and three exhibit evidence for time-variable RMs. Together this suggests that the emission sites responsible for the different components originate from different regions of a magnetosphere, and are potentially viewed along different sightlines through magnetized plasma.

    \item The origins of circular polarization and depolarization in FRBs remain mysterious. Faraday conversion in a highly magnetized near-source medium could explain the observed $|V|/I$ fractions. We disfavor bandwidth depolarization, RM scattering, and Faraday conversion as dominant depolarization mechanisms. We therefore have little evidence from polarimetry alone for the generic presence of dense, highly magnetized, inhomogeneous near-source plasma environments. Depolarization due to the superposition of multiple short bursts is possible. 

    \item The strong similarity between the $L/I$ distributions of FRBs and the average profiles (and likely also single pulses) of the youngest pulsars suggests that the radio emission from young pulsars may form a useful analog to probe at least the depolarization mechanisms of FRBs. We note that the abrupt jump in PA observed in FRB\,20220207C and the near-orthogonal positions on the Poincar\'{e} sphere of the two components of FRB\,20220310F are uncommon for young pulsars. 
    
\end{enumerate}

Polarization studies have great promise in identifying characteristics of FRB emission that may prove useful in testing emission-mechanism and progenitor models. At present, intrinsic emission-mechanism models favor high linear-polarization fractions, thus ascribing all other properties to propagation effects. In the future, studying FRB polarization properties in the context of other diagnostics of near-source environments, e.g., from accurate localizations, may be used to distinguish between intrinsic and extrinsic effects. At this point, we urge the further theoretical consideration of the intrinsic timescale(s) of FRB emission, and the causes of intrinsic depolarization. First, several of the DSA-110 FRBs are remarkably simple: the sample is characterized ($21/25$ events) by single emission components that are several tenths of a millisecond wide, with flat PPAs, and a wide range of linear-polarization fractions. We suggest that depolarization is likely intrinsic to the emission rather than a propagation effect; thus, the prospect of superposing several shorter bursts needs to be considered. At the same time, the existence of highly linearly polarized FRBs with flat PPAs over approx. millisecond timescales also requires an explanation. While not explored in this work, the scattering of FRBs over millisecond durations could be an alternate cause of flat PPAs. Third, our finding that multiple temporally separated emission components on approx. millisecond timescales tend to have different polarization properties poses an interesting additional challenge. 

\begin{acknowledgments}

The authors would like to thank Jim Cordes, Dongzhi Li, Bing Zhang, Yuanhong Qu, Joel Weisberg, and Alexandra Mannings, for insightful and essential conversations on polarization theory and direction on the analysis conducted, as well as Paul Bellan and Yang Zhang for a comprehensive Plasma Physics course. We also thank Yi Feng, Dipanjan Mitra, Yuan-Pei Yang, Dylan Nelson, Reshma Anna-Thomas, and an anonymous referee for useful comments and recommendations on the early draft. This material is based upon work supported by the National Science Foundation Graduate Research Fellowship under Grant No. DGE‐1745301. The authors thank staff members of the Owens Valley Radio Observatory and the Caltech radio group, including Kristen Bernasconi, Stephanie Cha-Ramos, Sarah Harnach, Tom Klinefelter, Lori McGraw, Corey Posner, Andres Rizo, Michael Virgin, Scott White, and Thomas Zentmyer. Their tireless efforts were instrumental to the success of the DSA-110. The DSA-110 is supported by the National Science Foundation Mid-Scale Innovations Program in Astronomical Sciences (MSIP) under grant AST-1836018.

\end{acknowledgments}

\bibliography{Refs}{}
\bibliographystyle{aasjournal}

\appendix

\section{Literature samples of polarization observations of FRBs and pulsars}\label{app_lit}

Previous analyses have yielded either RM or polarization measurements for at least 38 FRBs at time of writing; this sample was compiled using the Transient Name Server\footnote{\url{https://www.wis-tns.org/}} (TNS), the FRB  Catalog\footnote{\url{https://www.frbcat.org/}} (FRBcat), and the Canadian Hydrogen Intensity Mapping Experiment (CHIME) FRB Catalog 1\footnote{\url{https://www.chime-frb.ca/catalog}} \cite{2020TNSAN.160....1P}, \citep{petroff2016frbcat,Amiri_2021}. FRB polarization is often reported without uncertainties due to low resolution or low S/N that makes constraining the polarization difficult. In addition there is no standard method of computing the polarization fraction due to the wide range of morphologies in FRBs, and this should be held in mind when comparing polarimetry from multiple surveys. The FRBs in the literature sample compiled here were detected through the Apertif survey with the Westerbork Synthesis Radio Telescope (WSRT) and the Australian Square Kilometer Array Pathfinder (ASKAP) survey, as well as observations with the Parkes Telescope, CHIME, Green Bank Telescope, and Arecibo Observatory. Recent studies have observed frequency-dependent de-polarization thought to be the result of stochastic RM variations within a scattering screen \citep{feng2022frequency,mckinven2023revealing}. A subset of 11 FRBs in this sample have polarization measurements at lower ($< 1$\,GHz) frequencies.

Table~\ref{table:FRBSAMPLE} summarizes the polarization properties of each burst from the literature. Of the 38 bursts, 19 are repeaters, 34 have RM measurements, 15 have spectroscopic or photometric redshift measurements (9 repeaters, 6 non-repeaters), 25 have reported L-band linear polarization measurements, 11 have linear polarization measurements below 1\,GHz, and 26 have reported circular polarization measurements. Below we describe the selection criteria and process for the FRB RMs used here:

\begin{enumerate}
    \item Only RM measurements published before a 2023 April 10 cutoff date are considered for this sample.
    \item For non-repeaters and repeaters with only 1 RM measurement, we use the available measurement.
    \item For FRBs with RM measurements in the TNS or FRBcat databases, we defer to the catalogued value to maximize sample consistency.
    \item For repeaters with multiple RM measurements, we use the weighted mean, with the reciprocal of the errors as weights.
\end{enumerate}

\noindent We make three exceptions based on the observed RM behavior:

\begin{itemize}
    \item FRB\,20190208A shows a secular evolution in RM which follows a ‘U’ shaped curve \citep{mckinven2023revealing}. Therefore, we use the most recent RM value rather than the weighted mean. 
    \item FRB\,20190231B has multiple RM measurements from \citet{mckinven2023revealing}; however, CHIME reports difficulty in distinguishing RM detections near 0 from instrumental leakage. Therefore, we use the RM measurement with the lowest uncertainty to minimize propagation of any leakage errors.
    \item FRB\,20191117A has 2 RM measurements from \citet{mckinven2023revealing} separated by $\sim9\,$rad\,m$^{-2}$, but it is unclear with only 2 bursts whether this variation is stochastic or secular. Therefore, we use the value with the smallest error to avoid improperly combining secularly evolved RMs. 
\end{itemize}

\noindent We make three additional exceptions based on the conventions in the available literature; in cases where RM measurements are commonly compared to a well-constrained ‘accepted’ value which has not yet been superceded in accuracy, we use this accepted value. This affects the following FRBs:

\begin{itemize}
    \item FRB\,20121102A \citep[e.g.,][]{plavin2022frb,michilli2018extreme,hilmarsson2021rotation}; the value from \citet{plavin2022frb} is used as it pre-dates other RM measurements and has a low relative error. However, since \citet{sherman2023deep} compares the logarithm of the RM, deviations of the RM within an order of magnitude make little change to the analysis, and do not affect the analysis in this paper at all.
    \item FRB\,20200120E \citep[e.g.,][]{bhardwaj2021nearby,nimmo2021highly}; we use the value from the first detected burst in \citet{bhardwaj2021nearby}.
    \item FRB\,20201124A \citep[e.g.,][]{hilmarsson2021polarization,kumar2022circularly} we use the weighted mean value from \citet{hilmarsson2021polarization}
\end{itemize}

\noindent We acknowledge that these constraints are not complete; secularly time-varying RMs are not properly represented either by the weighted mean or the value with the lowest error. In addition, while the weighted mean incorporates the signal-to-noise of individual bursts via the uncertainties, limitations due to the intrinsic bandwidth of the FRB are not explicitly accounted for. We proceed with our analysis, reasonably assuming that the reported RM's uncertainty properly quantifies the limitations of the measurement system. We leave additional revisions for variable RMs to future work.

We furthermore describe the selection criteria and process for the FRB polarization fractions used here:

\begin{enumerate}
    \item Only polarization measurements published before the 2023 April 10 cutoff date are considered for this sample.
    \item For non-repeaters and repeaters with only 1 polarization fraction measurement, we use the available measurement. If the polarization is reported for individual sub-components of the same burst, we use the weighted mean of the sub-components.
    \item A minimum signal-to-noise $S/N > 8\sigma$ is imposed on all non-repeaters and individual bursts from repeaters included in this sample.
    \item For FRBs with polarization measurements in the TNS or FRBcat databases, we defer to the catalogued value to maximize sample consistency.
    \item For repeaters with multiple polarization measurements, we use the weighted mean, with the reciprocal of the errors as weights. If measurements are available at multiple frequencies, we use the L-band (1-2\,GHz) if available. Otherwise, the observation with the nearest frequency below 1\,GHz.
    \item If only the signed circular polarization ($V/I$) is reported, then the absolute value circular polarization ($|V|/I$) is taken as the absolute value ($|V/I|$) (for further details, see Section~\ref{analysis}). We make this condition given that most FRBs in the literature were reported with no $|V|/I$ measurement. If both $L/I$ and $V/I$ are reported but the total polarization $\sqrt{L^2 + V^2}/I$ is not explicitly stated, the total polarization is calculated as $\sqrt{(L/I)^2 + (V/I)^2}$.
\end{enumerate}

Expanding on criteria 3, the minimum $S/N > 8\sigma$ applies both to non-repeaters and to the individual bursts of repeaters included in the reported weighted average polarization fractions. Many studies fail to report the S/N of FRB detections \citep[e.g.][]{feng2022frequency,feng2022circular}. In these cases, we estimate the S/N as the reported flux density $S$ divided by the radiometer noise RMS $\sigma_S$ of the detector \citep[e.g.][]{condon2016essential}. If neither the S/N nor $S$ are reported, we have chosen to include the FRB in our sample under the assumption that a reasonable detection threshold was used in the original survey. 

As noted above, these selection criteria are not complete, as the weighted mean does not fully capture secularly time-varying polarization properties. We maintain the average FRB polarization fractions are sufficient for the cumulative distribution analysis presented here, and leave a fully time-dependent study to future work.

In the main text, we compare FRB properties to a sample of 1445 Galactic (and Magellanic Cloud) pulsars with published polarimetry of APs: 1305 canonical pulsars (CPs) with periods $P > 30\,$ms, 137 millisecond pulsars (MSPs,) with periods $P < 30\,$ms, and 3 magnetars and 22 RRATs with published polarization or RM data at L-band. These are compiled from \cite{johnston2022thousand}, \cite{serylak2021thousand}, \citet{wahl2022nanograv}, \citet{Spiewak_2022}, \citet{cameron2020depth}, \citet{olszanski2019arecibo}, \citet{weisberg1999arecibo}, \citet{1999ApJS..121..171W}, \citet{johnston2018polarimetry}, \citet{2011MNRAS.414.2087Y}, \citet{gould1998multifrequency}, \citet{kramer2007polarized}, \citet{camilo2007polarized}, \citet{camilo2008magnetar}, \citet{levin2012radio},  \citet{levin2010radio}, \citet{keane2011rotating}, \citet{caleb2019polarization}, \citet{karastergiou2009radio}, \citet{xie2022emission}. We use the following criteria and process for selecting the pulsar polarimetry and RM data:

\begin{enumerate}
    \item Only polarization measurements published before the 2023 April 10 cutoff date are considered for this sample.
    \item For each pulsar, we use the polarization fractions from the average profiles (APs) (except for the MSPES single-pulse sample discussed in Section~\ref{analysis}).
    \item Most pulsars are published with both $|V|/I$ and $V/I$; we record both values and use the absolute value of $V/I$ for consistency as discussed in Section~\ref{analysis}. 
    \item If multiple RM or polarization fractions are reported, the most recent measurement is used.
\end{enumerate}

\noindent We note here that, as for the FRB literature sample, these criteria are not complete, as they do not capture time-varying RMs or polarization fractions in pulsars. A weighted average of measurements is not used for pulsars given the large amount of available data. Additionally, the most recent average profile is more well-constrained, and using average profile polarization fractions consistent the current literature on pulsar polarimetry. The average profile polarization fractions are sufficient for the analysis presented in this work, and we leave a fully time-dependent discussion to future work.


\begin{deluxetable*}{  c    c  |c  c c |  c  c | c}[ht]
\setlength{\tabcolsep}{2pt}
\tabletypesize{\scriptsize}
\caption{Previously Published FRB Sample Polarization Properties}
\tablehead{\textbf{FRB Names} & \textbf{\begin{tabular}{@{}c@{}}Polarization \\ Telescope/ \\ Frequency \end{tabular}} &                 \textbf{$\mathbf{L/I}$} &             \textbf{$\mathbf{|V/I|}$ } & \textbf{\begin{tabular}{@{}c@{}}Polarization \\ Reference\end{tabular}} &  \begin{tabular}{@{}c@{}}$\rm \mathbf{RM}$ \\ $\rm \mathbf{(rad\,m^{-2})}$\end{tabular} & \textbf{\begin{tabular}{@{}c@{}}RM \\ Reference\end{tabular}} & \begin{tabular}{@{}c@{}}$\rm \mathbf{DM}^{\ddagger}
$ \\ $\rm \mathbf{(pc\,cm^{-3})}$\end{tabular}}
\startdata 
 FRB\,20201020A &                                            Apertif/1375.0 &                   -- &                                    -- & -- &                                 $110.0 \pm 69.0 $ & 1 & $398.59\pm0.08$\\
 \hline
 FRB\,20191108A &                                            Apertif/1375.0 &                      $70.0 \%$ &            $10.0 \%$&  2  &        $474.0 \pm 3.0 $ & 3 & $588.1\pm0.1$\\
 \hline
 \textbf{FRB\,20190711A}  &                                             ASKAP/1297.5 &  $97.48 \pm 1.39 \%$$^{\dagger}$ &   $0.96 \pm 1.15 \%$ $^{\dagger}$ & 4 & $9.0 \pm 2.0 $ & 4 & $593.1\pm0.4$\\ 
 \hline
 FRB\,20190611B &                                              ASKAP/1297.5 &    $81.5 \pm 2.1 \%$$^{\dagger}$ &    $31.8 \pm 1.7 \%$$^{\dagger}$ & 4 &$20.0 \pm 4.0 $ & 4 & $321.4\pm0.2$\\
 \hline
 FRB\,20190608B &                                              ASKAP/1297.5 &   $91.0 \pm 3.0 \%$ &     $9.0 \pm 2.0 \%$ & 4 &$353.0 \pm 2.0 $ & 4 & $338.7\pm0.5$ \\
 \hline
 FRB\,20181112A &                                              ASKAP/1297.5 &              $90.0 \%$ &            $10.0 \%$ & 5 &$10.9 \pm 0.9 $ & 5 & $589.27\pm0.03$\\
 \hline
 FRB\,20180924B &                                              ASKAP/1297.5 &    $90.2 \pm 2.0 \%$ &    $13.3 \pm 1.4 \%$ & 4 &$14.0 \pm 1.0 $ & 6 & $361.42\pm0.06$\\
 \hline
 FRB\,20180311A &                                            Parkes/1382.0 &    $75.0 \pm 3.0 \%$ &    $11.0 \pm 2.0 \%$ & 7&$4.8 \pm 7.3 $ & 7 & $1570.9\pm0.5$\\
 \hline
 FRB\,20171209A &                                             Parkes/1382.0 &   $100.0 \pm 1.0 \%$ &     $0.0 \pm 1.0 \%$ & 7 &$121.6 \pm 4.2 $ & 7 & $1457.4\pm0.03$\\
 \hline
 FRB\,20150807A &                                             Parkes/1382.0 &        $80.0 \pm 1.0 \%$ &                   -- & 8 &$12.0 \pm 0.7 $ & 8 & $226.5\pm0.1$\\
 \hline
 FRB\,20150418A &                                             Parkes/1382.0 &      $8.5 \pm 1.5 \%$ &   $0.0 \pm 4.5 \%$ & 9& $36.0 \pm 52.0 $ & 9 & $776.2\pm0.5$\\
 \hline
 FRB\,20150215A &                                             Parkes/1382.0 &     $43.0 \pm 5.0 \%$ &     $3.0 \pm 1.0 \%$ & 10 & $1.5 \pm 10.5 $ & 10 & $1105.6\pm0.8$\\
 \hline
 \textbf{FRB\,20190604A} &                                              CHIME/600.0  &          $91.2\pm8.1 \%$ &                   -- & 11 & $-20.0 \pm 1.0 $ & 12 & $553.18\pm3.23$\\
 \hline \textbf{FRB\,20190303A} &                                              FAST/1250.0&   $88.7\pm7.7\%$$^{\dagger}$  &                   $3.7\pm5.8\%$$^{\dagger}$& 13 &$-499.0 \pm 0.7 $ & 11 & $223.21\pm3.23$\\
 \hline
 FRB\,20190102C &                                             ASKAP/1297.5 &  $81.14 \pm 0.91 \%$$^{\dagger}$ &   $5.08 \pm 0.66 \%$$^{\dagger}$ &4 &$-110.0 \pm 1.0 $ & 14 & $363.6\pm0.3$\\
 \hline
\textbf{FRB\,20180916.J0158+65}  &                                              Effelsburg/1700.0 &          $102.1\pm4.5\%$$^{\dagger}$ &    $5.7\pm4.4\%$$^{\dagger}$ & 15&$-114.6 \pm 0.6 $ & 16  & $347.76\pm1.62$\\
\hline
 FRB\,20180714A &                                             Parkes/1382.0 &   $91.0 \pm 3.0 \%$ &     $5.0 \pm 2.0 \%$ &7 &$-25.9 \pm 5.9 $ & 7 & $1467.92\pm0.3$\\
 \hline
 FRB\,20160102A &                                             Parkes/1382.0 &    $84.0 \pm 15.0 \%$ &   $30.0 \pm 11.0 \%$ & 17&$-220.6 \pm 6.4 $ & 17 & $2596.1\pm0.3$\\
 \hline
 FRB\,20110523A &                                                GBT/800.0 &   $44.0 \pm 3.0 \%$ &   $23.0 \pm 30.0 \%$ & 18&$-186.1 \pm 1.4 $  & 18  & $623.3\pm0.06$\\
 \hline
 \textbf{FRB\,20121102A}&   \begin{tabular}{@{}c@{}}Effelsburg, \\ FAST/1700,1250.0 \end{tabular}       &    $13.0 \pm 3.0 \%$ &   $23.5 \pm 0.6 \%$ $^{\dagger}$ & 19,20  &$127000.0 \pm 800.0 $ & 19 & $557\pm2$\\
 \hline
 FRB\,20191001A &                                              ASKAP/1297.5 &     $57.0 \pm 1.0 \%$ &     $5.0 \pm 1.0 \%$ & 21&$55.5 \pm 9.0 $ & 21 & $506.92\pm0.04$\\
 \hline 
 \textbf{FRB\,20201124A} &  \begin{tabular}{@{}c@{}}Effelsburg, \\ FAST/1360.0,1250.0 \end{tabular}                                     &     $86.21\pm0.01\%$$^{\dagger}$&     $10.23\pm0.01\%$$^{\dagger}$& 13,22,23,31&$-601.0 \pm 11.1 $$^{\dagger}$ & 22 & $410.83\pm3.23$\\
 \hline
 FRB\,20180309A&                                             Parkes/ 1382.0 &   $45.56 \pm 0.06 \%$ &  $24.33 \pm 0.05 \%$ & 7&-- & -- & $263.42\pm0.01$\\
 \hline
 \textbf{FRB\,20200120E}  &                                              Effelsburg/1400.0 &   $99.3 \pm  1.3\%$$^{\dagger}$ &   $4.5 \pm 0.9 \%$$^{\dagger}$ & 30&$-29.8 \pm 5.0 $ & 24 & $88.96\pm1.62$\\
 \hline
 FRB\,20151230A &                                             Parkes/1352.0 &  $35.0 \pm 13.0 \%$ &    $6.0 \pm 11.0 \%$ & 25&--&--  & $960.4\pm0.5$\\
 \hline
 FRB\,20140514A &                                             Parkes/1352.0 &                      $0.0 \%$ &    $21.0 \pm 7.0 \%$ & 26&--&--  & $562.7\pm0.6$\\
 \hline
 \textbf{FRB\,20190520B} &                                                GBT/1450.0 &                     $9.0 \%$ &   $9.83 \pm 0.68 \%$ $^{\dagger}$ & 27,20&$-11000.0 \pm 13100.0 $ $^{\dagger}$ & 27 & $1202\pm10$\\
 \hline
 \textbf{FRB\,20220912A} &                                              FAST/1400.0 &   $80.3 \pm 24.3\%$$^{\dagger}$ & $5.2 \pm 0.2 \%$$^{\dagger}$ & 28&$0.6 \pm 0.1 $ & 29 & $219.46\pm0.04$\\
 \hline
 \textbf{FRB\,20180814A} &                                              CHIME/600.0 &       $69.1 \pm 4.7 \%$ &                   -- & 11 &$699.8 \pm 1.0 $ & 11 & $190.86\pm3.23$\\
 \hline
 \textbf{FRB\,20181030A} &                                              CHIME/600.0 &  $72.75 \pm 2.22 \%$$^{\dagger}$ &                   -- & 11 &$36.6 \pm 0.2 $$^{\dagger}$& 11 & $101.9\pm1.62$\\
 \hline
 \textbf{FRB\,20181119A}  &                                              CHIME/600.0 &   $67.58 \pm 2.76\%$$^{\dagger}$ &                   -- & 11&$1340.0 \pm 2.0 $ $^{\dagger}$& 11 & $368.79\pm6.47$\\
 \hline
 \textbf{FRB\,20190222A} &                                              CHIME/600.0 &   $24.61 \pm 0.58 \%$ &                   -- & 11&$571.0 \pm 1.7 $& 11 & $462.6\pm3.23$\\
 \hline
 \textbf{FRB\,20190208A} &                                              CHIME/600.0 &   $86.52 \pm 1.03 \%$ $^{\dagger}$ &                   -- & 11&$36.3 \pm 0.7 $& 11 & $579.06\pm3.23$\\
 \hline
 \textbf{FRB\,20190213B} &                                              CHIME/600.0 &   $90.19 \pm 2.68 \%$ $^{\dagger}$ &                   -- & 11&$-3.63 \pm 0.41 $ & 11 & $302.47\pm1.62$\\
 \hline
 \textbf{FRB\,20180908B} &                                              CHIME/600.0 &                       $0.0 \%$ &                   -- & 11&$0.18 $& 11 & $197.33\pm3.23$\\
 \hline
 \textbf{FRB\,20190117A} &                                              CHIME/600.0 &   $19.44 \pm 0.40 \%$$^{\dagger}$ &                   -- & 11&$76.31 \pm 0.40 $ & 11 & $393.05\pm1.62$\\
 \hline
 \textbf{FRB\,20190417A}&                                              FAST/1250.0 &  $71.5\pm2.5\%$ &    $11.7\pm2.0\%$ & 13& $ 4429.8 \pm 3.5 $ & 11 & $1378.11\pm6.47$\\
 \hline
 \textbf{FRB\,20190907A}  &                                              CHIME/600.0 &  $28.0 \%$                     &       -- & 11 &-- & --  & $307.32\pm3.23$\\
\hline
\enddata
\bigskip
\textbf{Notes:} Observation and polarization information for the compiled sample of previously published FRBs. The telescope and system observing frequency are those used for the polarization measurements. Names in boldface are repeaters, while non-boldfaced FRBs are non-repeating as of April 10th, 2023. For FRB\,20220912A, the RM is that reported by CHIME while the polarization was measured from the DSA-110 detections. For repeaters with multiple RM or polarization measurements, values are the weighted average of observations. The sample was compiled with the use of the Transient Name Server (TNS), the FRB  Catalog (FRBcat), and the Canadian Hydrogen Intensity Mapping Experiment (CHIME) FRB Catalog 1 (\cite{2020TNSAN.160....1P}, \cite{petroff2016frbcat}, \cite{Amiri_2021}). FRB\,20180301 was excluded from this literature sample due to its inconsistent RM and L-band polarization properties, which are not accurately described by average values \citep{luo2020diverse, kumar2023spectro}. While \citet{2023ApJ...955..142Z} and \citet{anna2023magnetic} (a later version of \citet{annathomas2023magnetic}) presented additional polarization data, they were released after the data cutoff date and are excluded from this sample.

$^{\dagger}$Weighted mean of multiple bursts; $^{\ddagger}$DM values are from the TNS catalog

Citations are listed below: (1)\citet{pastor2022fast} (2)\citet{connor2020bright} (3)\citet{van2022apertif} (4)\citet{day2020} (5)\citet{prochaska2019low} (6)\citet{bannister2019single} (7)\citet{oslowski2019commensal} (8)\citet{ravi2016magnetic} (9)\citet{keane2016host} (10)\citet{petroff2017polarized} (11)\citet{mckinven2023revealing} (12)\citet{fonseca2020nine} (13)\citet{feng2022frequency} (14)\citet{macquart2020census} (15)\citet{nimmo2021highly} (16)\citet{andersen2019chime} (17)\citet{caleb2018survey} (18)\citet{masui2015dense} (19)\citet{plavin2022frb} (20)\citet{feng2022circular} (21)\citet{bhandari2020limits} (22)\citet{hilmarsson2021polarization} (23)\citet{jiang2022fast} (24)\citet{bhardwaj2021nearby} (25)\citet{bhandari2018survey} (26)\citet{petroff2015real} (27)\citet{annathomas2023magnetic} (28)\citet{ravi2023deep} (29)\citet{2022ATel15679....1M} (30)\citet{nimmo2022burst} (31)\citet{xu2022fast}
\label{table:FRBSAMPLE}
\end{deluxetable*}

\section{DSA-110 voltage data model and polarization processing} \label{app_data_model}

All polarization analysis of DSA-110 FRBs detected during the commissioning observations reported on herein, and the calibration of these data, are accomplished using automatically triggered voltage dumps. This refers to the storage of 61440 contiguous post-channelization complex (i.e., undetected) spectra for each antenna, with a time-resolution of 32.768 $\mathrm{\mu}$s. The spectra consist of voltage measurements in two polarization channels for each of 63 antennas, in 6144 channels of 30.52\,kHz width centered on 1405\,MHz. Each complex sample is stored with 4-bit resolution for the real and imaginary parts. The four most significant bits are chosen using channel-dependent coefficients determined from the mean bandpass shapes of each antenna/polarization pair. For FRB data, the triggers are generated such that the arrival times of the bursts at 1498.75\,MHz (the upper limit of the sampled band) correspond to exactly 0.5\,s into the voltage dumps. Each voltage dump is stored in raw binary format, and is accompanied by a json header that includes all relevant descriptive information. 

The voltage dumps are processed using a custom full-polarization beamforming pipeline (\textit{dsa110-bbproc}). Post-correlation beamforming is implemented \citep{postcorrbeam} with the addition of both auto- and cross-correlation terms to mitigate quantization noise in applying per-antenna beamforming weights. For FRB data, the weights are determined from visibility data on bright bandpass calibrators typically within 12\,hr of detection. For voltage dumps on the continuum-source calibrators 3C48 and 3C286, the weights are determined from visibility data on the calibrators themselves. Ultimately, filterbank-format files are generated for each Stokes parameter by synthesizing beams in the directions of interest for each voltage dump. The polarization and Faraday-rotation definitions adopted here are summarized in Appendix~\ref{background}.

The \textit{dsa110-pol} module was developed to calculate FRB polarization and RM parameters offline from the polarized filterbank data. Major processing steps are as follows:
\begin{enumerate}

    \item Polarimetric calibration is applied using the formalism described in Appendix~\ref{app_derivation}. The stability of the calibration solutions is justified in Appendix~\ref{app_stability}, and critical assumptions on leakage are justified in Appendix~\ref{app_leakage}. 

    \item Full-polarization spectra for each FRB are generated by optimally averaging over the burst temporal profiles (Appendix~\ref{app_weights}). 

    \item RM estimation then proceeds on these spectra as described in Appendix~\ref{app_RMderivation}. If a significant RM is detected, the Stokes parameters are corrected for this RM. 
    
    \item Then, the total ($T$), linear ($L$) and circular ($C$) polarization spectra, as well as polarization position angle ($\chi$) are computed. The average polarization fractions, referred to as $L/I$, $|V|/I$ and $\sqrt{L^2 + V^2}/I$, are calculated as the weighted mean of $L_{\rm unbias}(t)/I(t)$, $C(t)/I(t)$, and $T(t)/I(t)$ across the FWHM of the burst using the optimal weights (Appendix~\ref{app_weights}). In addition, the sign-dependent circular polarization fraction $V/I$ is computed as the weighted mean of $V(t)/I(t)$ across the burst. The mean position angle $\chi$ is calculated with a similar weighted mean across the FWHM.
    
\end{enumerate}

\section{Polarization and Faraday Rotation Background} \label{background}

The polarization of light is defined by the correlation properties of components of the propagating electric field. In a linear basis, where $V_A$ and $V_B$ refer to orthogonal linear components of the electric field, the Stokes parameters are one representation of the polarization state: 

\begin{equation}
I = \frac{1}{2}(<V_AV^*_A> + <V_BV^*_B>)
\end{equation}

\begin{equation}
Q = \frac{1}{2}(<V_AV^*_A> - <V_BV^*_B>)
\end{equation}

\begin{equation}
U = {\rm Re}\{<V_AV^*_B>\}
\end{equation}

\begin{equation}
V = -{\rm Im}\{<V_AV^*_B>\}
\end{equation}

\begin{figure}[ht!]
     \centering
     \includegraphics[width=\linewidth]{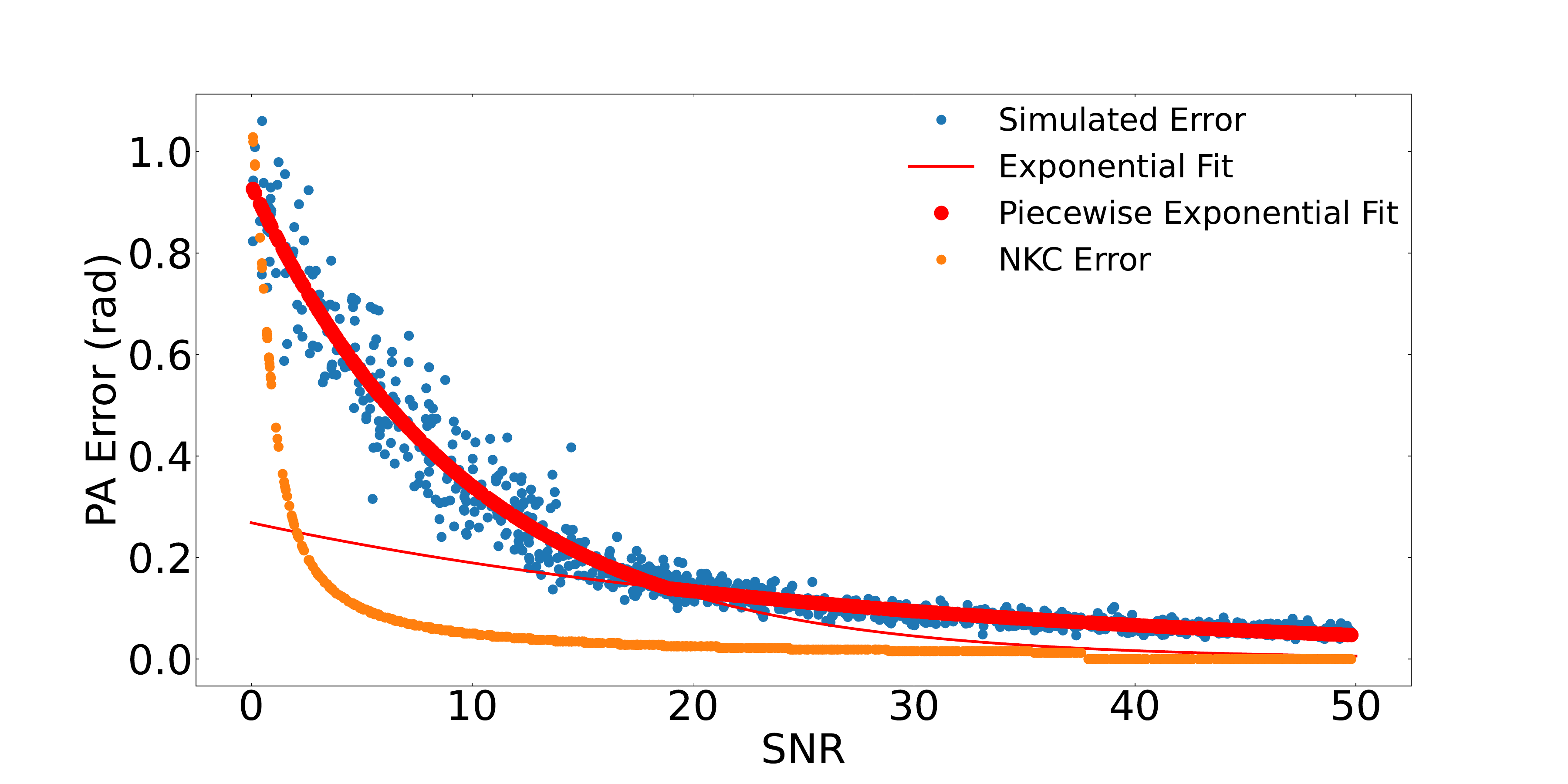}
     \hfill
     \caption{Simulated error in the linear polarization position angle (PA), as a function of linear-polarization S/N. See text for details of the simulation. The simulated error (blue) is taken as the standard deviation of the PA across the width of each boxcar pulse. The red lines indicate the piecewise exponential fit of the simulated error, while the red markers show the resulting error for each trial when computed using the fit. The error derived from the PA distribution of \citet{naghizadeh1993statistical}, shown in orange, appears to underestimate the error, especially at high S/N. }
     \label{fig:PAERR}
     
\end{figure}

\noindent The DSA-110 uses the IAU/IEEE convention that $V > 0$ is right-hand polarization, while $V < 0$ is left-hand polarization \citep{van2010psrchive}. The linear, circular, and total polarization fractions as a function of time are computed as:

\begin{equation}
\frac{L_{\rm unbias}(t)}{I(t)} = \begin{cases}
\sqrt{L(t)^2 - \sigma^2}/I(t),\text{ for } \frac{L(t)}{\sigma} > 1\\
0\text{, o/w}\\
\end{cases}
\end{equation}

\begin{equation}
\frac{C(t)}{I(t)} = \sqrt{V(t)^2}/I(t)
\end{equation}
\begin{equation}
\frac{T(t)}{I(t)} = \sqrt{L_{\rm unbias}(t) + V(t)^2}/I(t)
\end{equation}

\noindent where $L_{\rm unbias}(t)$ is the unbiased linear polarization following \citet{simmons1985point} and \citet{everett2001emission}, and $L(t) = \sqrt{Q(t)^2 + U(t)^2}$ \footnote{Note we adopt the convention used in equations 10 and 16 of \citet{simmons1985point} due to ambiguity in the unbiasing factor reported in \citet{everett2001emission}.}. These are averaged across a burst to estimate the average polarization fractions. For all sources, $T \le I$; in unpolarized sources, $T = 0$. In partially polarized sources,  $T = pI < I$, where $p$ is the polarization fraction.
Finally, the position angle $\chi$ defines the phase difference between linearly polarized components:

\begin{equation}
\chi = 0.5\tan^{-1}(\frac{U}{Q}) = 0.5\angle P
\end{equation}

\noindent where $P = Q + iU$ is the linear polarization vector. Errors in the position angle are estimated using a Monte Carlo simulation of 100 boxcar pulses with $\chi = 0$ and unity off-pulse standard deviation. The resulting error as a function of linear-polarization S/N is well-fit by a piecewise exponential with scale $28\sigma$ for $S/N > 19\sigma$, and $10\sigma$ for $S/N \le 19\sigma$. Figure~\ref{fig:PAERR} shows the results of the simulation and best fit piecewise exponential; note these results imply that the \cite{naghizadeh1993statistical} PA distribution significantly under-estimates the error, particularly for $S/N > 6\sigma$. 

A different dispersion relation applies to orthogonal circularly polarized components for propagation through a plasma-filled region with a magnetic field, resulting in a rotation of the position angle of linearly polarized radiation. The position angle changes as a function of squared wavelength such that the RM is defined as:

\begin{equation}
{\rm RM} = \frac{d\chi}{d\lambda^2}
\end{equation}

\noindent The RM characterizes the integrated line of sight magnetic field $\bar{B}_{||}$ that induces this rotation, weighted by the plasma electron number density $n_e$:

\begin{equation}
    {\rm RM} = 0.81\int \frac{n_e B_{||}}{(1+z(l))^2} dl
\end{equation}

\noindent where $z$ is the redshift of the dispersive material. Note that the $\bar{}$ indicates the field is averaged over the line-of-sight. In general, the RM has contributions from an FRB's host galaxy, the intergalactic medium (IGM), the Milky Way's circumgalactic medium (CGM) and interstellar medium (ISM), and the Earth's ionosphere due to the magnetic fields in these regions. This can, in addition to changing the magnitude of the RM, change the sign due to the direction dependence of the magnetic field. If the same plasma is assumed to contribute to both dispersion and RM:

\begin{equation}
    {\rm DM} = \int_l^0 \frac{n_e}{1 + z(l)} dl
\end{equation}

\noindent the average line-of-sight magnetic field can be estimated as:

\begin{equation}
    \bar{B}_{||}[\mu {\rm G}] \approx  \frac{\rm RM}{0.81 \cdot {\rm DM}}(1+z)
\end{equation}

\noindent Recent analyses imply a correlation between host RM (RM with contributions from the ionosphere, ISM, and IGM subtracted) and DM, implying that this approximation is reasonable \citep{mannings2022fast}. We refer to position angle after correcting for RM is as the polarization position angle (PPA; $\chi_0$), while PA, or $\chi$, refer to the position angle measured without RM correction.


\section{Derivation of Jones Matrix and Parallactic Angle Calibration for DSA-110}\label{app_derivation}

An effective Jones matrix for the coherently combined DSA-110 must be used to calibrate polarized observations, as described by, e.g., \cite{Robishaw_2021}. Calibration with a fully-defined Jones matrix accounts for cross-coupling between the orthogonal linearly polarized receptors, as well as gain and phase differences between the receptors  \citep{heiles2001mueller}. Here we will choose a Cartesian basis aligned with the linearly polarized receptors. For the DSA-110, we operate in the `ideal feed' assumption, that is, that there is zero cross-coupling between receptors. This requires that differences in gain or phase between the X and Y feeds are the dominant instrumental effect, which results in mixing among Stokes parameters. A difference between the X and Y gain will mix Stokes I and Q creating apparent linear polarization. A difference in phase corresponds to a time delay $\Delta t \approx \phi(\nu)(2\pi\nu)^{-1}$ between the X and Y feeds. This mixes Stokes U and V and creates residual circular polarization. Quantitatively, for the arbitrary Jones matrix below consisting of four complex terms:

\begin{equation}
     J(\nu) = \begin{bmatrix} 
    g_{xx}(\nu) &  g_{xy}(\nu) \\
     g_{yx}(\nu) &  g_{yy}(\nu)
    \end{bmatrix}
\end{equation}

\noindent we assume $|g_{xy}(\nu)| = |g_{yx}(\nu)| = 0$. We further simplify this by assuming $g_{yy}(\nu)$ is real and parameterize in terms of the absolute gain in the Y receptor, $ |g_{yy}(\nu)|$, the ratio of the magnitudes of X and Y gains, $r(\nu) = |g_{xx}(\nu)|/|g_{yy}(\nu)|$, and the phase difference between the X and Y receptors, $\phi(\nu)$. The Jones matrix is then simplified to:

\begin{equation}
    J(\nu) = \begin{bmatrix}
    r(\nu)|g_{yy}(\nu)|e^{i\phi(\nu)} & 0 \\
    0 & |g_{yy}(\nu)|
    \end{bmatrix}
\end{equation}
\bigskip

One can estimate $r$ by observing a known unpolarized source and assuming $Q = U = V = 0$. Similarly, one estimates $\phi$ by observing a known linearly polarized source and assuming $V = 0$. For DSA-110 observations, 3C48 was periodically observed as the unpolarized calibrator to estimate $r(\nu)$, while 3C286 was observed as the linearly polarized calibrator to estimate $\phi(\nu)$. Given the fixed declination (71.6$^\circ$) of the DSA-110 commissioning survey, the affect of sky position on polarization calibration is left to future work. For this work, it is reasonable to assume that a single set of calibrators, 3C48 and 3C286, is sufficient for a constant pointing.  At our 1.4 GHz observing frequency, 3C48 is known to be weakly polarized ($\sim 0.5\%$), which we assume is acceptable for use as an unpolarized calibrator within the error of the DSA-110\footnote{In Appendix~\ref{app_leakage} we verify this assumption through leakage polarization tests. For additional information on polarized calibrators, see documentation for the Very Large Array (VLA) at \url{https://science.nrao.edu/facilities/vla/docs/manuals/obsguide/modes/pol}.} \citep{Perley_2013}. 3C286 is known to have 9.5\% linear polarization at 1.4 GHz, making it suitable as a linearly polarized calibrator \citep{Perley_2013}. The absolute gain in the Y receptor, $|g_{yy}(\nu)|$, is estimated using 3C48 by comparing the observed flux to the polynomial fit (after correcting for the measured RM\footnote{Note the RM of 3C48 is measured at higher frequencies ($1.6-30$\,GHz, or $1-18$\, cm wavelengths) where the polarization fraction is higher \citep{Perley_2013}.} of $-68$\,rad\,m$^{-2}$ and estimated parallactic angle) defined by \cite{perley2017accurate}. $r(\nu)$, $\phi(\nu)$, and $|g_{yy}(\nu)|$ were determined after averaging over spurious peaks in Stokes parameters for each source as function of frequency, downsampling in frequency, then interpolating back to native resolution. For FRB observations, we calibrate by applying the inverse Jones matrix to the observed Stokes parameters, effectively setting the gain and phase of the X and Y feeds equal and minimizing the mixing of Stokes parameters.

Absolute position angles cannot be estimated with the current DSA-110 pipeline, as it would require tracking a calibrator source over a wide range of parallactic angles which the DSA-110, as a transit instrument, is not currently equipped for. 
In this work, an attempt to derive absolute position angle estimates by treating the DSA-110 as a single antenna; following \cite{thompson2017interferometry} (Chapter 4), the elevation and azimuth of the array for a given FRB are estimated using the RA, $\alpha$, and declination, $\delta$, of the detected beam and observation time $\rm LST$:

\begin{equation}
    \sin(\theta ) = \sin(b_{\rm DSA})\sin(\delta) + \cos(b_{\rm DSA})\cos(\delta)\cos({\rm LST} - \alpha)
\end{equation}

\begin{equation}
    \sin(\phi + (b - 125)\delta \theta_b) = -\frac{\cos(\delta)\sin({\rm LST} - \alpha)}{\cos(\theta)}
\end{equation}

\noindent where $l_{\rm DSA}$ and $b_{\rm DSA}$ are the longitude and latitude of the DSA-110 (Owens Valley). $b$ is the detected beam and $\delta \theta_b = 14"$ is the synthesized beamwidth. The parallactic angle $\chi_{\rm par}$ is then estimated as:

\begin{equation}
    \sin(\chi_{\rm par}) = -\frac{\sin(\phi)\cos(b_{\rm DSA})}{\cos(\delta)}
\end{equation}

\noindent and can be applied as a correction $\chi_{\rm cal} =\chi_0 + \chi_{\rm par}$.

\section{Calibration Stability with Time and Primary-Beam Location}\label{app_stability}

\begin{figure*}[ht]
\begin{center}
\begin{tabular}{cc}
\includegraphics[width=0.5\linewidth]{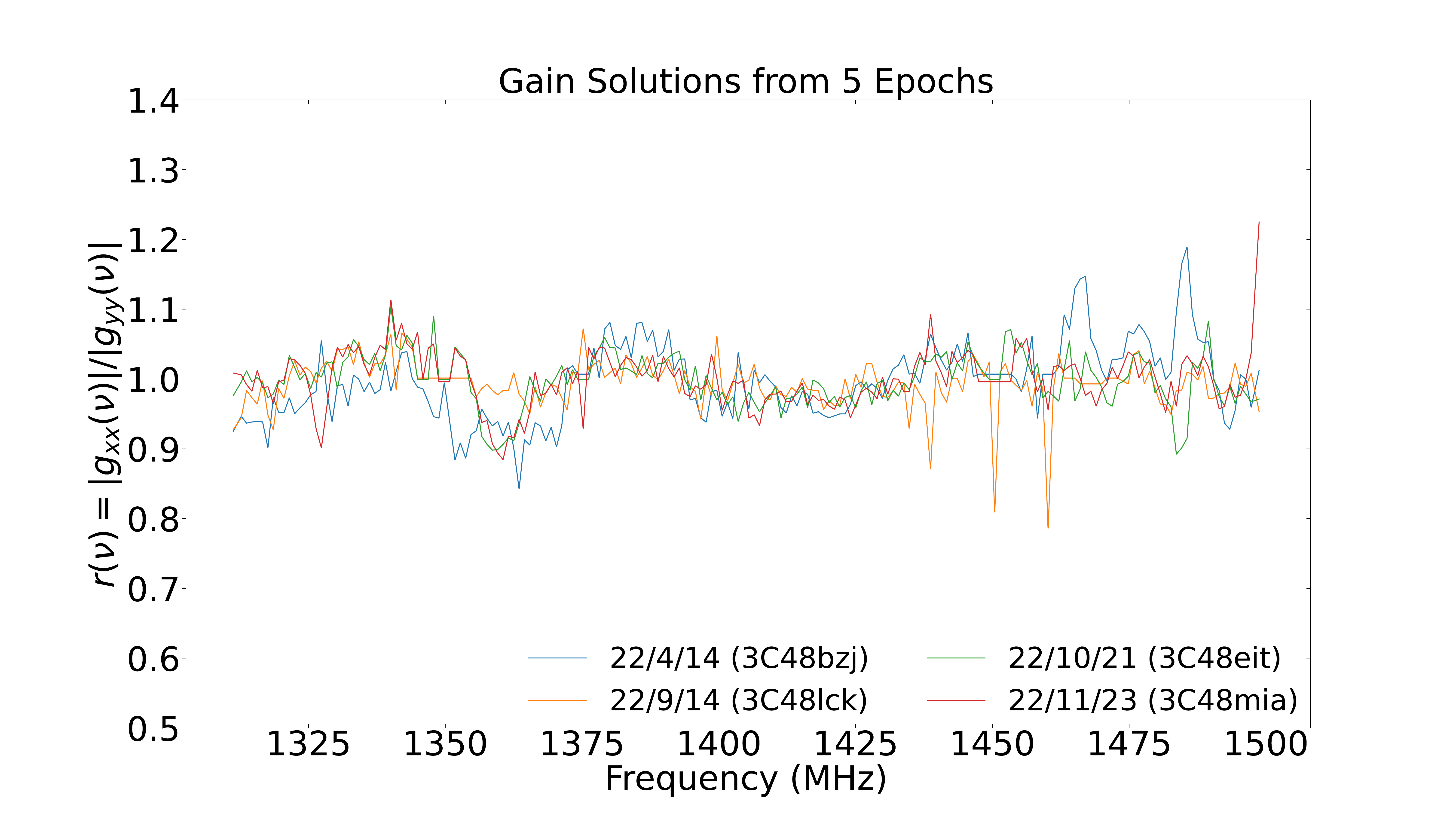} & \includegraphics[width=0.5\linewidth]{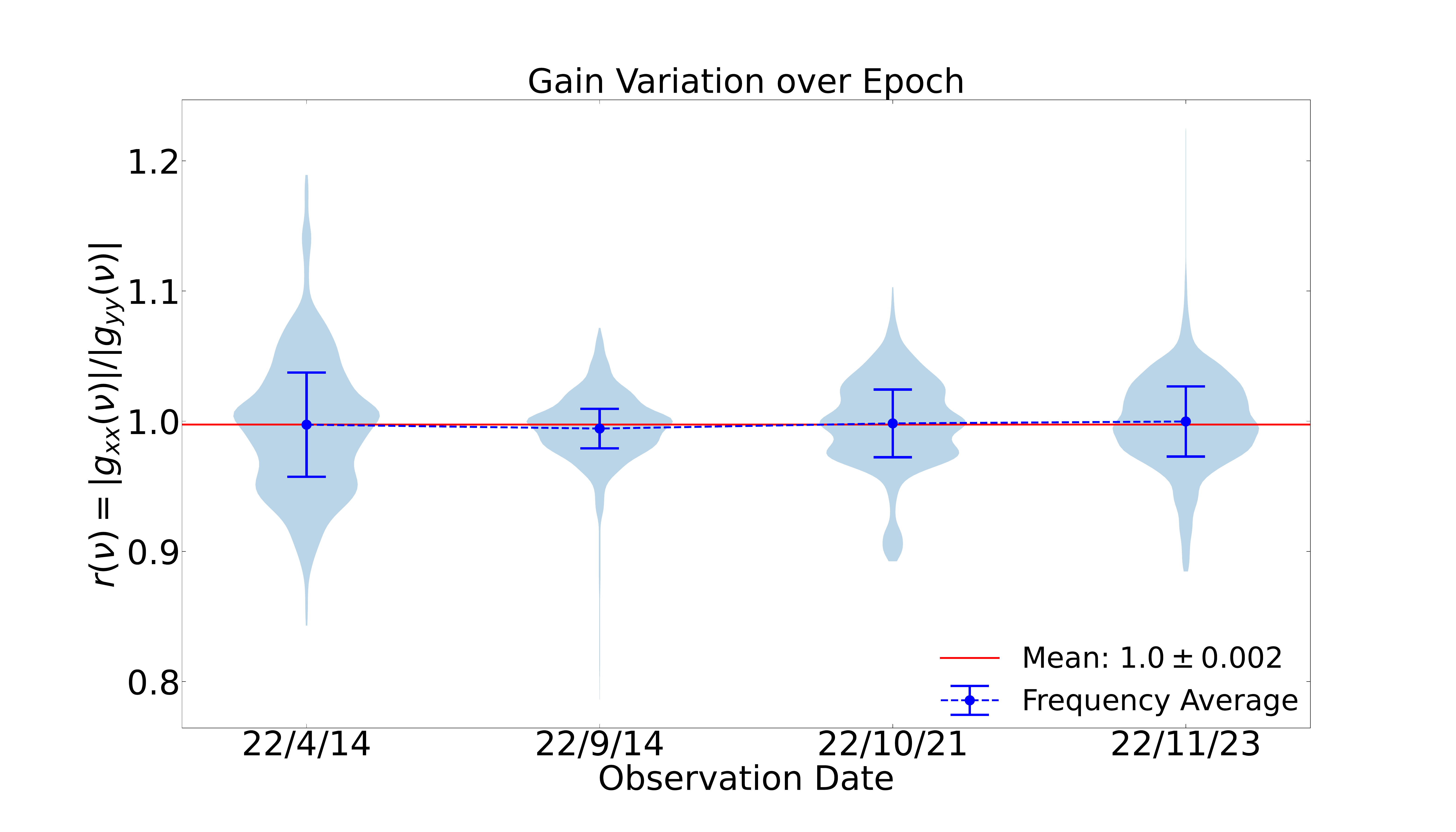}   \\
\includegraphics[width=0.5\linewidth]{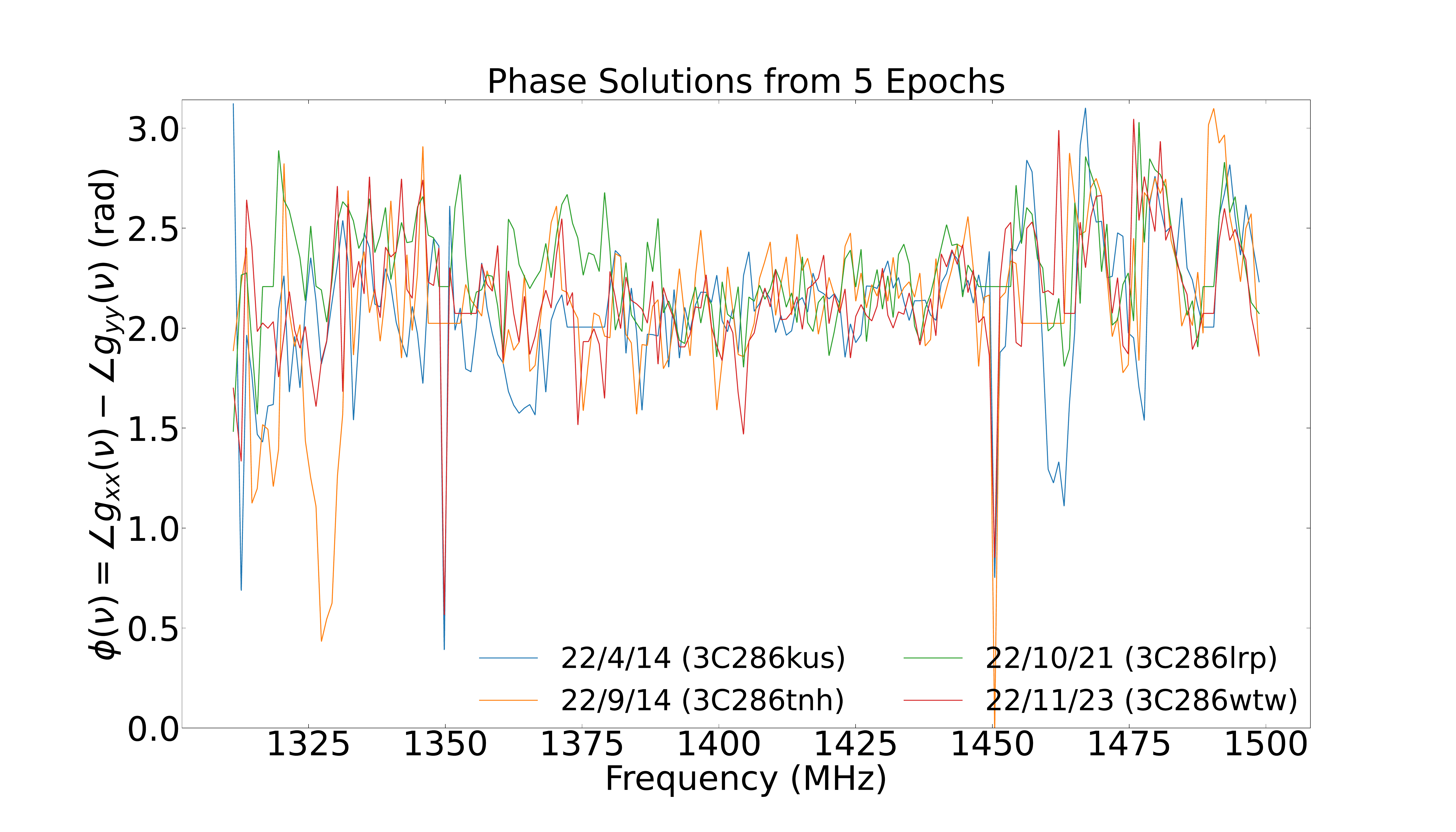} & \includegraphics[width=0.5\linewidth]{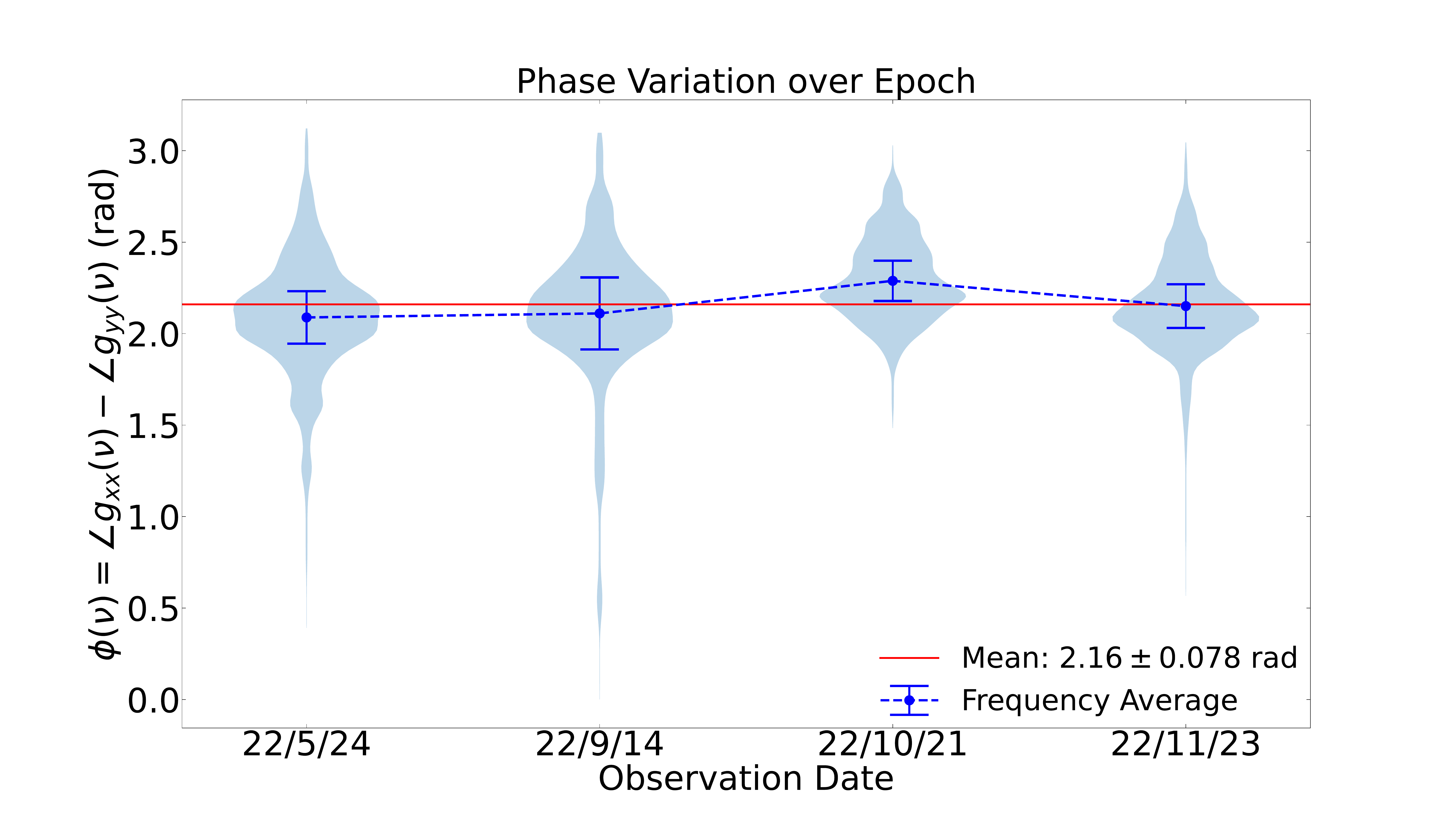}  \\
\includegraphics[width=0.5\linewidth]{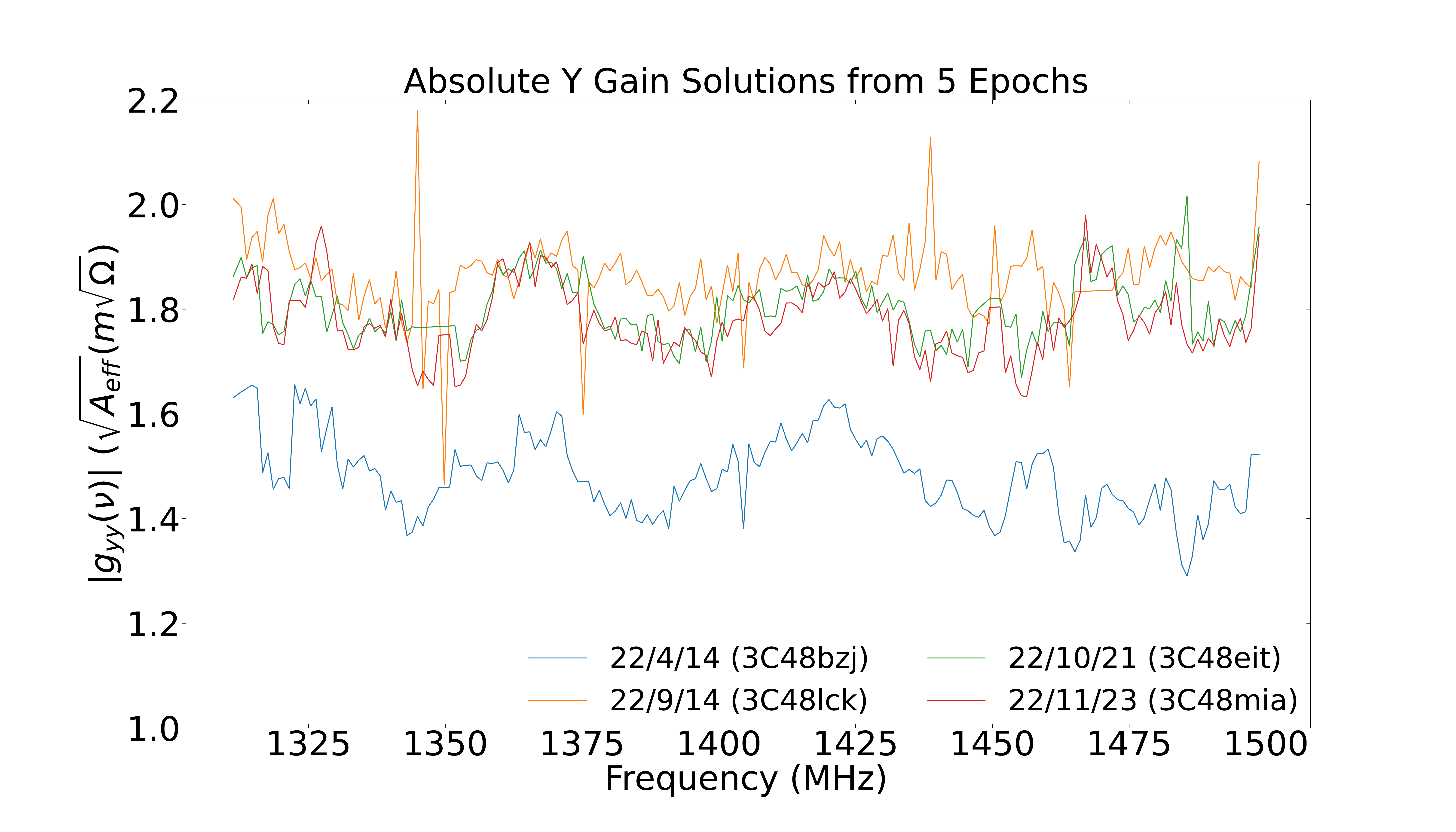} & \includegraphics[width=0.5\linewidth]{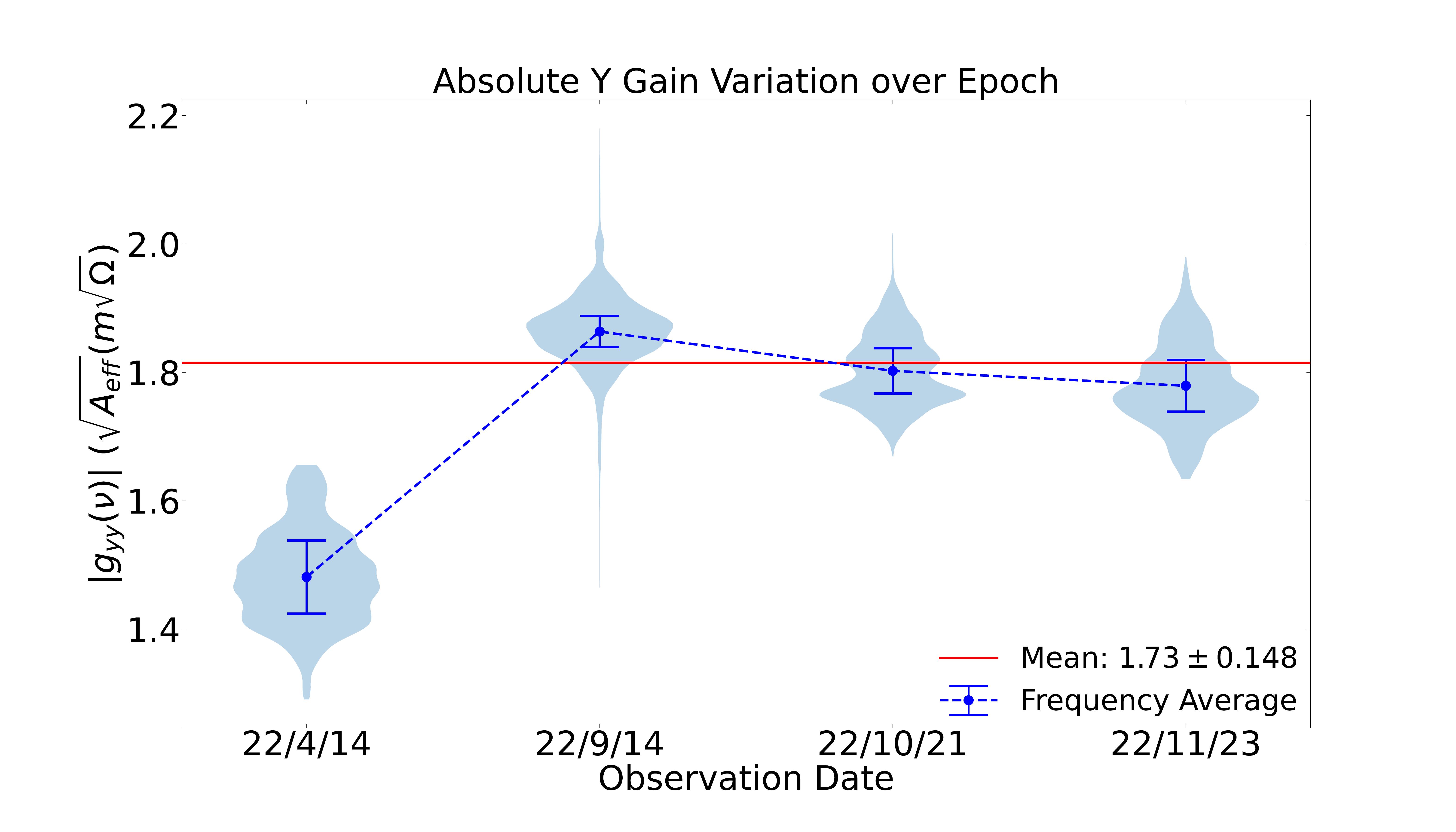} \\
\end{tabular}
\caption{Jones matrix solutions on four epochs. \textit{Left:} Frequency dependent $r(\nu)$, $\phi(\nu)$, and $|g_{yy}(\nu)|$ after removing birdies, downsampling and interpolating. \textit{Right:} Solutions averaged over frequency for each observation. The violin plots show the distribution of the frequency dependent points while the red line shows the average value over all epochs.}
\label{fig:allJones}
\end{center}
\end{figure*}

Four epochs of calibration observations were taken during the survey to apply to the sample of FRBs reported in this paper. Table~\ref{table:CalObsTable} summarizes each observation. Each observation consists of multiple voltage dumps during the calibration-source transit. The transits nearest the center of the primary beam were used to derive Jones matrix parameters at each epoch. Figure~\ref{fig:allJones} shows the Jones matrix solutions as a function of frequency for each epoch. From this, it is clear that the solution is stable, with standard deviation across epoch of 0.2\%, 0.1 rad, and 3.6\% for $r(\nu),\, \phi(\nu), $ and $|g_{yy}(\nu)|$, respectively. Given this stability, the solutions were averaged together over all epochs and filtered in the frequency domain with a first order Savitsky-Golay filter, and this average solution was then used to calibrate all FRBs detected. The outlier $|g_{yy}(\nu)|$ solution from the 2022 April 14 observation was investigated and attributed to a software error which prevented the measurement set from being properly saved. The beamformer weights for this observation were therefore not gain calibrated resulting in a vertical offset in the measured Stokes parameters. Following the correction of this issue, no outliers were identified in subsequent calibration solutions, as shown in Figure~\ref{fig:allJones}. We therefore do not include the $|g_{yy}(\nu)|$ value from the 2022 April 14 observation in the average $|g_{yy}(\nu)|$ solution or the standard deviation reported above.

\begin{deluxetable*}{c | c | c | c | c | c | c  }[ht]
\tabletypesize{\scriptsize}
\tablewidth{0pt} 
\tablecaption{Polarization Calibration Observations and Jones Matrix Parameters}
\label{table:CalObsTable}
\tablehead{
\textbf{\begin{tabular}{@{}c@{}} 3C48 observation \\ name (Beam)\end{tabular} }& \textbf{MJD} & \textbf{\begin{tabular}{@{}c@{}}3C286 observation \\ name (Beam)\end{tabular}}& \textbf{MJD} & $\mathbf{< r( \nu ) >}$ & $\mathbf{<\phi(\nu)>}$ & $\mathbf{<|g_{yy}(\nu)|>}$}
\startdata 
\multicolumn{7}{c}{\textbf{2022/04/14, 2022/05/24 Observation}} \tabularnewline
\hline
kkz (46) & 59683.829 & dnz (18) & 59723.215 & $1.00 \pm 0.04$ & $2.09 \pm 0.14$ & $1.48 \pm0.06$ \\
ane (197) & 59683.837& fvn (57) & 59723.217  &  &  & \\
\textbf{bzj} (110) &59572.309 &  \textbf{kus} (135) & 59723.221 & &  & \\
& & oxr (174) & 59723.223 & & &   \\
\hline
\multicolumn{7}{c}{\textbf{2022/09/14 Observation}} \\
\hline
\textbf{lck} (107) & 59835.417 & mph (44) & 59835.908 & $0.99 \pm 0.02$ & $2.11 \pm 0.20$ & $1.86 \pm0.02$ \\
bxc (182) & 59835.422 & \textbf{tnh} (122) & 59835.912  &  &  & \\
& &  ezb (200) & 59835.916 & &  & \\
\hline
\multicolumn{7}{c}{\textbf{2022/10/21 Observation}} \\
\hline
xtu (47) & 59872.313 & gbf (45) & 59872.807 & $1.00 \pm 0.03$ & $2.15 \pm 0.12$ & $1.78 \pm0.04$  \\
\textbf{eit} (123) & 59872.317 & \textbf{lrp} (123) & 59872.811 &  &  &  \\
jvp (198) & 59872.321 &  fml (201) & 59872.815 & &  &  \\
\hline
\multicolumn{7}{c}{\textbf{2022/11/23 Observation}} \\
\hline
xxc (22) & 59906.219 & quh (19)  &59906.713 & $1.00 \pm 0.03$ & $2.29 \pm 0.11$ & $1.80 \pm0.04$\\
fjr (72) & 59906.222 & xnr (71) & 59906.716 &  &  & \\
\textbf{mia} (123) & 59906.224 &  \textbf{wtw} (123) & 59906.718 & &  &   \\
xol (173) & 59906.227 &  vxo (175)& 59906.721 & &  &  \\
wsj (223) & 59906.230 &  hwj (227) &59906.724 & &  &  \\
\enddata
\bigskip
\textbf{Notes:} Labels in bold are those nearest the center beam and are used to derive the Jones parameters in the last three columns. $<>$ indicates the average value over frequency.
\end{deluxetable*}

In addition to long-term stability, stability across the primary beam was estimated by using multiple Jones solutions from each epoch to calibrate the adjacent voltage dumps during the transits of 3C48 and 3C286. For each epoch, the gain solutions $r(\nu)$ and $|g_{yy}(\nu)|$ were taken as piecewise polynomial fits, while the phase solution $\phi(\nu)$ was taken as the median value across the band in order to reduce the impact of noise. The calibration is deemed stable if there is no systematic change in the resulting polarization fraction and position angle as a function of each source's location within the primary beam. No significant variations in the recovered polarization properties, including RM, for 3C48 and 3C286 were observed. We therefore proceeded to use the beam-center voltage dumps to derive calibration parameters.

\section{Calibration Leakage Estimates}\label{app_leakage}

In the above method, the off-diagonal terms characterizing leakage between the orthogonal receptors are assumed negligible, meaning that calibrated data will be uncertain up to the true magnitude of these terms. The above calibration corrects leakage between Stokes I and Q and between U and V. The leakage terms cannot be explicitly solved for without the observation of additional polarized sources over a wide range of position angles; this is not feasible with the DSA-110. Leakage between the other pairs of Stokes parameters is instead estimated by again calibrating adjacent voltage dumps on 3C48 and 3C286 on each observation epoch as described in the previous section. The remaining Q, U, and V fractions in calibrated 3C48 observations then estimate the remaining I leakage. Subtracting these leakages and the expected $9.6\%$ linear polarization at position angle of $\chi = 33^\circ$ from calibrated 3C286 then allows for other leakage terms to be estimated. Specifically, the residual linear polarization above the expected 9.6\% then corresponds to leakage from V into Q or U; the residual circular polarization is likewise leakage from Q and U into V. One can visualize this with the Mueller polarization matrix formalism \citep{heiles2001mueller}. The I-Q, I-U, I-V leakages then roughly estimate the first row and first column of the Mueller matrix, while QU-V and V-QU roughly estimate the other off-diagonal terms, and the diagonal terms are derived from $r(\nu)$, $\phi(\nu)$, and $|g_{yy}(\nu)|$. While the matrix cannot be explicitly solved based on the 3C48 and 3C286 observations available, the leakage estimates reported here offer order of magnitude estimates.

\begin{figure*}[ht]
\begin{center}
\begin{tabular}{c c}
\includegraphics[width=0.45\linewidth]{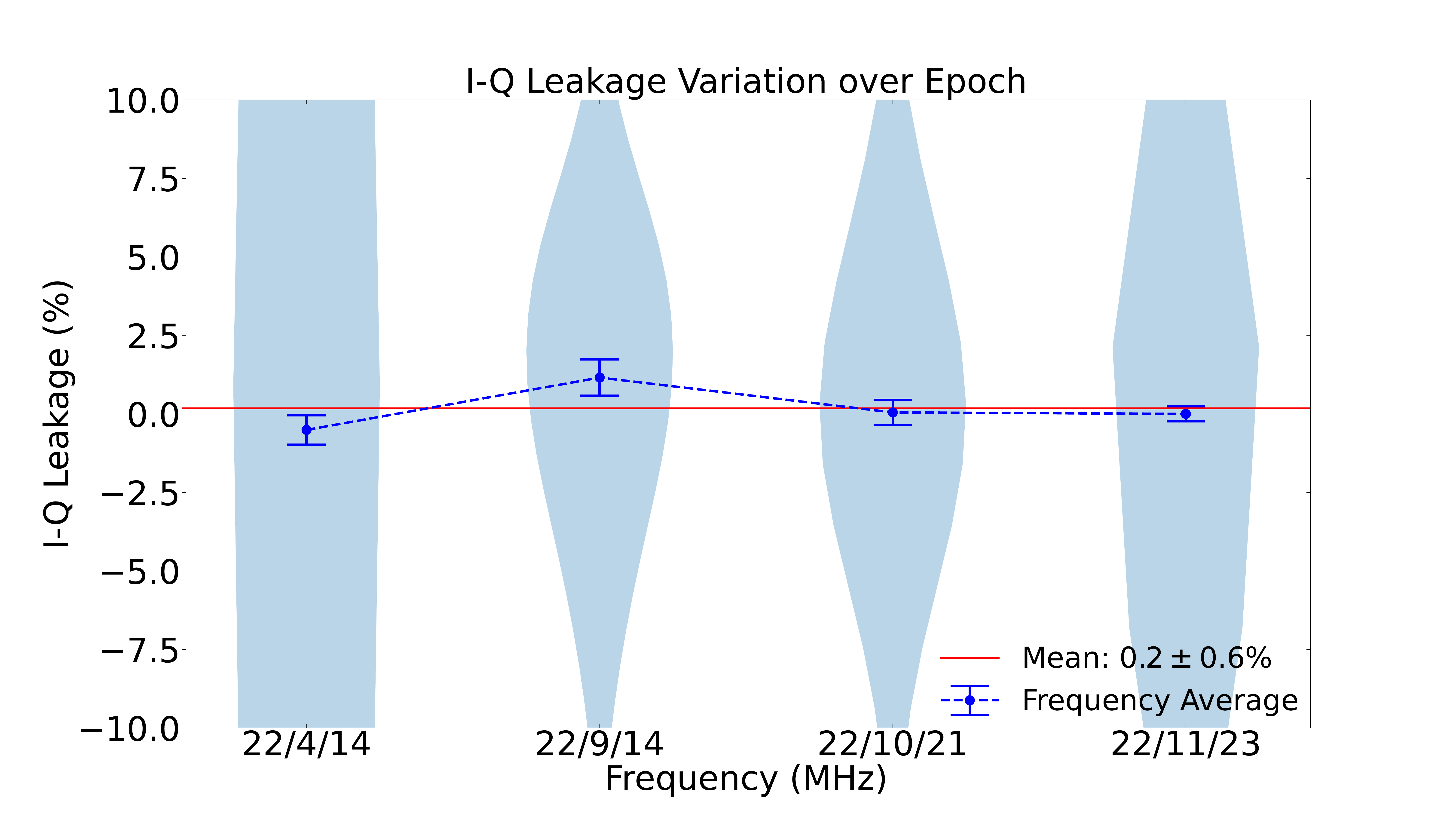}  & \includegraphics[width=0.45\linewidth]{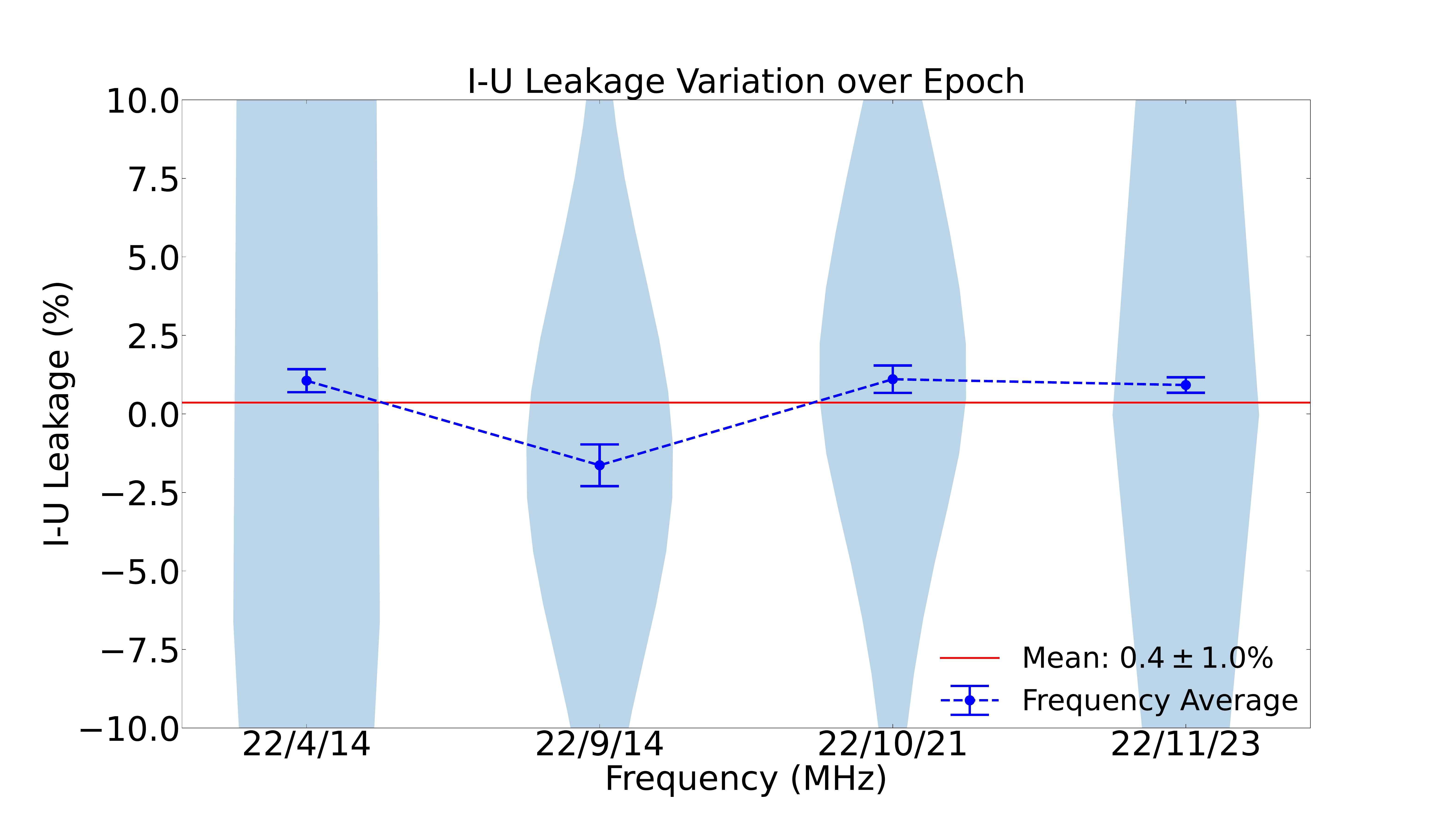} \\
\includegraphics[width=0.45\linewidth]{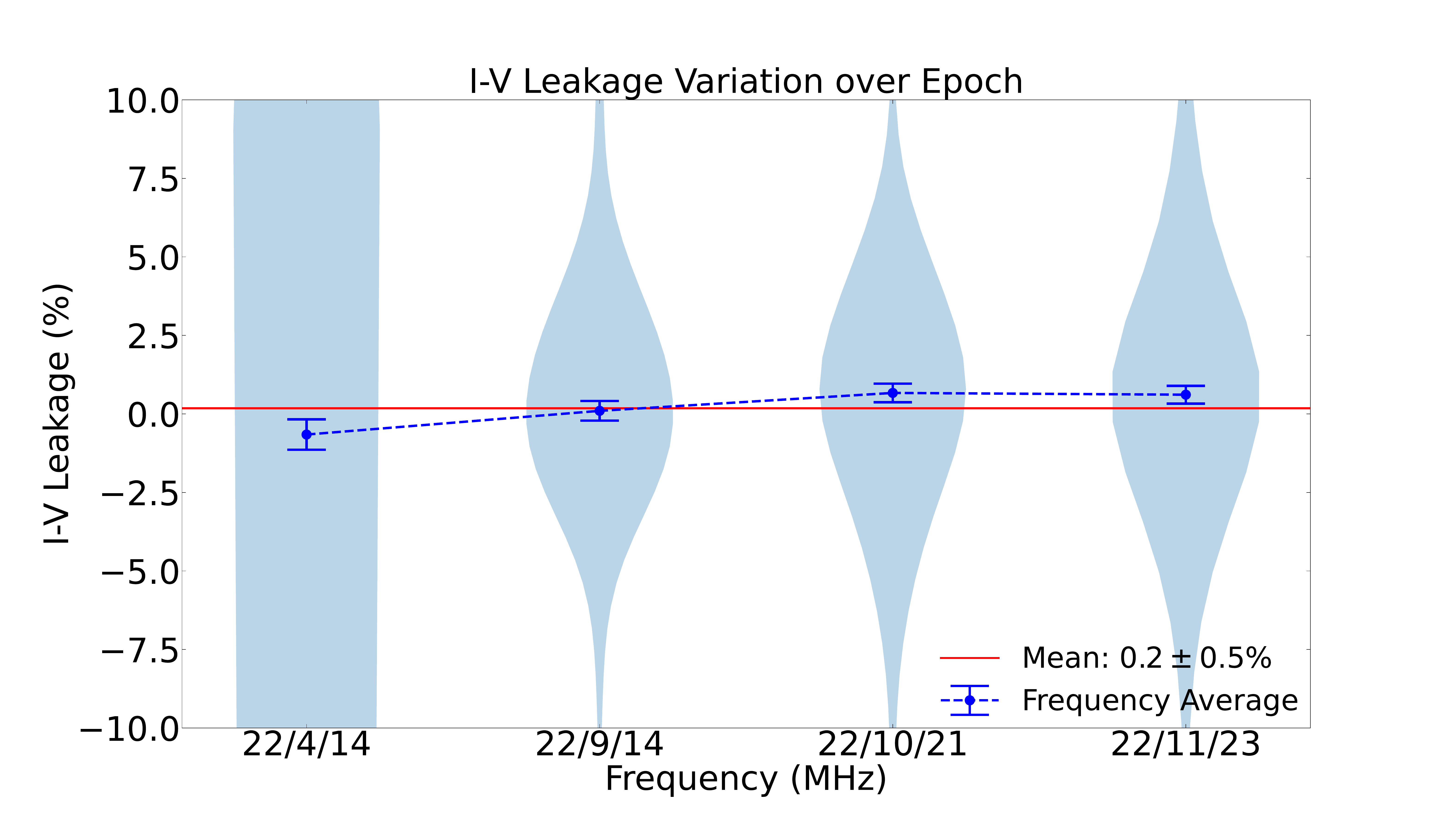} & \includegraphics[width=0.45\linewidth]{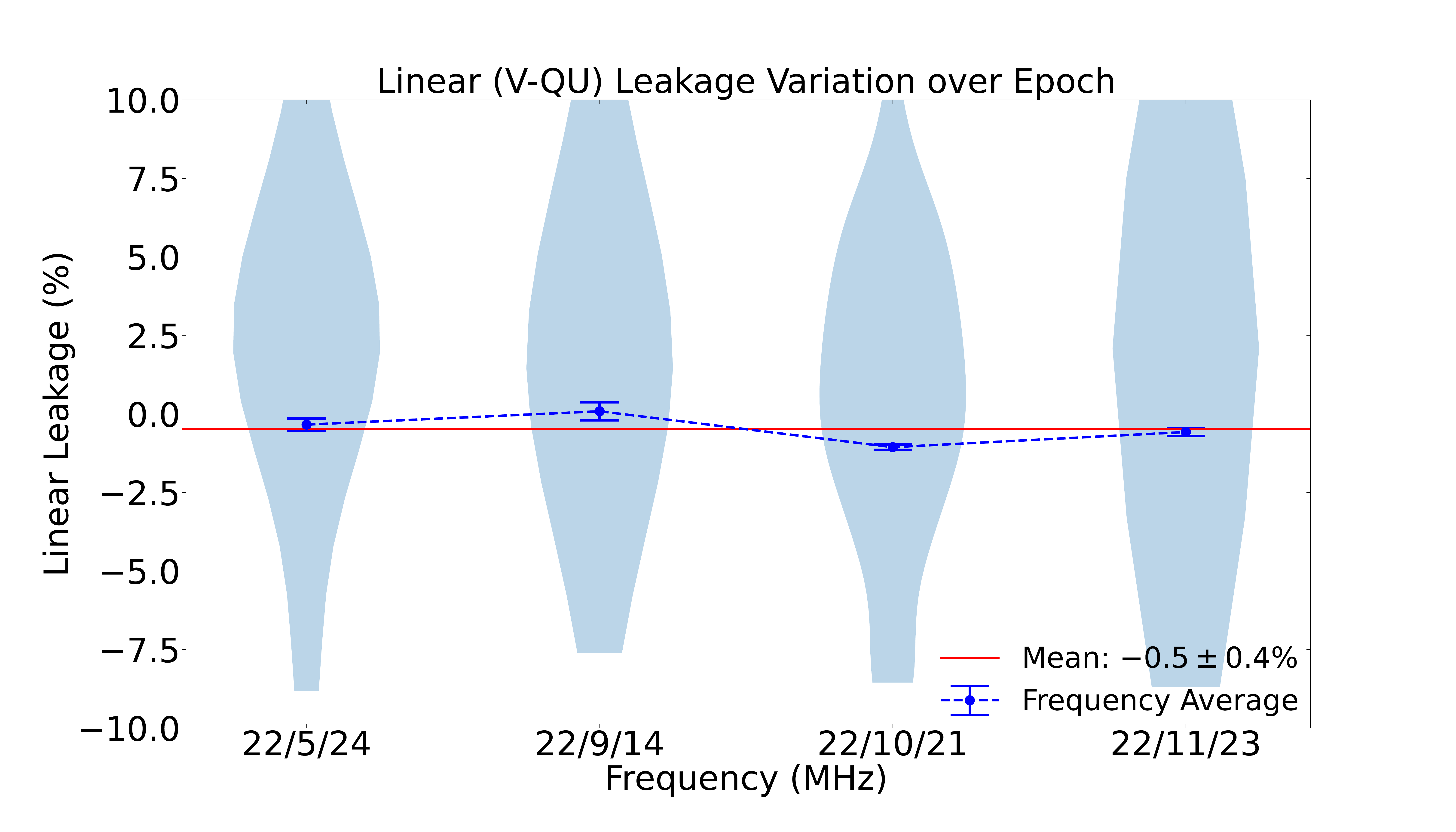} \\
\multicolumn{2}{c}{\includegraphics[width=0.5\linewidth]{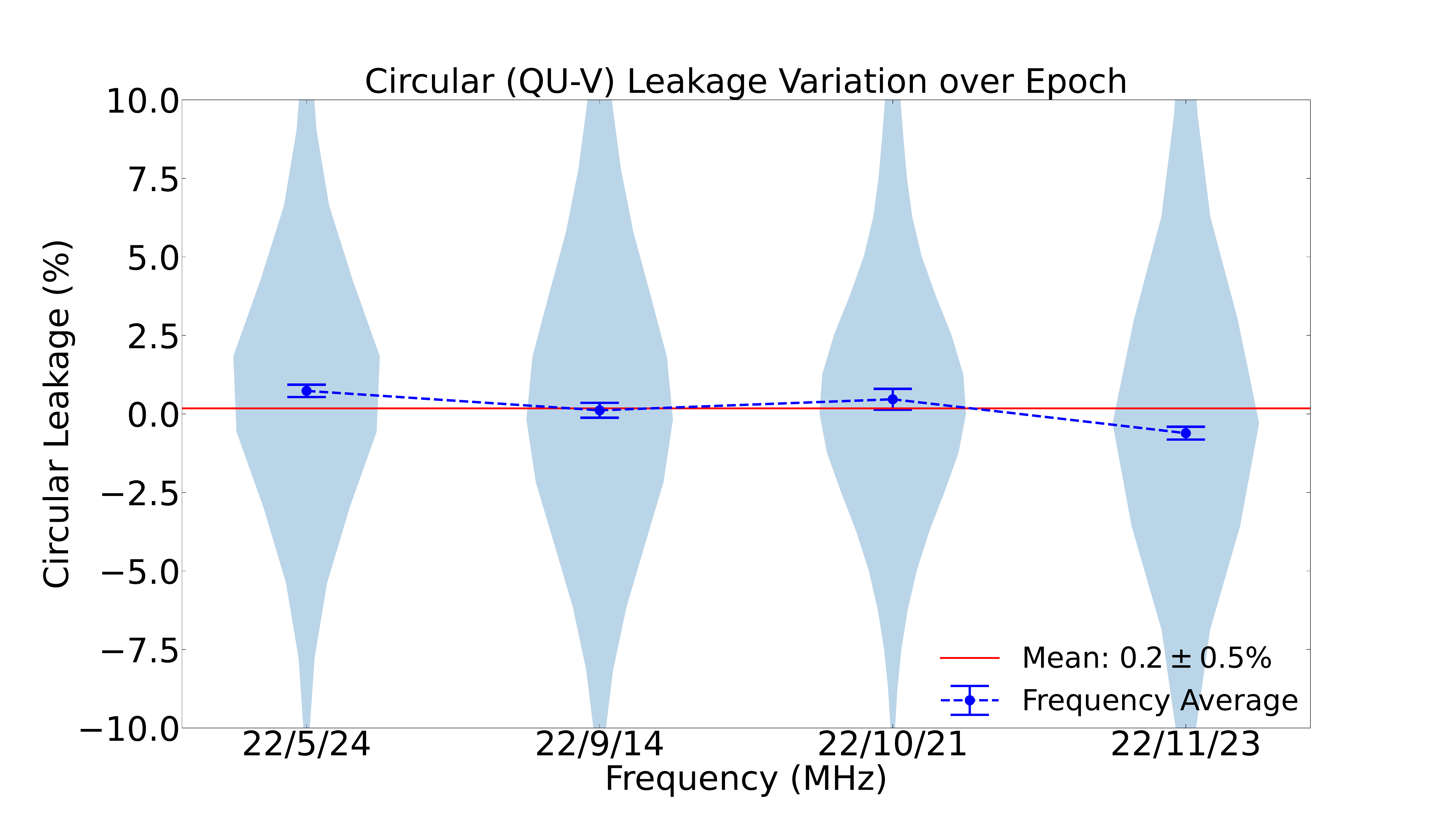}}  \\
\end{tabular}
\caption{Violin plots of leakage estimates. The mean values are shown as markers, while the distributions are values from all calibrated voltage dumps and all frequency channels in each observation. }
\label{fig:LeakageFig}
\end{center}
\end{figure*}

Error bars are estimated by taking the standard error and subtracting the systematic standard error from beams within each voltage dump in which the source was not present, which characterizes the thermal noise. Leakage values for each epoch are summarized in Figure~\ref{fig:LeakageFig}; the maximum leakage observed is $1.6\%$ between I and U. This is within acceptable limits and justifies the ideal feed assumption above.

\section{Example of Polarization Calibration with Pulsar J1935+1616}\label{app_J1935}

\begin{figure*}[ht]
\begin{center}
\begin{tabular}{c}
\includegraphics[width=0.75\linewidth]{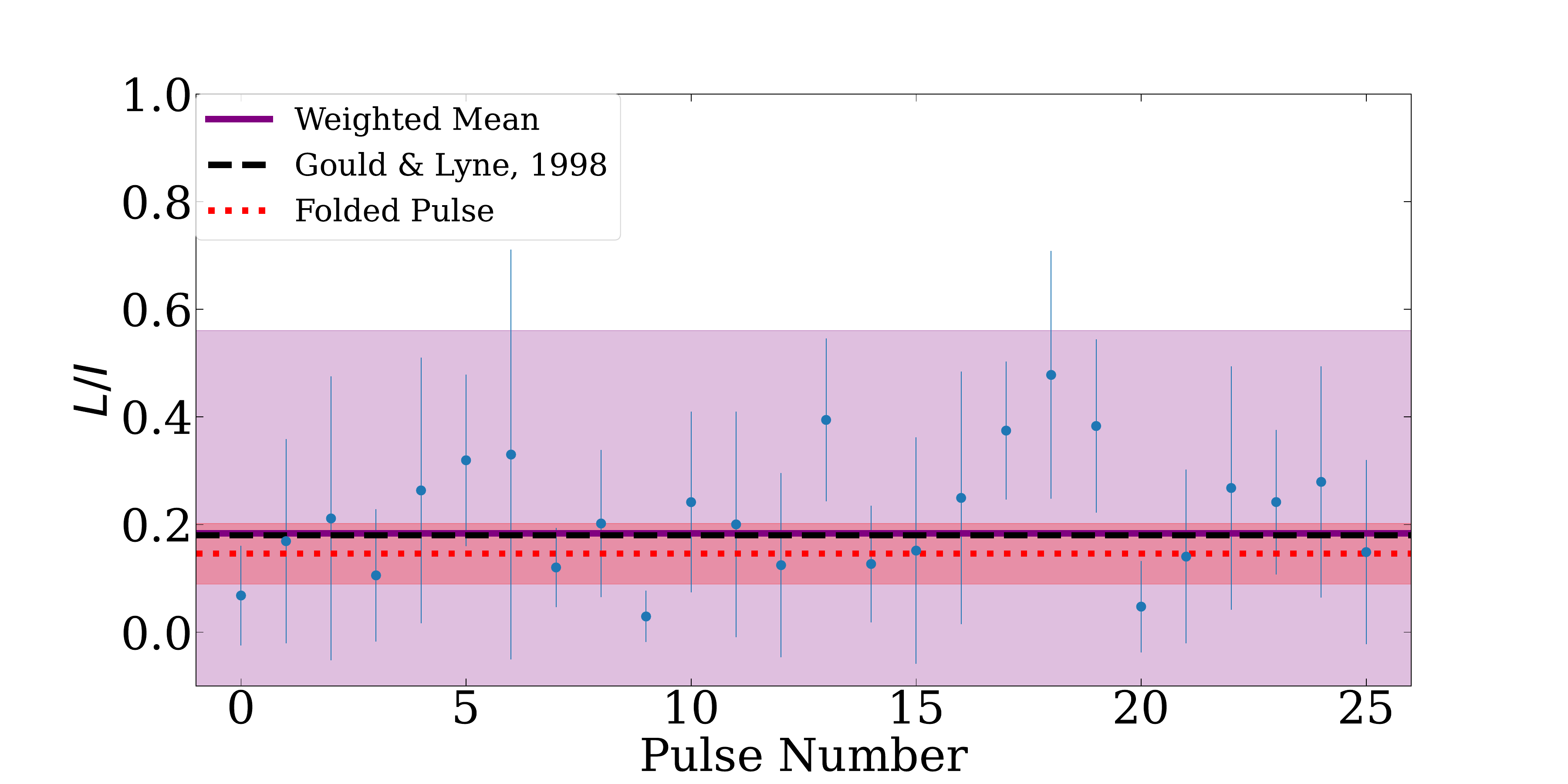}    \\
\includegraphics[width=0.75\linewidth]{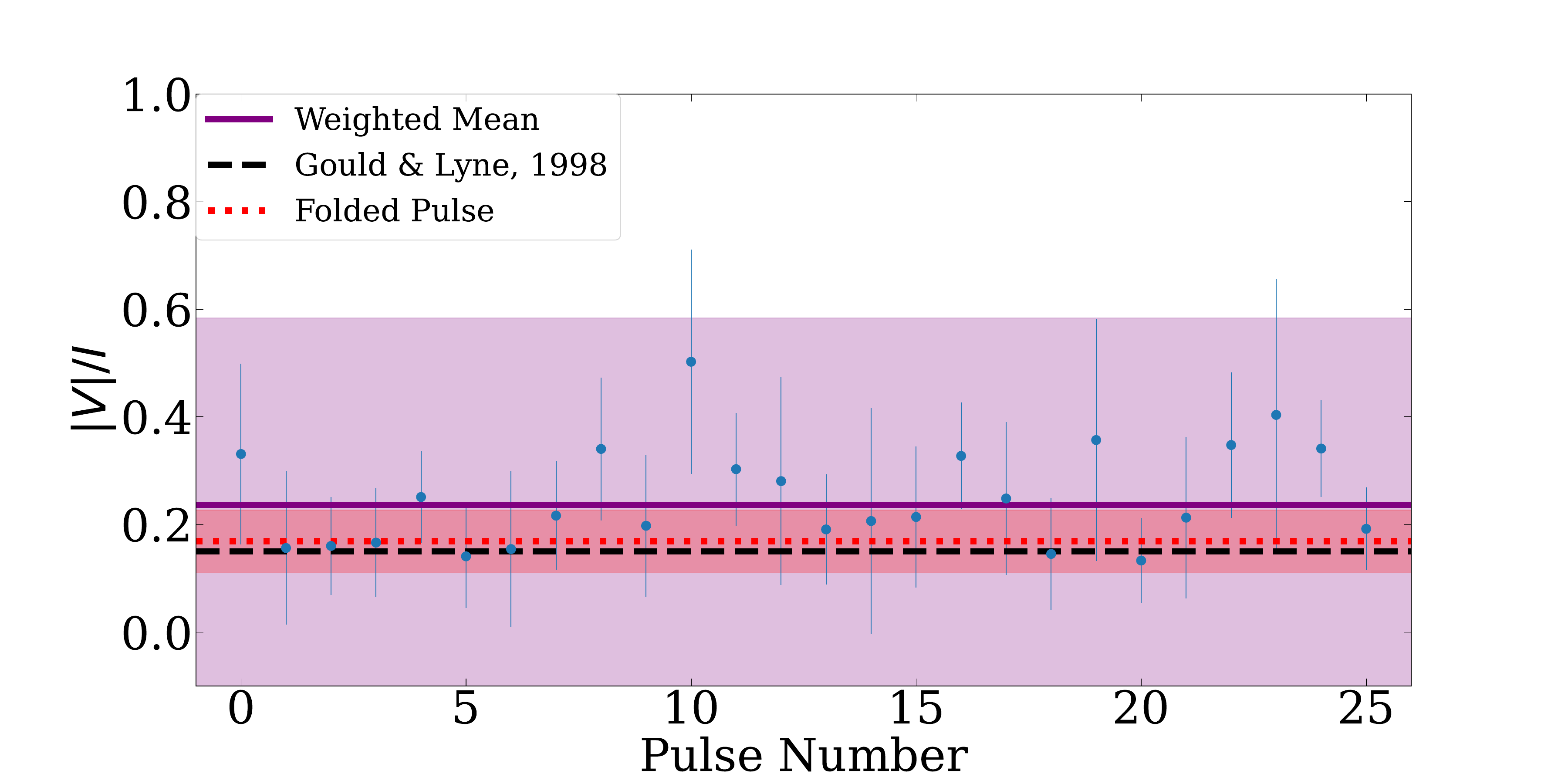} \\ \includegraphics[width=0.75\linewidth]{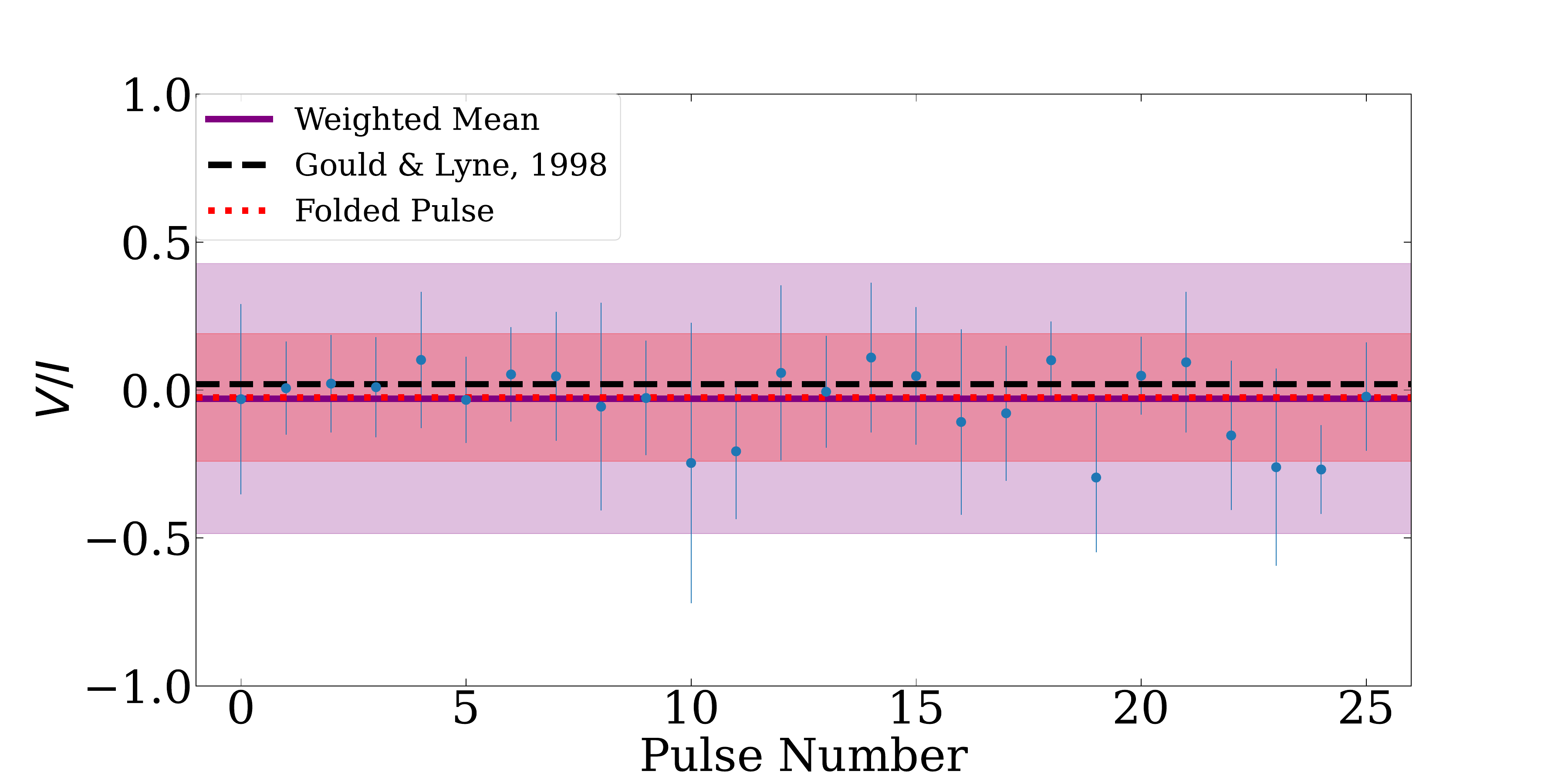}   \\
\end{tabular}
\caption{Linear (\textit{top}), absolute value circular (\textit{middle}), and signed circular polarization (\textit{bottom}) measurements of J1935+1616. Polarization fractions for individual pulses are shown in blue, and the weighted mean is shown in purple, with the shaded area indicating the weighted error. The expected polarization fractions from \citet{gould1998multifrequency} ($18\%$, $15\%$, and $2\%$ from top to bottom) are shown in black. The linear and absolute value circular polarization fractions measured from the folded profiles, shown in red, are in good agreement with expectation within the $1\sigma$ errorbar.}
\label{fig:J1935_ALL}
\end{center}
\end{figure*}

In Section~\ref{observations}, we presented the functioning polarization pipeline using the stacked profile of J1935+1616 shown in Figure~\ref{fig:J1935}. Here we further compute $L/I$, $|V|/I$, and $V/I$ for each of the 26 pulses and for the folded profile to compare to the measured values from \citet{gould1998multifrequency}. The linear polarization of the folded profile is $L/I=14.6\pm5.6\%$, which is lower than the expected $18\%$ but agrees within the $1\sigma$ errorbar. The absolute value circular polarization is $|V|/I=16.9\pm5.7\%$ which is also within $1\sigma$ of the expected $15\%$. The signed circular polarization, $-2.5\pm21.6\%$ agrees with expected $2\%$ within $1\sigma$ despite the opposite sign. From the pulse profiles in Figure 1 it is clear that this difference in sign is stochastic rather than a sign error in calibration\footnote{In this paper we adopt the IAU/IEE convention for Stokes V, in which $V>0$ is right-handed and $V<0$ is left-handed. \citet{gould1998multifrequency} use the PSR/IEE convention which has the opposite definition. Therefore, we have flipped the sign of Stokes V in both the EPN profile in Figure 1 and the expected $V/I$ so that both are in the IAU/IEE convention for more accurate comparison.}. From this we conclude that there is good agreement between the DSA-110 polarization measurement of J1935+16 and expectation and find no significant systematic errors to apply. Figure~\ref{fig:J1935_ALL} shows the polarization fractions of each of the 26 pulses as well as those of the average profiles. Note the low expected RM, $-2.3\pm0.5\,$rad\,m$^{-2}$, and the low detected linear-polarization S/N prevent a confident RM measurement for verification.

\section{Optimal Spectral Estimation}\label{app_weights}

After calibration of each FRB, full-Stokes spectra are calculated by averaging over the burst time series. In order to prevent inclusion of noise in the spectrum, a window around each burst is first manually defined to contain all apparent emission components, over which the average spectrum is computed. Ideally, the pulse would then be weighted using the variance at each time step. These variances are estimated as the inverse of the frequency averaged S/N at each time step, i.e., the temporal profile of the burst. Previous studies have used a Gaussian fit and similar functions to estimate pulse profiles. However, the large variety in burst morphology among FRBs implies a Gaussian may not offer a reasonable fit. The pulse profile is determined by first averaging the total-intensity dynamic spectrum over frequency, then downsampling and linear interpolating to the desired time resolution to eliminate stochasticity and insignificant mid-pulse structure. Finally, to remove artifacts in due to the interpolation, the profile is filtered with a third-order Savitsky-Golay filter, with the filter width chosen specific to each burst. The profile is set to zero outside of the burst window defined above. The spectrum can then be computed as a weighted sum of the time series within the defined window using the profile as weights.

\section{Detailed RM Synthesis Derivation and S/N Method Error}\label{app_RMderivation}

What follows is a detailed derivation of the RM Synthesis method and its variation, the S/N Method, utilized in this paper \citep{heald_2008,brentjens2005faraday,burn1966depolarization}. In the discrete case, \citet{burn1966depolarization} estimates the Faraday dispersion function (FDF) as the Fourier transform of the observed linear polarization vector $P_{\rm obs}(\lambda^2) = Q_{\rm obs}(\lambda^2) + iU_{\rm obs}(\lambda^2)$ from $\lambda^2$ space to RM space:

\begin{equation}
    \hat{F}({\rm RM}) = \frac{1}{N_{\lambda^2}}\sum_{\lambda^2}{P_{\rm obs}(\lambda^2)W(\lambda^2)(\lambda^2)e^{-2i{\rm RM}(\lambda^2 - \lambda_0^2})}
\end{equation}

\noindent where $\lambda_0$ is the mean of the $\lambda^2$ axis, and $W(\lambda^2)$ are weights applied to the spectra. Note that $Q_{\rm obs}(\lambda^2)$ and $U_{\rm obs}(\lambda^2)$ are the optimally weighted spectra of Q and U as described above. $\hat{F}({\rm RM})$ is proportional to the linear polarization for a given RM correction, and this can therefore be interpreted as a linear polarization maximization method. The most probable value of the RM then occurs where $\hat{F}({\rm RM})$ peaks; a parabola is fit to an oversampled RM range around the peak and its maximum taken to be the RM estimate. An error on this value is estimated as the full-width at half-max (FWHM) of the RM spread function (RMSF), divided by the linear-polarization S/N:

\begin{equation}
    \sigma_{\rm RM1} = \frac{{\rm FWHM} \sigma_{L}}{{\rm max}(\hat{F}({\rm RM}))}
\end{equation}

\noindent where $\sigma_L$ is the off-peak standard deviation in the linear-polarization signal, $|P(t)|$, after applying the ideal spectral weights described in the previous section. Sources here are assumed to be `Faraday thin', meaning that for a small emitting volume, the amount of Faraday rotation that can occur is strictly limited; each FRB with significant RM is then assumed to have a dispersion function that peaks at only one value \citep[e.g.,][]{brentjens2005faraday}.

RM synthesis is first performed with the \textit{RM-tools} package\footnote{\url{https://github.com/CIRADA-Tools/RM-Tools}}, a Python implementation which also uses `cleaning' to minimize sidelobes in the RM spectrum \citep{2020ascl.soft05003P}. In order to exercise more control over the trial RM range, resolution, and synthesis method, an RM synthesis pipeline was implemented within the \textit{dsa110-pol} library as \textit{faradaycal}. In this function, the RM range and resolution can be specified, and since synthesis is implemented by brute force without optimization, the trial RM range can be arbitrarily wide and fine, and within arbitrary limits. Following synthesis with \textit{RM-tools}, \textit{faradaycal} is run on a fine RM trial grid ($10^6$ trial RMs from $\pm10^6\,\rm{rad}\,m^{-2}$ for a resolution $2\,\rm{rad}\,m^{-2}$). A smaller region of $\pm 1000\, \rm{rad}\,m^{-2}$ is then defined around the peak in the FDF and run with 5000 trial RMs for a resolution of $0.4\,\rm{rad}\,m^{-2}$ to obtain a better estimate. 

The traditional RM synthesis methods described above operate on the time-averaged spectrum of an FRB, and therefore neglect any time-variation in polarization properties that could warp the measured RM. To address this, an additional RM synthesis function, \textit{faradaycal\_SNR} has been implemented within the \textit{dsa110-pol} library. Rather than operating on the time-averaged spectrum, \textit{faradaycal\_SNR} instead applies for each trial RM de-rotation at each time-step within the burst as shown below:

\begin{equation}
    P_{\rm RM}(t,\lambda^2) = P_{\rm obs}(t,\lambda^2)(\lambda^2)e^{-2i{\rm RM}(\lambda^2 - \lambda_0^2})
\end{equation}

\noindent The linear-polarization S/N for each trial RM, $\hat{F}_{\rm S/N}$, is then calculated by averaging $|P_{\rm RM}(t,\lambda^2)|$ over frequency, then taking the weighted average over the pulse, using the ideal spectral weights. The detected RM is taken to be the RM such that $\hat{F}_{\rm S/N}$ is maximized. This not only includes the time dependence of polarization properties, but also more closely matches how RM calibration is applied, i.e. de-rotation is applied at each time step. However, this method, referred to as the `S/N Method' is more computationally expensive, and can therefore only be run on the smaller RM range of $\rm \pm 1000 \,\rm rad\,m^{-2}$ around the peak determined from running \textit{faradaycal}. In addition to obtaining the RM through synthesis with the time averaged S/N, the S/N method is performed on the spectrum at each timestep within the burst to investigate any time dependence on sub-millisecond timescales.

\begin{figure}[ht!]
     \centering
     \includegraphics[width=\linewidth]{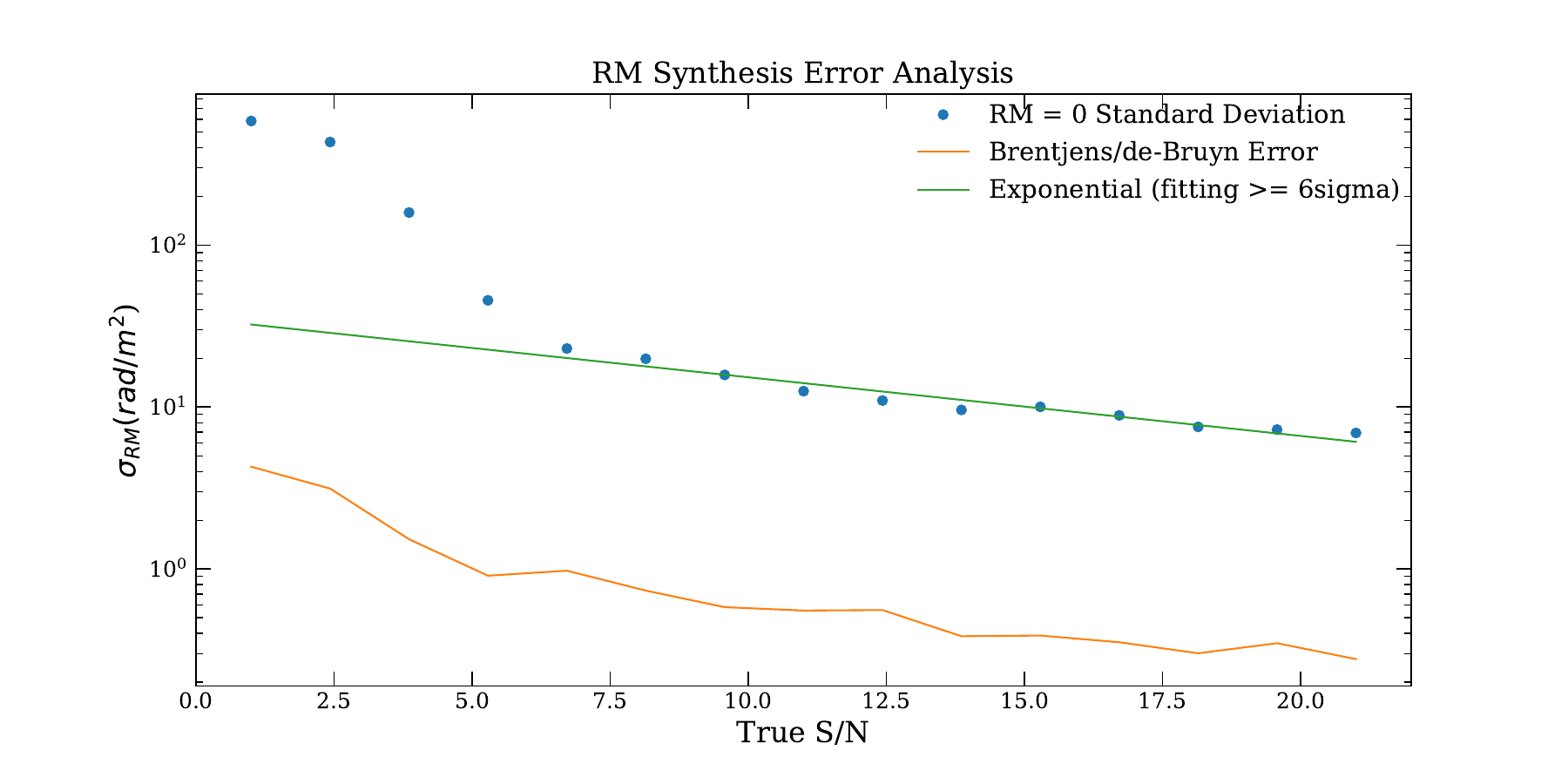}
     \hfill
     \caption{Simulated error in the recovered RM through RM Synthesis, as a function of linear-polarization S/N. See text for details of the simulation. For each true S/N, the mean analytic RM-uncertainty estimate from \citet{brentjens2005faraday} is shown as an orange curve. The blue points show the standard deviations of the recovered RMs, which are fit by the green line.}
     \label{fig:RMERR}
     
\end{figure}

\cite{brentjens2005faraday} estimate the error in RM determined from RM synthesis to be equivalent to that estimated from QU-fitting, shown below, in the limit of high S/N:

\begin{equation}
    \sigma_{\rm RM} = \frac{\sqrt{3}\sigma}{4|P|(\lambda^2_{\rm max} - \lambda^2_{\rm min})}
\end{equation}

\noindent where $|P|$ is the linear polarization at the peak RM, and $\sigma$ is the root-mean-squared (RMS) noise in the Q and U maps. However, many of the FRBs detected in this sample have low S/N, and this error is likely an underestimate. A more conservative error is estimated through a small simulation, in which 100 top-hat pulses are simulated at linear-polarization S/N trials between 1,2,...,20. The RM is then estimated for each trial and the error taken to be the standard deviation of all trials. The results of this simulation, as well as the error estimate from \cite{brentjens2005faraday} are shown in Figure~\ref{fig:RMERR}. From this, it is clear that there is higher scatter in RM than implied by the formula above. Since a $9\sigma$ threshold is imposed on the peak $\hat{F}_{\rm S/N}$ based on the linear-polarization S/N needed to clearly resolve a peak in the RM spectrum\footnote{Note this threshold is determined through trial and error and was still subject to visual scrutiny in the reported results.}, an exponential of the form:

\begin{equation}
    \sigma_{\rm RM2} = Ae^{B\hat{F}_{\rm S/N}}
\end{equation}

\noindent is fit to the simulated error above $\sim6\sigma$. A least-squares fit results in best-fit parameters $A = 35.17$, $B = -0.08$. This exponential is shown in Figure~\ref{fig:RMERR} .

We apply the standard RM synthesis methods to each FRB on the full range of sensitivity from $\pm 10^6$\,rad\,m$^{-2}$. This range was estimated following the method outlined by \cite{mckinven2021polarization}, which computes the maximum RM before the PA, $\chi$ can change significantly due to Faraday rotation. The change in PA that can occur within a frequency channel is given by \cite{burke2019introduction}:

\begin{equation}
    \Delta \chi = \frac{-2{\rm RM}c^2\Delta\nu}{\nu_c^3}
\end{equation}

\noindent where $\Delta\nu$ is the width of each frequency channel and $\nu_c$ is the center frequency. The fraction of the initial polarization as a function of RM is then \cite{schnitzeler2015rotation}:

\begin{equation}
    f = \frac{\sin(\Delta \chi)}{\Delta \chi}
\end{equation}

\begin{figure}[ht!]
     \centering
     \includegraphics[width=\linewidth]{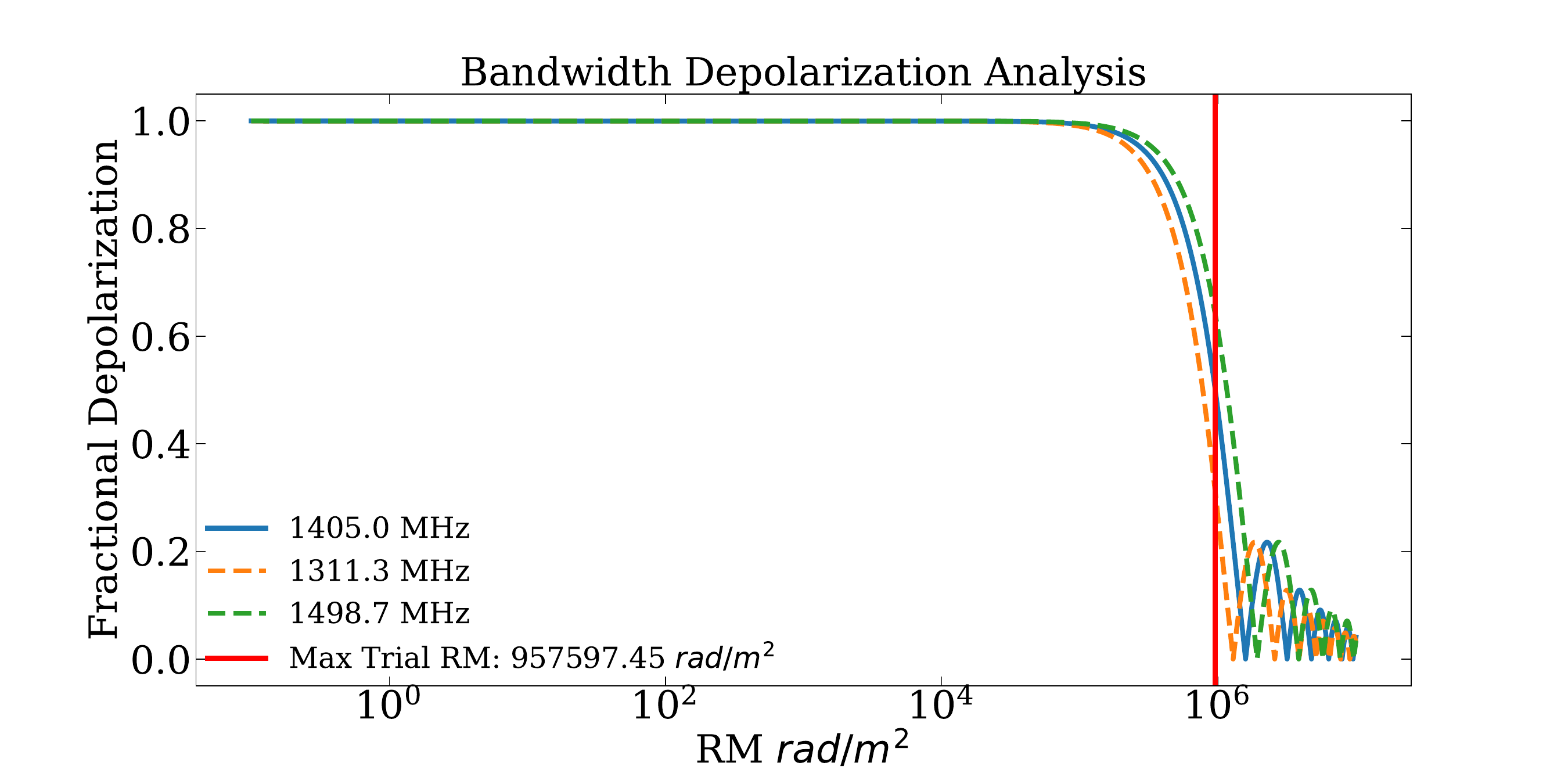}
     \hfill
     \caption{Fractional depolarization the DSA-110 as a function of RM, for different frequencies within the DSA-110 band.}
     \label{fig:BWdepol}
     
\end{figure}

\noindent Figure~\ref{fig:BWdepol} shows a plot of $f$ as a function of trial RM. The DSA-110's high frequency resolution means that it is sensitive to RM rotation up to $\mathrm{RM}_{\rm max} = 9.6^{+2.0}_{-1.8}\times 10^5$\,rad\,m$^{-2}$, making it well-equipped to detect RM's larger than the current maximum (FRB\,20121102, which has an RM on the order of $\sim 10^5$\,rad\,m$^{-2}$) \citep{hilmarsson2021rotation}.

\section{An RM-scattering scenario for FRB 20220319D (Mark)}\label{app_mark}

Recall that for FRB\,20220319D, $\rm DM_{\rm host}$ and therefore $\bar{B}_{||}$ could not be accurately estimated given its close proximity to the Milky Way \citep{ravi2023deep}. One can instead place limits on the host magnetic field by assuming the derived $\sigma_{\rm RM}$ and using a scattering timescale, $\tau_s$, assuming that all depolarization is due to stochastic RM variation. A detailed derivation is described in the supplementary material of \citet{feng2022frequency}, and a short summary is provided here.

If all refractive scattering is contributed by a screen within the host galaxy at a distance $R$ from the source, the scattering timescale $\tau_s$ is related to the refraction angle $\theta_s$ by:
\begin{equation}
    \tau_{s} \approx \frac{R\theta_{s}^2}{2c}
\end{equation}
The spatial scale of fluctuations in the electron density causing this refraction is then:
\begin{equation}
    l_{\rm scat} \approx \theta_s R \approx \sqrt{2c\tau_s R}
\end{equation}
If it is further assumed that the screen has stochastic RM variations with standard deviation $\sigma_{\rm RM}$ around mean $|{\rm RM}_{\rm scat}|$, they can be written as:
\begin{equation}
    \sigma_{\rm RM} \approx \sqrt{\frac{R}{l_{\rm scat}}}{|{\rm RM}_{\rm scat}|}. 
\end{equation}
$|{\rm RM}_{\rm scat}|$ is created by a perturbation in the electron-density or magnetic field. Assume for simplicity that only the electron density $n_e$ is perturbed so that:
\begin{equation}
    \sigma_{\rm RM} \approx \sqrt{R l_{\rm scat}}\frac{r_e^2}{2\pi e}\bar{B}_{||}\delta n_e
\end{equation}
where $r_e$ is the classical electron radius, and $e$ is the electron charge. Solving for $\bar{B}_{||}\delta n_e$,

\begin{equation}
    \bar{B}_{||}\delta n_e \approx \frac{2\pi \sigma_{\rm RM}e}{r_e^2 (2c\tau_s R^3)^{1/4}}.
\end{equation}

In FRB\,20220319D's case, the low ${\rm DM} = 110.95$\,pc\,cm$^{-3}$ suggests the scattering screen is not located near the source. Assume instead that $R \sim 1$\,kpc. Consider first taking the estimated $\sigma_{\rm RM} = 20.81$\,rad\,m$^{-2}$, and the upper limit on scattering derived for FRB\,20220319D, $\tau_s \le 0.061$\,ms \citep{ravi2023deep}. This results in a lower limit of $\bar{B}_{||}\delta n_e \gtrsim 139.9\,\mu$G\,cm$^{-3}$. For electron-density variations of order $\delta n_e \sim 0.01 - 0.1$ cm$^{-3}$ (estimated from $DM$ and $R$), this would require a magnetic field $\bar{B}_{||} \lesssim 1399.0 - 13989.5\,\mu$G, which would be relatively high compared to the typical ISM magnetic field, but consistent with a dense (e.g., star-forming) intervening environment. 

Alternatively, \cite{feng2022frequency} identifies a correlation of $\sigma_{\rm RM}$ with $|{\rm RM}|$ that can be approximated by $|{\rm RM}| \propto 100\sigma_{\rm RM}$. Following this rule, a more likely value for FRB\,20220319D would be $\sigma_{\rm RM} \approx 0.73\,$\,rad\,m$^{-2}$, implying the intrinsic emission is not $100\%$ polarized\footnote{The RM value used here is derived from as host galaxy contribution to the RM, which is derived in the companion paper, \citet{sherman2023deep}}.. Under this assumption, the upper limit becomes $\bar{B}_{||}\delta n_e \gtrsim 4.9\,\mu$G\,cm$^{-3}$. Again using electron-density variations of order  $\delta n_e \sim 0.01 - 0.1$\,cm$^{-3}$, this would require a magnetic field $\bar{B}_{||} \lesssim 73.9 - 739.5\,\mu$G. Although this is still large, it is more easily achievable in a Milky Way-like environment. The two scenarios give insignificant mean $|{\rm RM}_{\rm scat}|<<{\rm RM}$, equal to $3.9\times10^{-3}$\,rad\,m$^{-2}$ and $0.1\times10^{-3}$\,rad\,m$^{-2}$, respectively. In summary, an RM scattering scenario for the depolarization of FRB\,20220319D is likely more plausible than for other DSA-110 FRBs, but would nonetheless require an unusually dense and magnetized intervening system. 

\section{PPA Variation Analysis}\label{app_chi}

To estimate the significance of PPA ($\chi_0$) variations, we conduct a variation of the $\chi^2$ test that can be applied to circular data. For this we follow methods described in  \citet{cit.oai.edge.caltech.folio.ebsco.com.fs00001057.7399d9f8.d802.4ce0.8701.5afe7b79117220000101}, which we refer to as MJ2000 \citep[see also][and NCSS Technical Procedures Documentation\footnote{https://www.ncss.com/wp-content/themes/ncss/pdf/Procedures/NCSS/Circular\_Data\_Analysis.pdf}]{cit.oai.edge.caltech.folio.ebsco.com.fs00001057.73b39503.802d.4a13.939f.ea01212e06c519720101, edssjs.DCE18EFE19720301, bc1d15dd-524c-38f4-a83a-d55cfe9db0a2, 6b74fb8c-2780-33df-866f-0b025406bc3f}. To determine if the PPA is constant, we assume a null hypothesis (H0) that the PPA is distributed as a von Mises random variable with mean angle $\hat{\mu}$ and concentration parameter $\hat{\kappa}$. For each FRB, we estimate $\hat{\mu}$ from the set of PPA $\{\chi_{0,i}\}$ in time bins with linear S/N $>3\sigma$ using equations 2.2.1-2.2.4 in MJ2000:

\begin{equation}
    \hat{\mu} = \begin{cases}
        {\rm tan}^{-1}(\frac{\bar{S}}{\bar{C}}), & \text{for } \bar{C} \ge 0\\
        {\rm tan}^{-1}(\frac{\bar{S}}{\bar{C}}) + \pi, & \text{for } \bar{C} < 0
        \end{cases}
\end{equation}

\noindent where $\bar{C}$ and $\bar{S}$ are defined as:

\begin{equation}
    \bar{C} = \frac{\sum{\sigma_{\chi_{0,i}}^{-2}{\rm cos}(\chi_{0,i})}}{\sum{\sigma_{\chi_{0,i}}^{-2}}}
\end{equation}

\begin{equation}
    \bar{S} = \frac{\sum{\sigma_{\chi_{0,i}}^{-2}{\rm sin}(\chi_{0,i})}}{\sum{\sigma_{\chi_{0,i}}^{-2}}}
\end{equation}

\noindent where $\{\sigma_{\chi_{0,i}}\}$ are the uncertainties on each PA sample. From MJ2000 equation 5.3.5, the maximum likelihood estimate of $\hat{\kappa}$ is defined by the condition:

\begin{equation}
    \frac{I_0(\hat{\kappa})}{I_1(\hat{\kappa})} = \bar{R}
\end{equation}

\noindent where $\bar{R} = \sqrt{\bar{C}^2 + \bar{S}^2}$ and $I_d$ is the $d^{\rm th}$ order modified Bessel function of the first kind. We use a spline linear interpolation to numerically solve for $\hat{\kappa}$. Finally the Score test statistic, which we refer to as $\chi^2_{\angle}$, is given by:

\begin{equation}
    \chi^2_{\angle} = \begin{cases}
        2N\hat{\kappa}(\bar{R}_{\hat{\mu}}-\bar{C}_{\hat{\mu}})(1 - (4N\hat{\kappa}\frac{I_0(\hat{\kappa})}{I_1(\hat{\kappa})})^{-1}), & \text{for } N < 5\\
        4N\frac{\bar{R}_{\hat{\mu}}^2 - \bar{C}_{\hat{\mu}}^2}{2 - \bar{C}^2}, & \text{for } N \ge 5, \bar{C}_{\hat{\mu}} \le 2/3\\
        \frac{2N}{1 + \bar{C}_{\hat{\mu}}^2 + 3n^{-1}}{\rm ln}(\frac{1-\bar{C}^2}{1-\bar{R}^2}), & \text{for } N \ge 5, \bar{C}_{\hat{\mu}} > 2/3\\
        \end{cases}
\end{equation}

\noindent where N is the number of PPA samples and we define:

\begin{equation}
    \bar{C}_{\hat{\mu}} = N^{-1}\sum{{\rm cos}(\chi_{0,i} - \hat{\mu})}
\end{equation}

\begin{equation}
    \bar{S}_{\hat{\mu}} = N^{-1}\sum{{\rm sin}(\chi_{0,i} - \hat{\mu})}
\end{equation}

\begin{equation}
    \bar{R}_{\hat{\mu}} = \sqrt{\bar{S}_{\hat{\mu}}^2 + \bar{C}_{\hat{\mu}}^2}
\end{equation}

\noindent Through Monte Carlo simulations, we verify that $\chi^2_{\angle}$ follows a $\chi^2$ distribution with one degree of freedom, and use this to compute a p-value for each FRB's $\chi^2_{\angle}$ estimate.

\begin{figure*}[ht]
    \centering
    \includegraphics[width=0.67\linewidth]{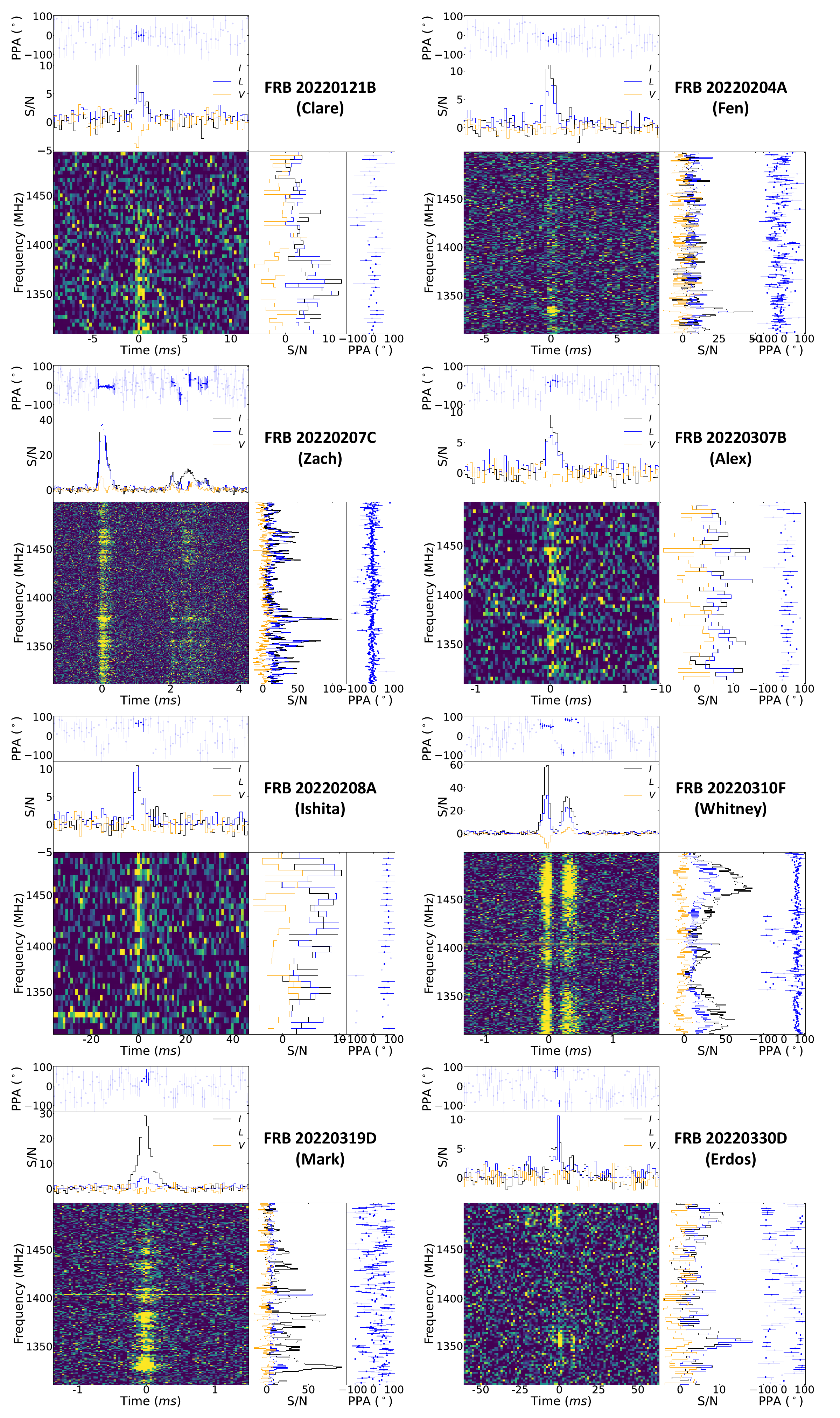} \\
    \caption{Summary of polarization data for each DSA-110 FRB. For each FRB, from top, we show the PPA (with $1\sigma$ errorbars) measured after correcting for RM, and frequency averaged time series of Stokes $I$, $L$, and $V$. The PPA is highlighted in bins with linear-polarization S/N $>3\sigma$. The dynamic spectrum is shown at center, to the right of which are shown optimally summed spectra (Appendix~\ref{app_weights}) and the spectrally resolved PPA with $>3\sigma$ bins highlighted. For FRBs with no measured RM, the PA without RM correction, rather than the PPA is shown.}
    \label{fig:FRBStokes1}
\end{figure*}

\begin{figure*}[ht]
\begin{center}
\begin{tabular}{c}
\includegraphics[width=0.67\linewidth]{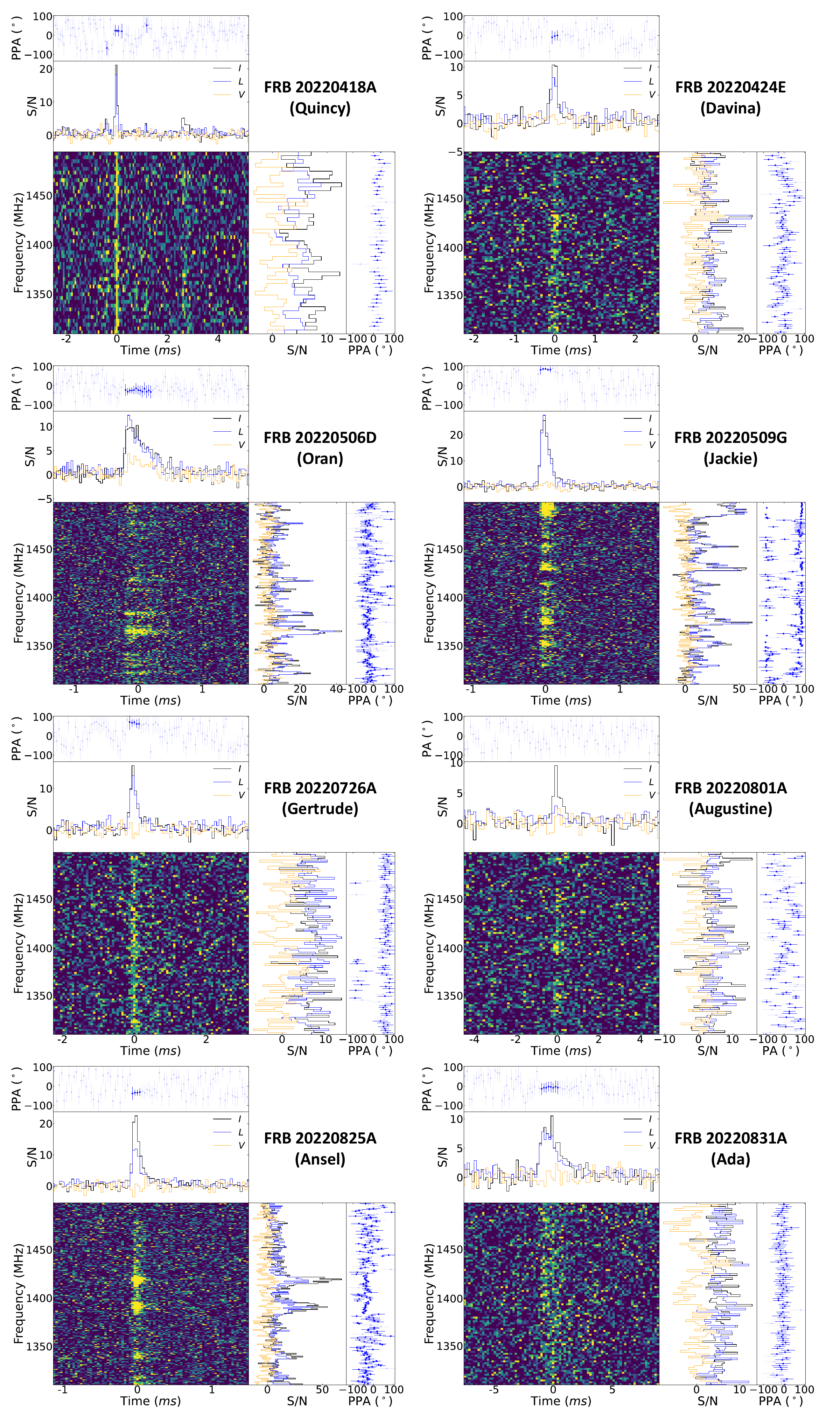} \\
\end{tabular}
\end{center}
\caption{Same as Figure~\ref{fig:FRBStokes1}}
\label{fig:FRBStokes2}
\end{figure*}

\begin{figure*}[ht]
\begin{center}
\begin{tabular}{c}
\includegraphics[width=0.67\linewidth]{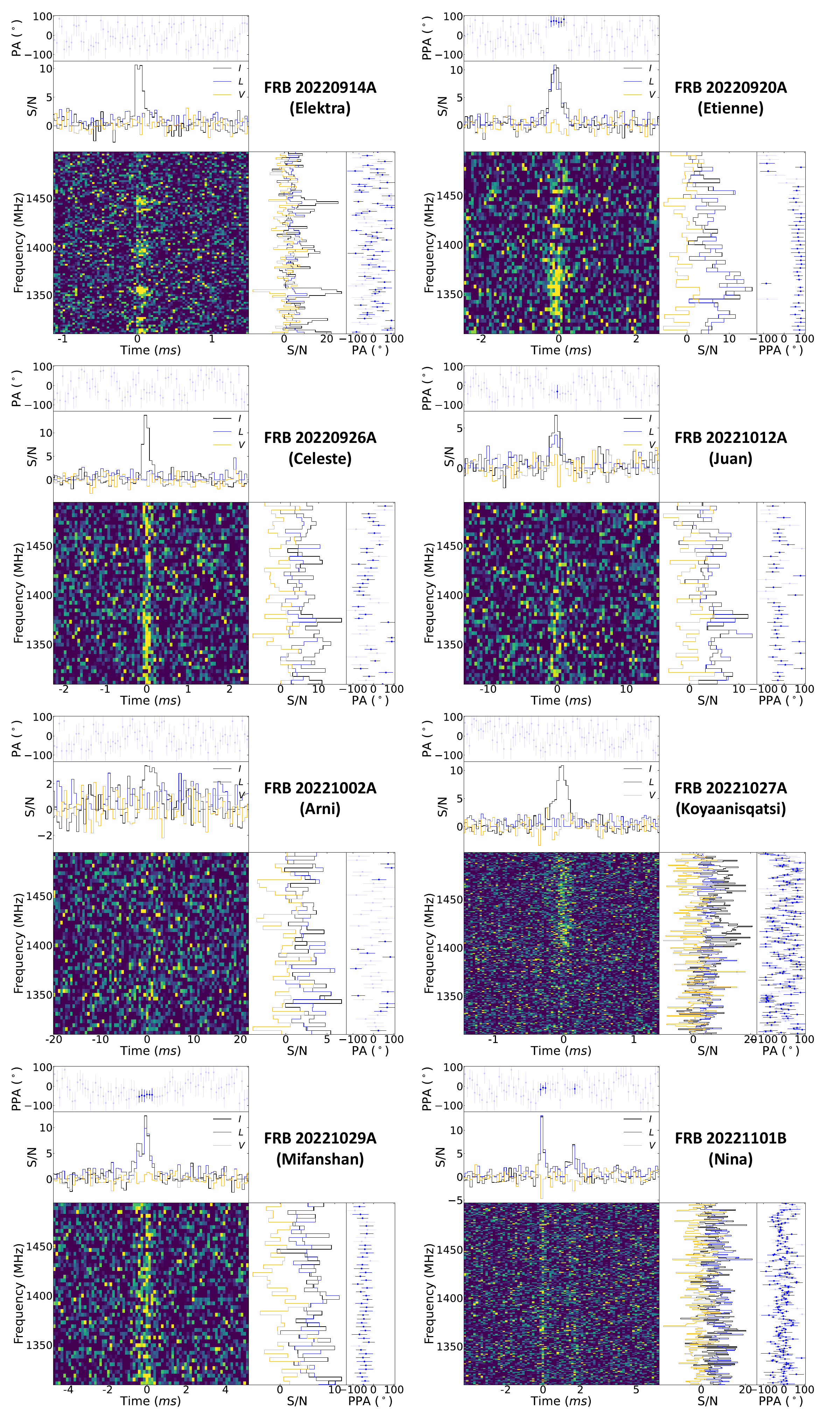} \\
\end{tabular}
\end{center}
\caption{Same as Figure~\ref{fig:FRBStokes1}}
\label{fig:FRBStokes3}
\end{figure*}

\begin{figure*}[ht]
\begin{center}
\begin{tabular}{c}
\includegraphics[width=0.67\linewidth]{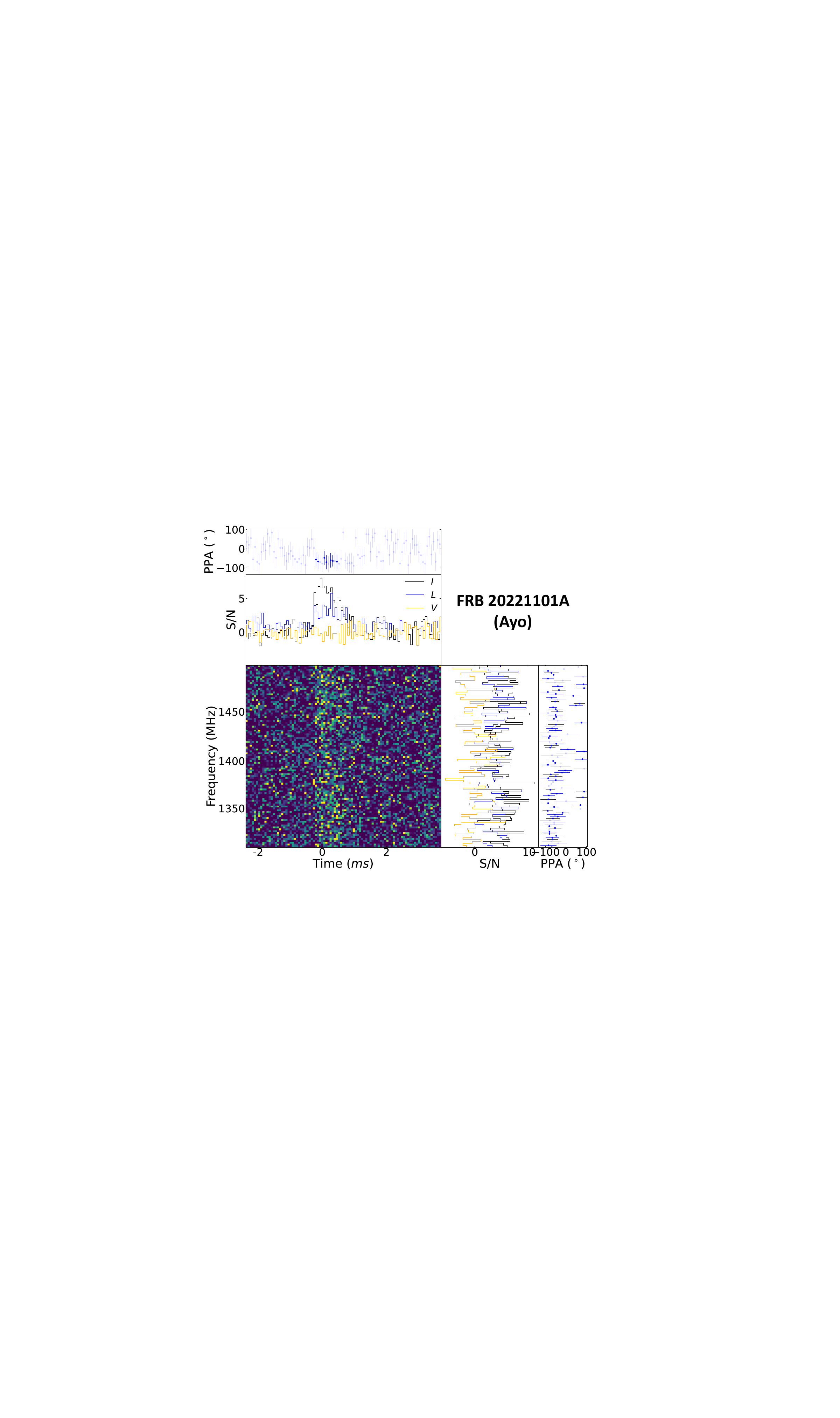} \\
\end{tabular}
\end{center}
\caption{Same as Figure~\ref{fig:FRBStokes1}}
\label{fig:FRBStokes4}
\end{figure*}

After computing $\chi^2_{\angle}$ for each FRB, we find no significant PPA variations at the $90\%$ level or beyond. From visual inspection of some bursts, for example FRB\,20220207C and FRB\,20220310F (see Figures~\ref{fig:FRBStokes1}-\ref{fig:FRBStokes4}), that show evidence of variation without being identified as significant, we infer that the large PPA uncertainties and small N prevent rejection of the null hypothesis. We follow-up this analysis by computing the maximum difference in PPA across each FRB and its signficance, following a similar method to \citet[][(Appendix C)]{jiang2022fast}. With this we find a maximum $3.8\sigma$ PPA change across FRB\,20220310F, which supports the mode change discussed in Section~\ref{discussion}. A less significant $2\sigma$ PPA change is observeed for FRB\,20220207C, supporting the $\chi^2_\angle$ result that its PPA variation is not significant.

\section{Summary Plots for 25 DSA-110 FRBs}\label{app_data}

\begin{figure*}[ht]
\begin{center}
\begin{tabular}{c}
\includegraphics[width=0.77\linewidth]{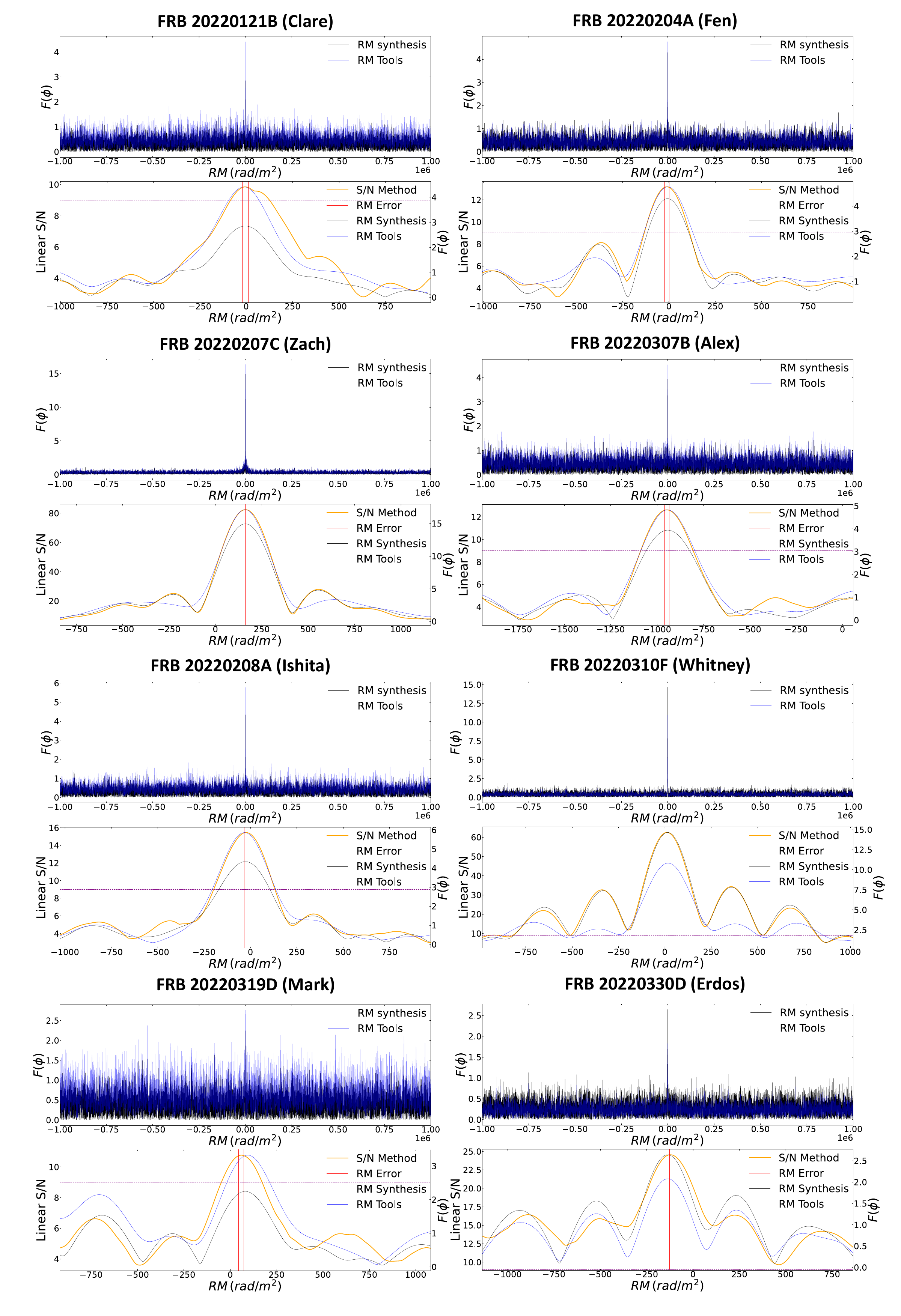} \\
\end{tabular}
\end{center}
\caption{Summary of RM data on each FRB with significant detection. For each FRB, the top panel shows the RM spectra evaluated over the full range of sensitivity using two different methods (Appendix~\ref{app_RMderivation}), and the bottom panel shows a detailed analysis of the S/N of the peak. The horizontal purple line indicates the $9\sigma$ significance threshold applied to the S/N Method spectrum (yellow).}
\label{fig:FRBRMs1}
\end{figure*}

\begin{figure*}[ht]
\begin{center}
\begin{tabular}{c}
\includegraphics[width=0.77\linewidth]{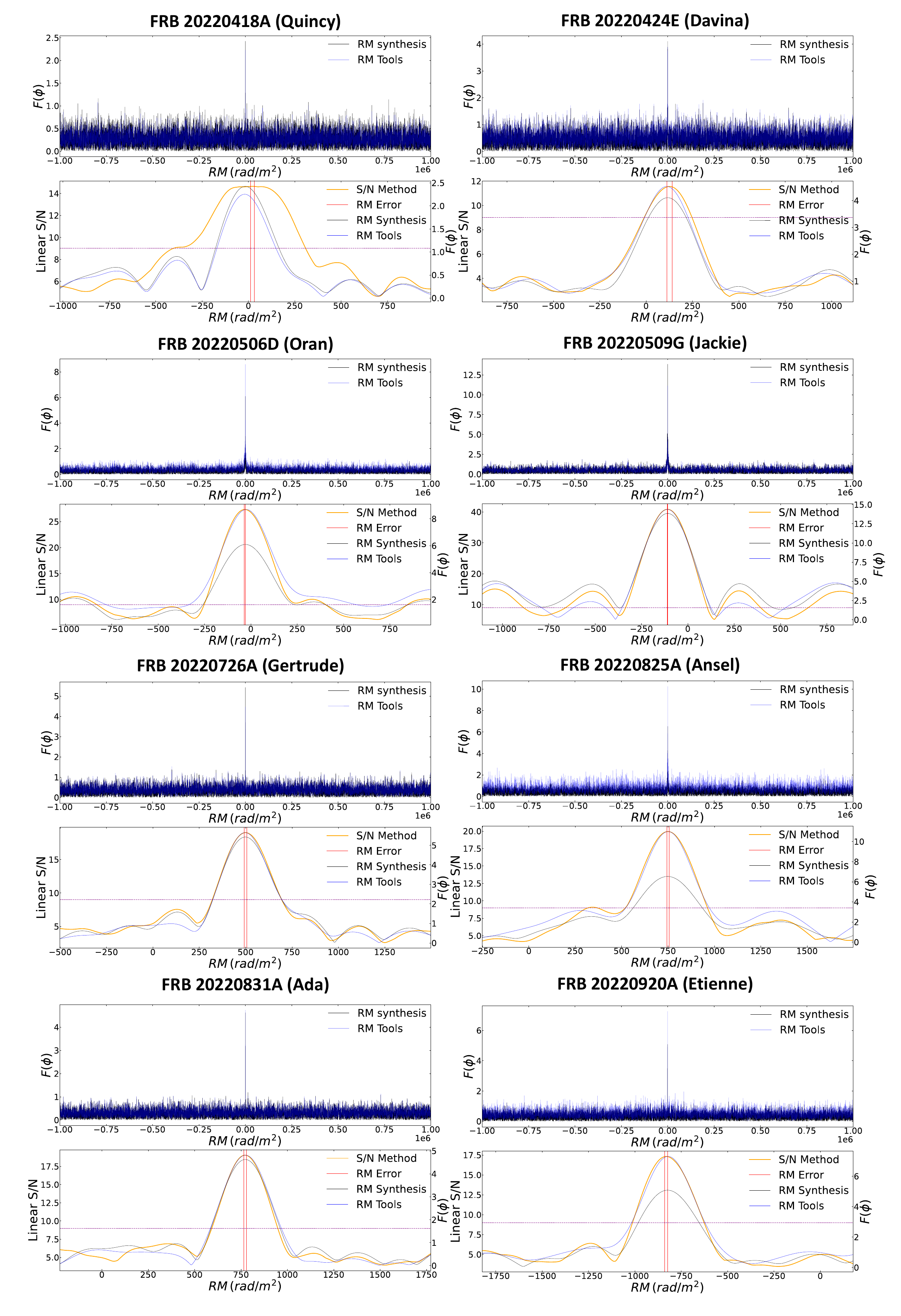} \\
\end{tabular}
\end{center}
\caption{Same as Figure~\ref{fig:FRBRMs1}}
\label{fig:FRBRMs2}
\end{figure*}

\begin{figure*}[ht]
\begin{center}
\begin{tabular}{c}
\includegraphics[width=0.77\linewidth]{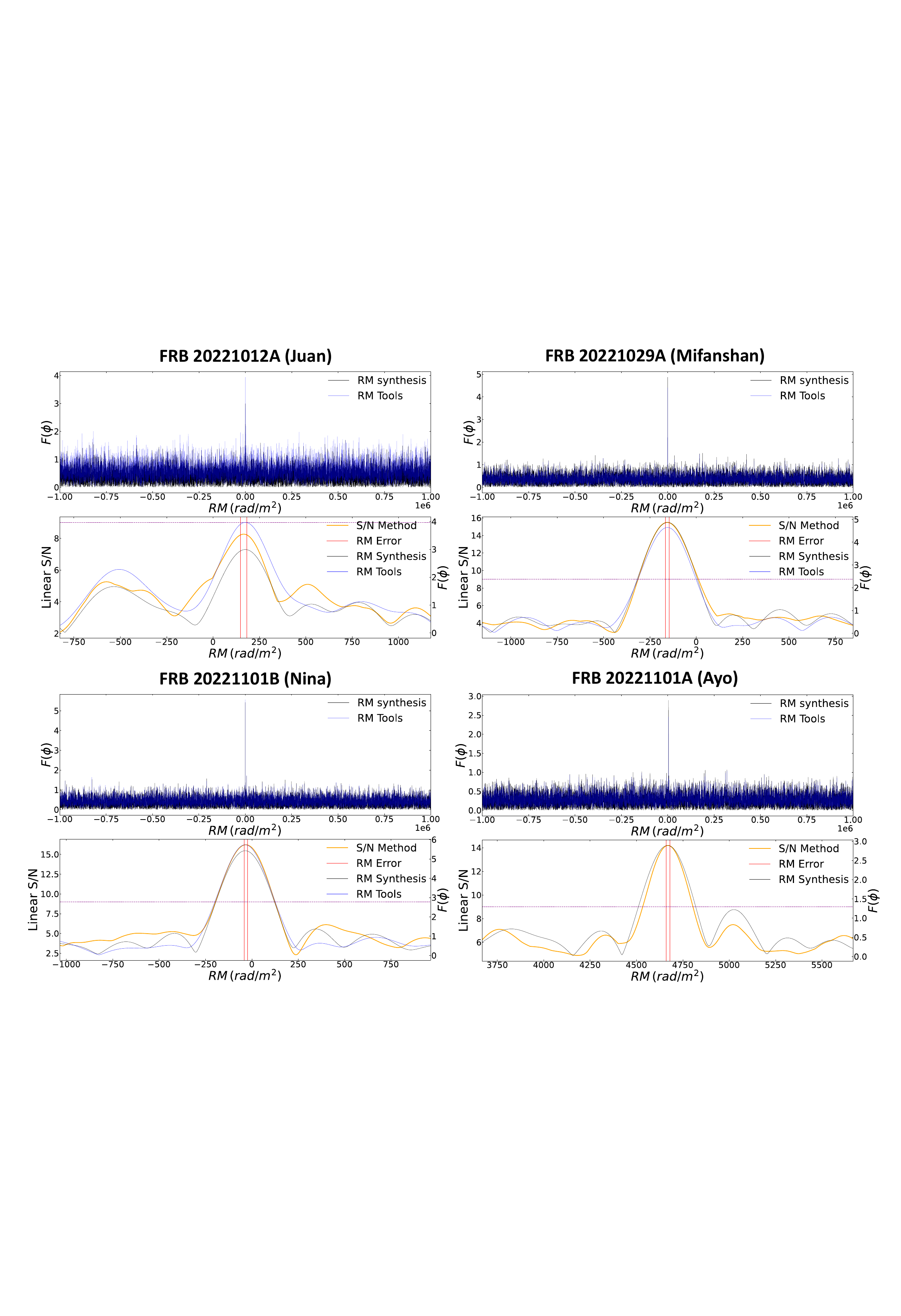} \\
\end{tabular}
\end{center}
\caption{Same as Figure~\ref{fig:FRBRMs1}}
\label{fig:FRBRMs3}
\end{figure*}

\section{RM Time Evolution Plots for Multicomponent FRBs}\label{app_rmtime}

\begin{figure*}[ht]
\begin{center}
\begin{tabular}{c}
\includegraphics[width=1\linewidth]{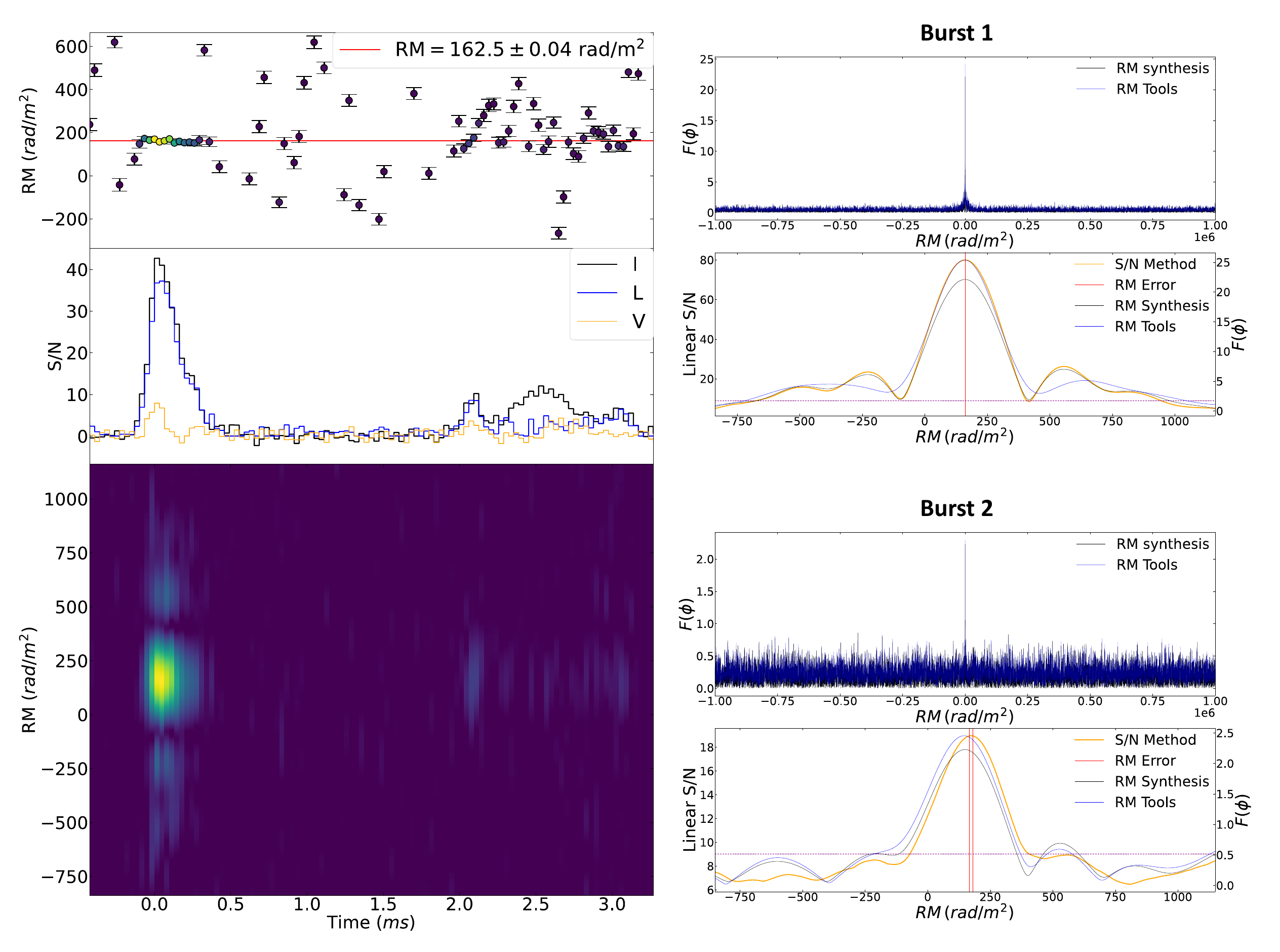} \\
\end{tabular}
\end{center}
\caption{Time evolution of RM in FRB\,20220207C; 
\textit{Right:} The top plot shows the measured RM from the S/N method for each time bin; the polarized profile showing Stokes I (black), linear polarization (blue) and circular polarization (orange) is shown in the middle plot. The bottom plot shows the RM spectrum in each time bin, referred to as the Rotation Measure Transform. In the top plot, the color of each point corresponds to the peak S/N in the RM spectrum for that time bin, and correlates with the colormap in the RM transform. The red line is drawn at the average RM value from both sub-components, $162.48 \pm 0.04\,$rad\,m$^{-2}$. \textit{Right:} The RM synthesis summary plots are shown for sub-bursts 1 
(top) and 2 (bottom) individually. See the caption for Figure~\ref{fig:FRBRMs1} for a description of each trace in the RM synthesis plots.}
\label{fig:rmtime1}
\end{figure*}

\begin{figure*}[ht]
\begin{center}
\begin{tabular}{c}
\includegraphics[width=1\linewidth]{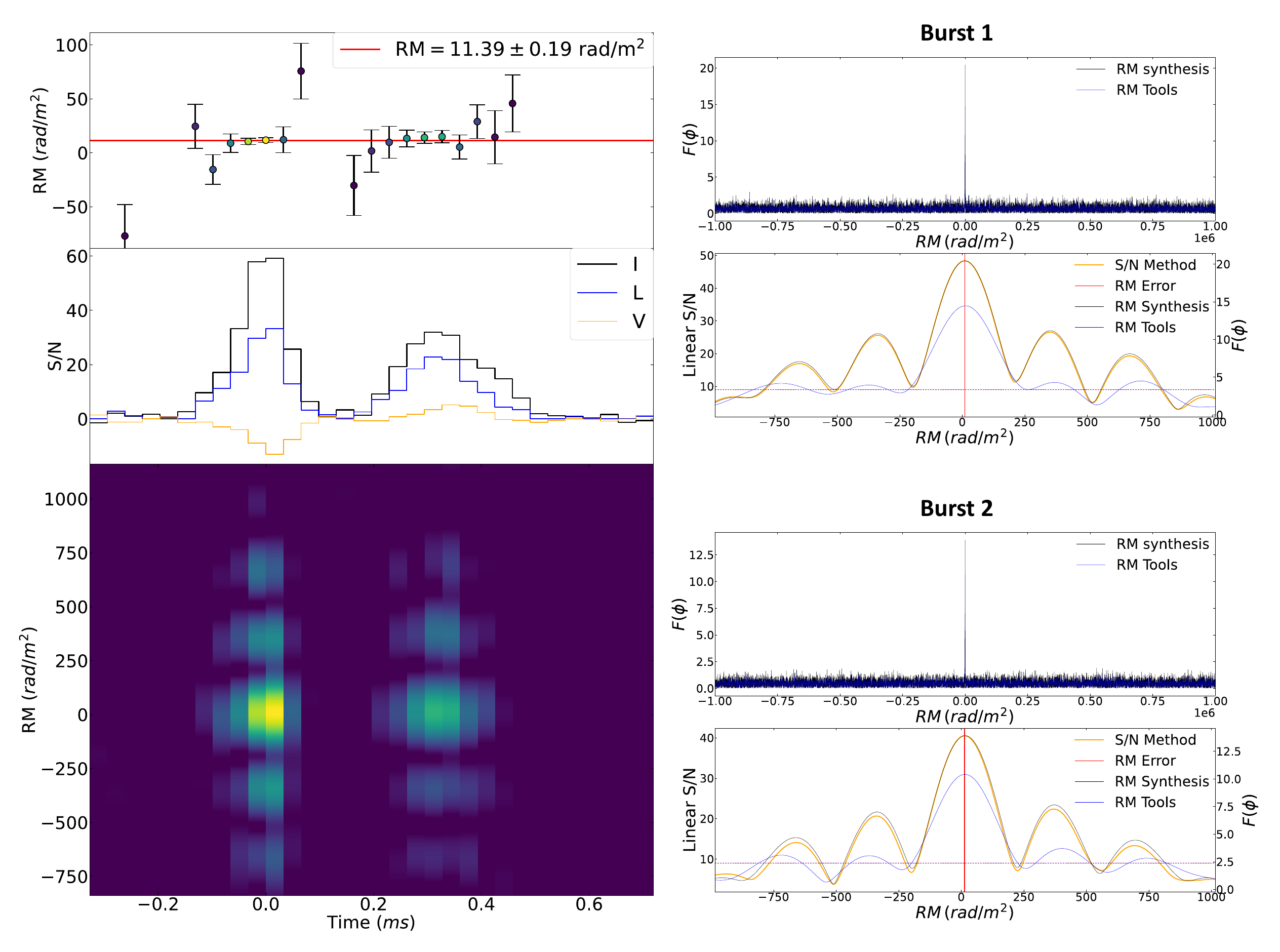} \\
\end{tabular}
\end{center}
\caption{Same as Figure~\ref{fig:rmtime1} for FRB\,20220310F. The red line is drawn at the average RM value from both sub-components, $11.4 \pm 0.2\,$rad\,m$^{-2}$.}
\label{fig:rmtime2}
\end{figure*}

\begin{figure*}[ht]
\begin{center}
\begin{tabular}{c}
\includegraphics[width=1\linewidth]{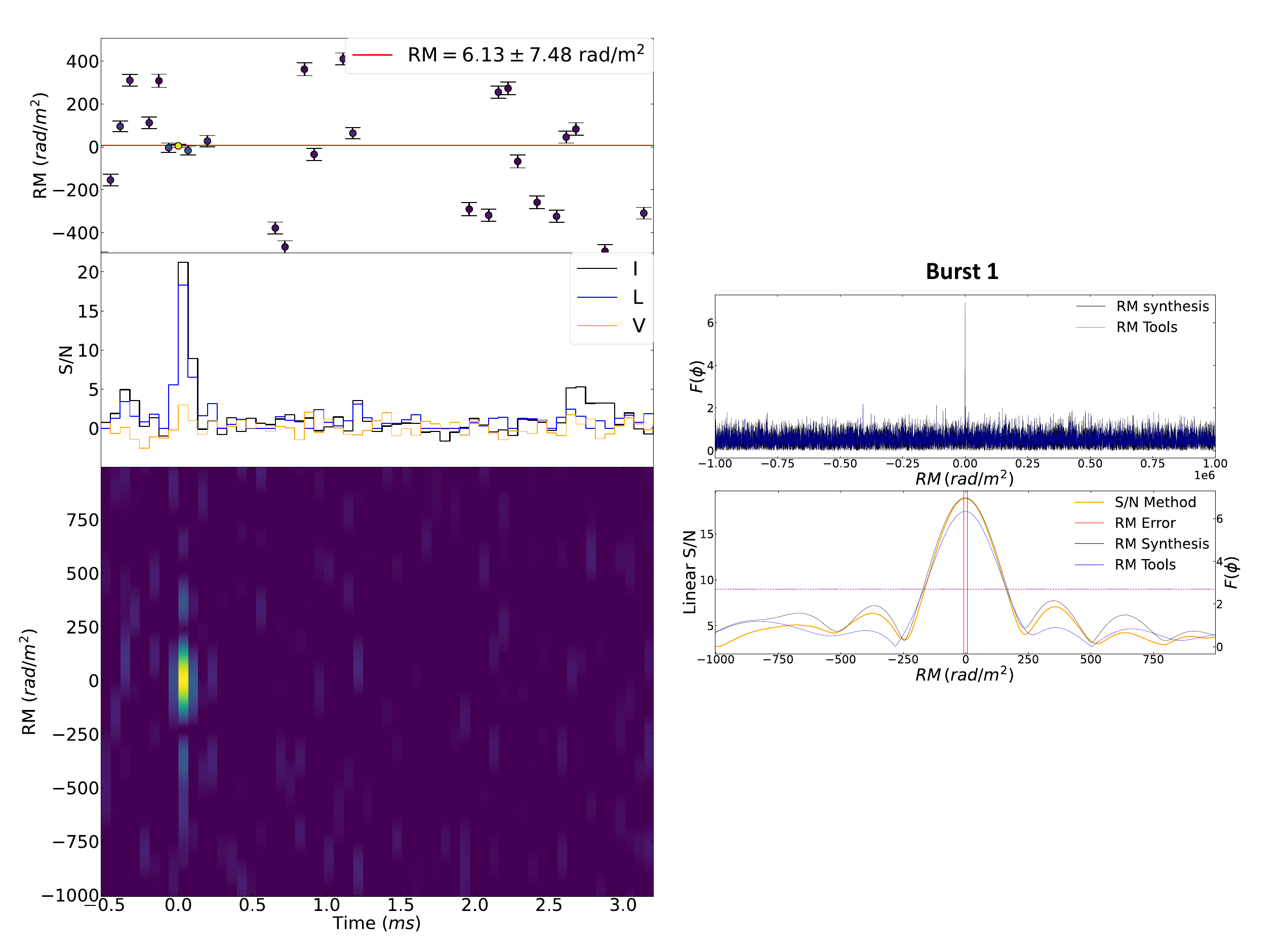} \\ 
\end{tabular}
\end{center}
\caption{Same as Figure~\ref{fig:rmtime1} for FRB\,20220418A. The red line is drawn at the average RM value from all sub-components, $6.1 \pm 7.5\,$rad\,m$^{-2}$. For the individual components, a significant RM is found only for the first.}
\label{fig:rmtime3}
\end{figure*}

\begin{figure*}[ht]
\begin{center}
\begin{tabular}{c}
\includegraphics[width=1\linewidth]{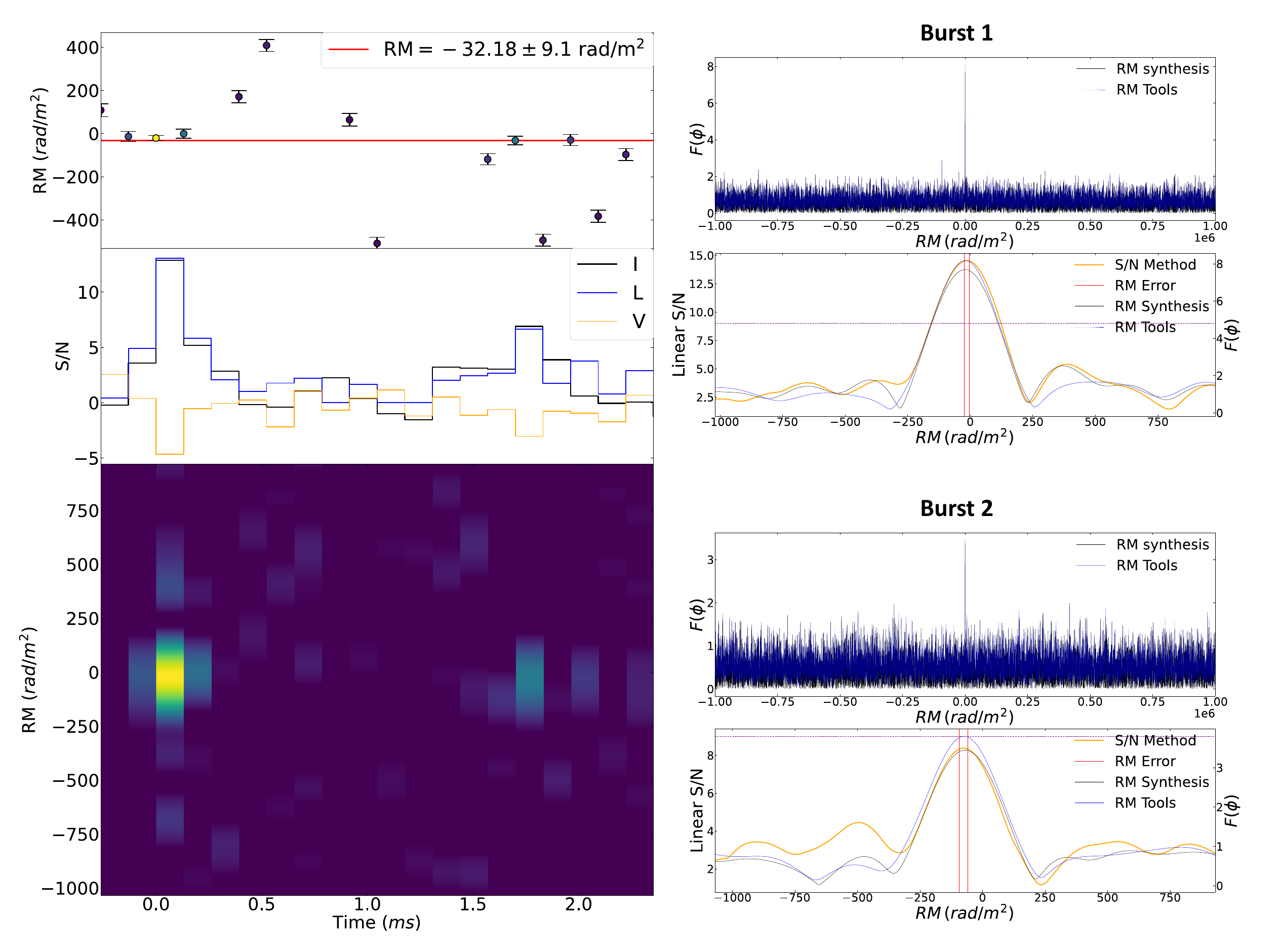} \\ 
\end{tabular}
\end{center}
\caption{Same as Figure~\ref{fig:rmtime1} for FRB\,20221101B. The red line is drawn at the average RM value from both sub-components, $-13.1 \pm 10.5\,$rad\,m$^{-2}$.}
\label{fig:rmtime4}
\end{figure*}

\end{document}